\documentclass[10pt,twocolumn,aps,pra,balancelastpage,longbibliography,eqsecnum]{revtex4-1}
\usepackage[latin9]{inputenc}
\usepackage{color}
\usepackage{verbatim}
\usepackage{amsmath}
\usepackage{manfnt}
\usepackage{amssymb,ulem}
\usepackage{graphicx,cellspace,booktabs}
\usepackage{esint}
\usepackage[unicode=true,pdfusetitle,
 bookmarks=true,bookmarksnumbered=false,bookmarksopen=false,
 breaklinks=false,pdfborder={0 0 1},backref=false,colorlinks=true,urlcolor=blue,citecolor=blue,linkcolor=blue]{hyperref}
\usepackage{breakurl}
\usepackage{epstopdf}
\usepackage{feynmp-auto}
\usepackage{youngtab}
\usepackage{ytableau}
\usepackage[caption=false]{subfig}

\makeatletter

\@ifundefined{textcolor}{}
{%
 \definecolor{BLACK}{gray}{0}
 \definecolor{WHITE}{gray}{1}
 \definecolor{RED}{rgb}{1,0,0}
 \definecolor{GREEN}{rgb}{0,1,0}
 \definecolor{BLUE}{rgb}{0,0,1}
 \definecolor{CYAN}{cmyk}{1,0,0,0}
 \definecolor{MAGENTA}{cmyk}{0,1,0,0}
 \definecolor{YELLOW}{cmyk}{0,0,1,0}
}

\usepackage{bm}

\newcommand{\ket}[1]{|#1\rangle}

\newcommand{\bv}[1]{\mathbf{#1}}

\newcommand{\tabimage}[2]{\raisebox{-.5\height}{\includegraphics[width=#2]{#1}}}

\newcommand{\cube}{\mbox{\mancube}}

\def\l@subsubsection#1#2{}
\makeatother

\begin{document}

\title{Braiding and Gapped Boundaries in Fracton Topological Phases}

\author{Daniel Bulmash}
\author{Thomas Iadecola}

\affiliation{Condensed Matter Theory Center and Joint Quantum Institute, Department of Physics, University of Maryland, College Park, Maryland 20472 USA}

\date{\today}
\begin{abstract}
We study gapped boundaries of Abelian type-I fracton systems in three spatial dimensions. Using the X-cube model as our motivating example, we give a conjecture, with partial proof, of the conditions for a boundary to be gapped. In order to state our conjecture, we use a precise definition of fracton braiding and show that bulk braiding of fractons has several features that make it \textit{insufficient} to classify gapped boundaries.  Most notable among these is that bulk braiding is sensitive to geometry and is ``nonreciprocal," that is, braiding an excitation $a$ around $b$ need not yield the same phase as braiding $b$ around $a$. Instead, we define fractonic ``boundary braiding," which resolves these difficulties in the presence of a boundary. We then conjecture that a boundary of an Abelian fracton system is gapped if and only if a ``boundary Lagrangian subgroup" of excitations is condensed at the boundary; this is a generalization of the condition for a gapped boundary in two spatial dimensions, but it relies on boundary braiding instead of bulk braiding. We also discuss the distinctness of gapped boundaries and transitions between different topological orders on gapped boundaries.
\end{abstract}
\maketitle

\section{Introduction}

Topological states of matter are often characterized by nontrivial physics on their boundaries.  For instance, the first experimentally realized example of topological order in two spatial dimensions [(2+1)D], namely the fractional quantum Hall effect, is characterized in the simplest cases by gapless boundary modes that carry fractional charge.  However, in many cases it is possible for topologically ordered phases to support gapped boundary states of various types~\cite{BeigiQuantumDoubleBoundary,BravyiSurfaceCodes,KitaevKong}.  For 2+1D topological phases, the physics of anyon condensation~\cite{BaisAnyonCondensation} and the bulk-boundary correspondence can be used to fully classify such gapped boundaries, using Lagrangian subgroups~\cite{KapustinSaulina,LevinGappedBoundaries,BarkeshliLagrangianSubgroups} for Abelian topological order and more generally Lagrangian algebra objects~\cite{KongAnyonCondensation,LanGappedDomainWalls,Hung2015,Hu2018}. Some progress has been made along these lines in (3+1)D as well~\cite{KongBoundaryBulk,Wang3DBoundaries}. Gapped boundaries are intrinsically interesting in this context for several reasons---for example, they play a key role in quantum information applications such as surface codes~\cite{FreedmanPlanarCodes,BravyiSurfaceCodes,DennisQuantumMemory}, and they can be used to understand defects in the underlying topological order~\cite{BarkeshliLagrangianSubgroups}.

In (3+1)D, a new generalization of topological order, termed ``fracton order," has been recently discovered and is under intense study\cite{ChamonGlass,CastelnovoFirstXCubePaper, BravyiChamonModel, HaahsCode, YoshidaFractal, VijayFractons, NandkishoreFractonReview, VijayGaugedSubsystem, RegnaultLayer, VijayNonAbelianFractons, SlagleGenericLattices, ShirleyXCubeFoliations, SlagleFieldTheory, SongTwisted, MaLayer, VijayLayer, PremCageNet, FractonCorrFunctions, ShiFractonEntanglement, FractonEntanglement, ShirleyEntanglement, RecoverableInformation, Williamson2016, HalaszFractons, HsiehFractonsPartons, EmergentPhasesFractons, YouSymmetricFracton, BravyiHaahSelfCorrection, SivaBravyiMemory, PremDynamics, ShirleyFoliatedGauge, ShirleyFoliatedCheckerboard, ShirleyFoliatedFractional}. 
In these gapped systems, the excitations are ``subdimensional," in that isolated pointlike quasiparticles can only move without energy cost in a lower-dimensional submanifold of the system. For example, such models can support two-dimensional (2D) excitations that are mobile along planar subsystems, one-dimensional (1D) excitations that are mobile only along lines, and zero-dimensional (0D) excitations, or ``fractons," that are completely immobile in isolation. 

In this paper we use the term ``fracton order" to refer to gapped systems and exclude gapless generalized $U(1)$ gauge theories~\cite{XuFractons1, XuFractons2, RasmussenFractons, PretkoSubdimensional, PretkoElectromagnetism, BulmashHiggs, MaHiggs, BulmashGeneralizedGauge}, despite the fact that these bear some relation to the gapped systems~\cite{BulmashHiggs, MaHiggs}. With that definition, in the language of Ref.~\cite{VijayFractons}, there are two different classes of fracton systems: ``type-I" fracton systems have subdimensional but mobile point-like excitations, whereas all point-like excitations are immobile in ``type-II" fracton systems. Furthermore, fracton order seems to support generalizations of some concepts from standard topological order, such as braiding~\cite{MaLayer,SongTwisted,SlagleFieldTheory} and a topology-dependent ground state degeneracy. However, in fracton order, these concepts also depend on the \textit{geometry} of the underlying spatial manifold~\cite{ShirleyXCubeFoliations,SlagleFieldTheory,SlagleGenericLattices}; for example, their ground state manifold on an $L\times L \times L$ three-torus typically encodes a subextensive ($\sim L$) number of qubits.

Recent work has greatly improved our understanding of fracton order, but there is not yet a description of their properties that is known to be generic, even for type-I phases (which seem considerably easier to study). As a result, our work essentially reverses the approach taken in (2+1)D topological order---by studying gapped boundaries, we hope to provide clues to the important physical properties that characterize fracton systems, in particular to understand the physics of fracton condensation and fracton braiding. Gapped boundaries of fracton systems may also be of interest for quantum information purposes.

In this paper, our main result is a conjecture about which gapped boundary states are allowed in Abelian type-I fracton models.  By studying the X-cube model~\cite{CastelnovoFirstXCubePaper,VijayGaugedSubsystem} as our primary example, we show that the most natural definition of bulk braiding for subdimensional particles is \textit{insufficient} to fully characterize the various possible gapped boundaries. Instead, we introduce a notion of ``boundary braiding" for fractons that we conjecture is sufficient to determine the allowed gapped boundaries (modulo standard anyon condensation in the surface topological order). In the process, we develop a physical understanding of gapped boundaries in fracton models in terms of quasiparticle condensation. This leads to the curious property that particles can change mobility at the boundary---for example, particles that are immobile in the bulk can become 2D particles at the boundary. We also show that bulk braiding as it has been discussed thus far in the fracton literature suffers from several undesirable properties, but that this is not the case for boundary braiding.

The paper is organized as follows. We begin in Sec.~\ref{sec:2DReview} with a review of the classification of gapped boundaries in (2+1)D topological order. Sec.~\ref{sec:bulkXCube} reviews the X-cube model and studies its bulk braiding properties, illustrating the non-reciprocity and geometry-dependence of bulk braiding. In Sec.~\ref{sec:XCubeBoundaries}, we examine some example boundaries of the X-cube model. We show that nontrivial excitations condense at gapped boundaries and explain how this leads to changes in the mobility of uncondensed excitations at the surface. We also show that some gapped boundaries can be characterized using (2+1)D topological orders with some additional structure.  In Sec.~\ref{sec: boundary braiding}, we define the notion of boundary braiding and provide some examples of it in the X-cube model. We then promote in Sec.~\ref{sec: Boundary Lagrangian Subgroup Conjecture} the intuition gained from the X-cube model to a conjecture of the desired form. In particular, we partially prove a conjecture that a gapped boundary exists if and only if there is a ``boundary Lagrangian subgroup," which defines the set of excitations that can be condensed at a given boundary and includes information about the subdimensional nature of these excitations. In the process, we show how our conjecture explains the phenomenology of the X-cube model. We also argue that boundary Lagrangian subgroups provide a full characterization of the possible gapped boundaries of a fracton phase modulo the condensation of excitations that are mobile only on the boundary.  In other words, any two gapped boundaries with the same boundary Lagrangian subgroup can be related to one another by such a purely (2+1)D condensation process, which we discuss in detail in Sec.~\ref{sec: Surface Condensation}. We conclude with a discussion of open questions and the prospects for including type-II fracton models in our framework.

\section{Review of Gapped Boundaries in (2+1)D Abelian topological phases}
\label{sec:2DReview}

The bulk-boundary correspondence for topological order in (2+1)D suggests that the properties of bulk anyons should be sufficient to classify gapped boundaries. In fact, anyons are condensed at gapped boundaries; here and in the remainder of this paper, we use the term ``condensed" to mean that a single excitation can be created by a local unitary operator at the boundary. For Abelian anyons,
Refs.~\cite{LevinGappedBoundaries,BarkeshliLagrangianSubgroups} show that a gapped boundary of a topological phase exists if and only if the set $S$ of anyons condensed at the boundary is a ``Lagrangian subgroup," that is,
\begin{enumerate}
\item $S$ is closed under fusion
\item All anyons in $S$ braid trivially with each other
\item Every anyon not in $S$ braids nontrivially with some anyon in $S$
\end{enumerate}
The proof of condition (2) is important for the development of the fractonic analogue, so we review it now along with the intuition for these statements; see Refs.~\cite{LevinGappedBoundaries,BarkeshliLagrangianSubgroups} for the rest of the details. We will use the standard notation whereby $\bar a$ denotes the antiparticle of the anyon $a$.

Condition (1) is essentially obvious: if local operators create anyons $a$ and $b$ at nearby points on the boundary, then the product of those operators is a local operator which creates the fusion product $a \times b$ at the boundary.  

\begin{figure}
\centering
\subfloat[\label{fig: 2D boundary a}]{\includegraphics[width=.85\columnwidth]{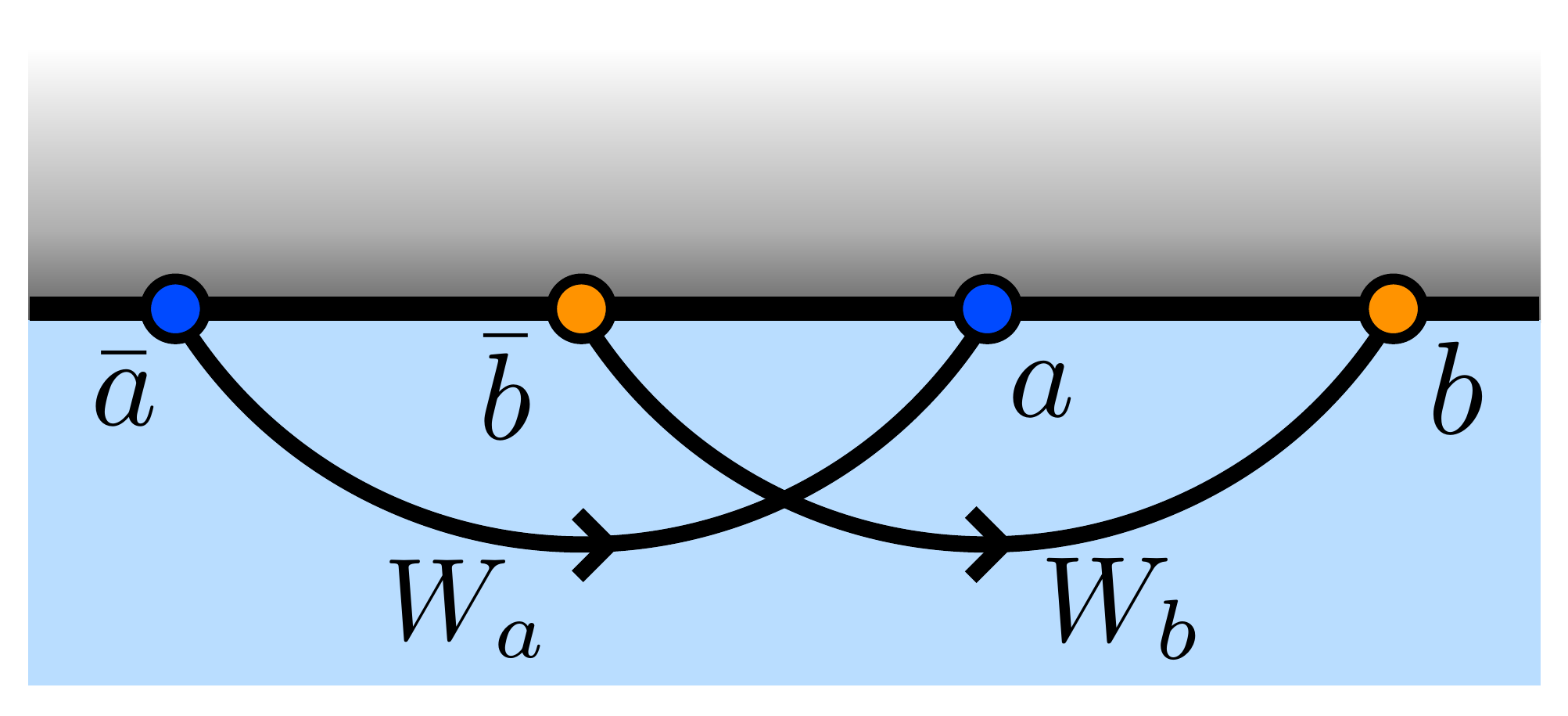}}\\
\vspace{3mm}
\subfloat[\label{fig: 2D boundary b}]{\includegraphics[width=.85\columnwidth]{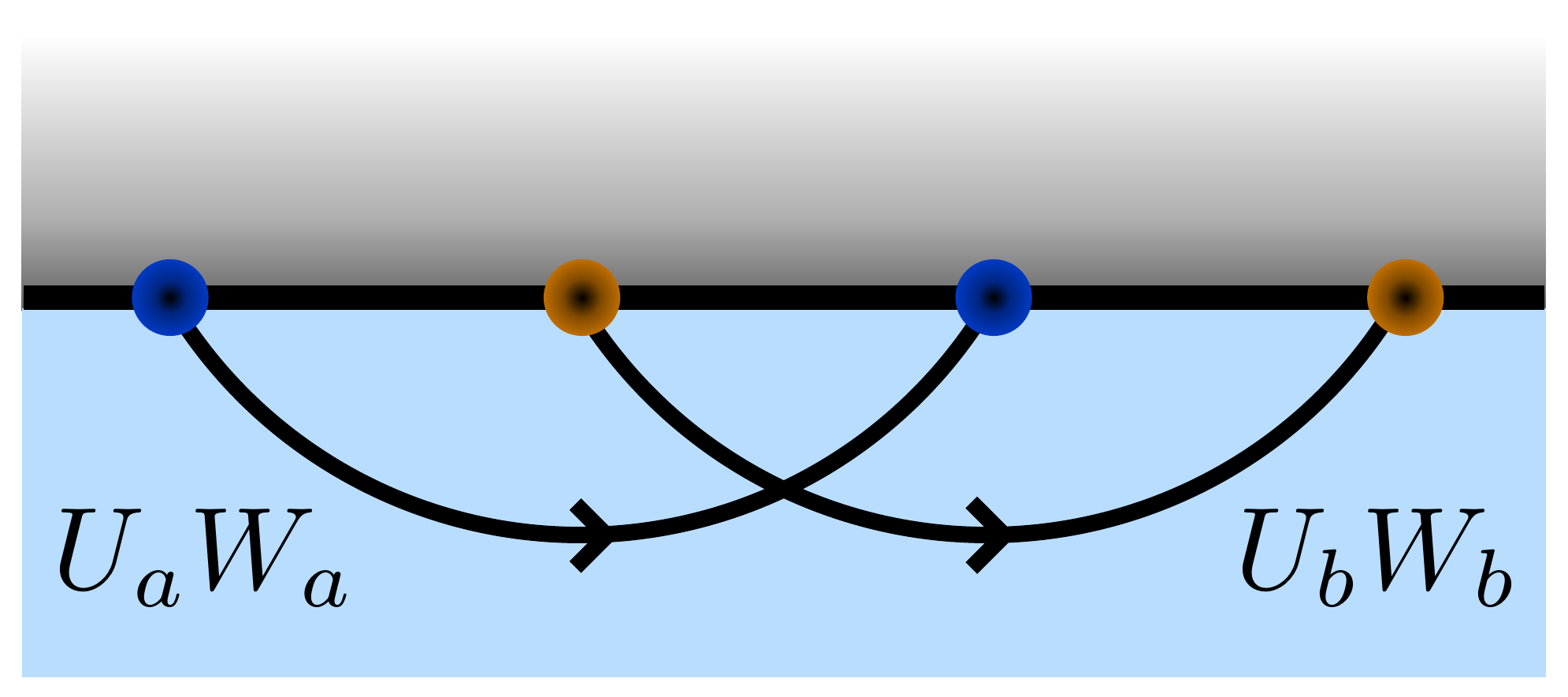}}
\caption{Braiding anyons $a$ and $b$ at a gapped boundary.  (a) A pair of open Wilson loops for $a$ and $b$ that leave $a$, $\bar a$ and $b$, $\bar b$ excitations on the boundary.  (b) If $a$ and $b$ are condensed at the boundary, one can apply operators $U_a$ and $U_{b}$ that delete the $a$, $\bar a$ and $b$, $\bar b$ excitations at the boundary and have support only in the vicinity of these excitations.}
\label{fig: 2D boundary}
\end{figure}

Condition (3) says that the Lagrangian subgroup is in some sense ``maximal." The intuition is simply that any nontrivial uncondensed anyon $a$ at the boundary should be detectable by an interference (braiding) experiment. Since uncondensed excitations are unable to encircle $a$ due to the presence of the boundary, the only way to braid with a particle $a$ at the boundary is to locally create a condensed anyon $b$ at the boundary, move into the bulk and around $a$, then bring it back to the boundary and annihilate it locally, as shown in Fig.~\ref{fig: 2D boundary a}. Some such braiding process must yield a nontrivial phase in order for $a$ to be detectable by braiding. 

Condition (2) motivates much of the formal part of this paper. The proof that this condition is necessary is fairly simple; we review Levin's proof \cite{LevinGappedBoundaries} here. Suppose that two anyons $a$ and $b$ are condensed at the boundary. Let $W_a$ and $W_b$ be open Wilson loop operators for $a$ and $b$ that terminate at the boundary and are arranged in space as in Fig.~\ref{fig: 2D boundary a}.  (Physically, an open Wilson loop operator $W_a$ is a unitary operator that moves an anyon of type $a$ from one place to another.  Equivalently, it can be viewed as creating an $a$-$\bar{a}$  anyon-antianyon pair out of the vacuum and separating the $a$ and $\bar{a}$ excitations in space.) Since $a$ and $b$ are condensed, there exists a unitary operator $U_a$ (resp. $U_b$), localized at the endpoints of $W_a$, which annihilates the $a$ and $\bar{a}$ excitations created by $W_a$, see Fig.~\ref{fig: 2D boundary b}. Acting on the ground state $\ket{G}$, $U_aW_a \ket{G} = U_bW_b\ket{G} = \ket{G}$ (absorbing any extraneous phases into the definitions of $U_a$ and $U_b$). Note that this step requires the boundary to be gapped. From the action on the ground state, we deduce that
\begin{equation}
U_aW_a U_bW_b\ket{G} = U_bW_b U_aW_a\ket{G} = \ket{G}.
\label{eqn:trivialWilson}
\end{equation}
Separately, note that
\begin{align}
[U_a,W_b]=[U_b,W_a]=[U_a,U_b]=0,
\label{eq: 2D trivial exchange}
\end{align}
since the supports of these pairs of operators are assumed to be disjoint. However, $W_a$ moves an $a$ anyon from one of its ends to the other---therefore, in the arrangement shown in Fig.~\ref{fig: 2D boundary a}, the operator $W_aW_b$ creates a pair $b$ and $\bar{b}$, then creates a pair $a$ and $\bar{a}$ and exchanges the $a$ around the $b$ in a counterclockwise direction. Likewise, $W_bW_a$ creates the same set of anyons and exchanges $b$ counterclockwise around $a$. Hence,
\begin{align}
W_a\, W_b = e^{i\theta_{ab}}\,W_b\, W_a,
\label{eq: 2D nontrivial exchange}
\end{align}
where the phase $\theta_{ab}$ is the mutual statistics of $a$ and $b$. Applying Eqs.~\eqref{eq: 2D trivial exchange} and \eqref{eq: 2D nontrivial exchange}, we find
\begin{equation}
U_aW_a U_bW_b\ket{G} = e^{i\theta_{ab}}U_bW_b U_aW_a\ket{G} = e^{i\theta_{ab}}\ket{G}.
\label{eqn:nontrivialWilson}
\end{equation}
That Eqs.~\eqref{eqn:trivialWilson} and \eqref{eqn:nontrivialWilson} must hold simultaneously forces $\theta_{ab} = 0$ (mod $2\pi$), that is, $a$ and $b$ must braid trivially.

\section{X-cube Model and Bulk ``Braiding" and ``Fusion" for Fractons}
\label{sec:bulkXCube}

\begin{figure}
\centering
\includegraphics[width=\columnwidth]{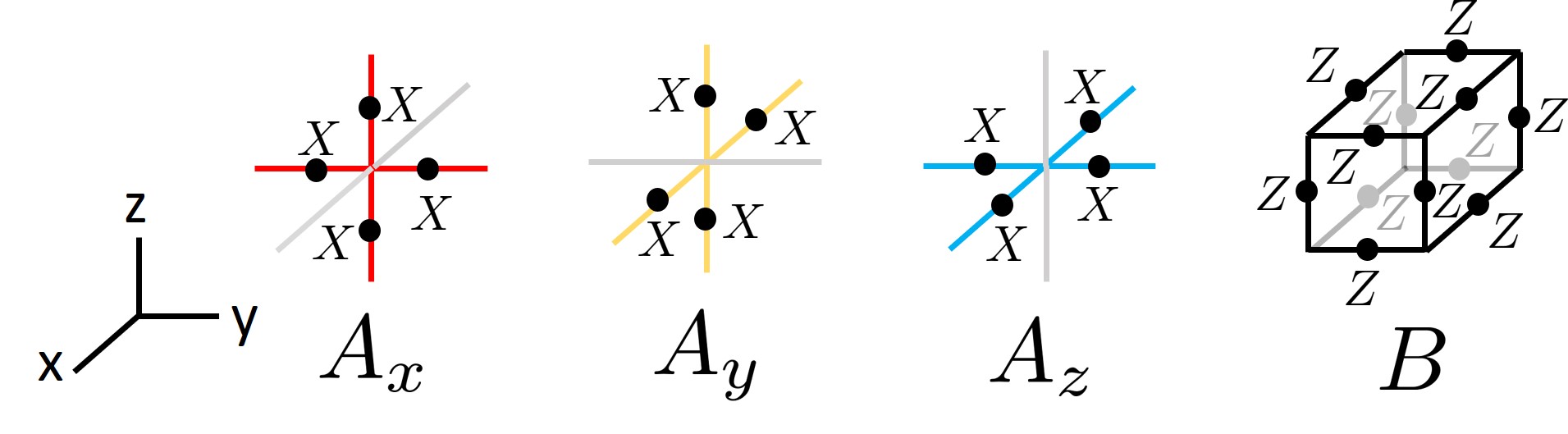}
\caption{Terms in the X-cube model Hamiltonian \eqref{eqn:XCubeH}. Colors are for emphasis and (later) to indicate pictorially which terms are affected at a site. The label $i$ in the star operator $A_i$ denotes the plane perpendicular to the star operator.}
\label{fig:XCubeTerms}
\end{figure}

Although the general features of our discussion apply equally well to other type-I fracton models, we will use the X-cube model~\cite{CastelnovoFirstXCubePaper,VijayGaugedSubsystem}, which we presently define, as our primary example. Its Hilbert space consists of spin-1/2 objects on the links of a cubic lattice, and its Hamiltonian is
\begin{align}
\begin{split}
H &= -\sum_{+}\prod_{l\in +} X_l -\sum_{\cube} \prod_{l\, \in\, \cube} Z_l\\
&\equiv -\sum_{\text{sites}}\sum_{i=x,y,z} A_i - \sum_{\cube} B 
\end{split}
\label{eqn:XCubeH}
\end{align} 
Here $+$ denotes a ``star" consisting of 4 coplanar vertex-sharing links of the cubic lattice, and $\cube$ denotes a ``cube" consisting of 12 links in the cubic lattice that surround a point in the dual cubic lattice.  $X_l$ and $Z_l$ are Pauli operators defined on the link $l$. The operators $A_i$ and $B$ are shown in Fig.~\ref{fig:XCubeTerms}.  This model is a commuting projector model, and its ground states satisfy the conditions
\begin{align}
\label{eq: ground state}
A_i=B=+1
\end{align}
for all $+$ and $\cube$.

The excitations of this model can be constructed by acting with Pauli operators $X_l$ or $Z_l$ on any link, which leads to violations of the ground-state constraint~\eqref{eq: ground state}. For later purposes and to establish notation, we tabulate the excitations and the operators that create them in Table~\ref{tab:XCubeExcitations}. The X-cube model harbors immobile excitations (fractons) as well as 1D and 2D excitations.
The subscripts on the particle labels in Table~\ref{tab:XCubeExcitations} refer to the directions in which the particles are mobile.  
\begin{table*}
\centering
\renewcommand{\arraystretch}{1.3}
\begin{tabular}{@{}clp{4cm}lp{2.5cm}lp{4cm}}
\toprule[2pt]
\textbf{Excitation}&&\textbf{Hamiltonian terms $=-1$}&&\textbf{Pictorial representation} && \textbf{Creation operator}\\ \hline\\
$f_0$ && Single $B$ term && \tabimage{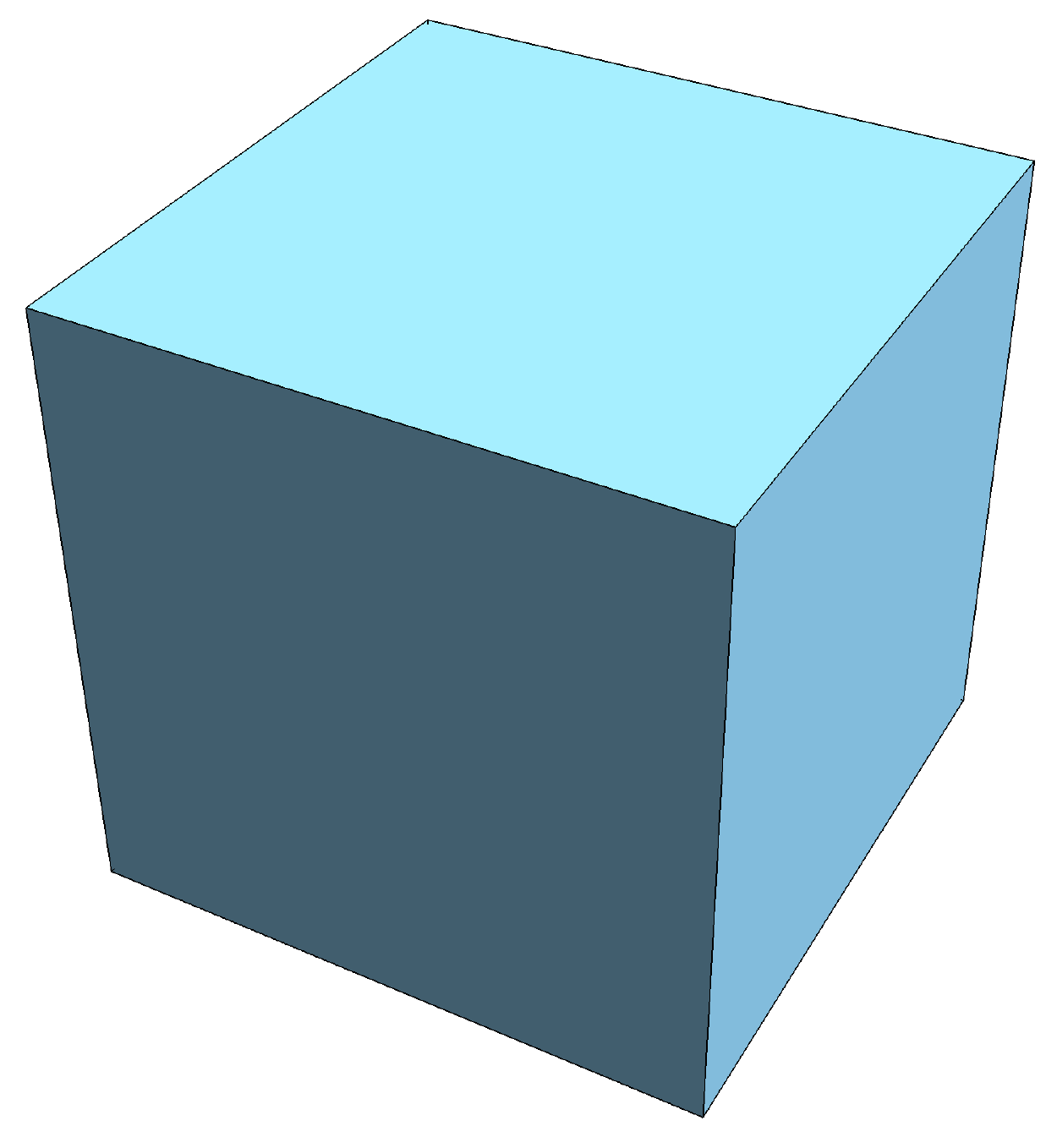}{1.5cm} && \tabimage{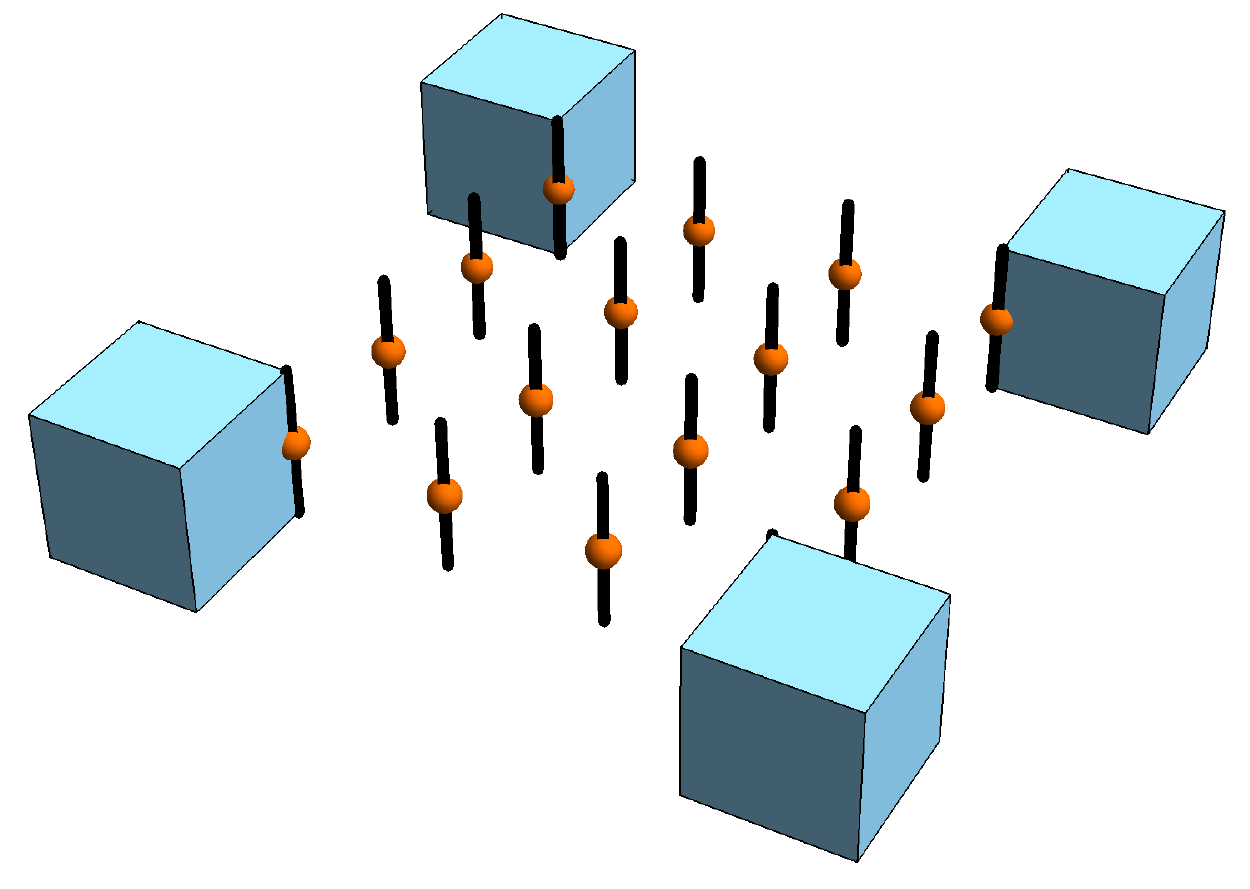}{2.5cm}\\
$m_{ij}$ && Bound state of two face-sharing $B$ terms separated in the $k\neq i,j$ direction ($m_{yz}$ shown) && \tabimage{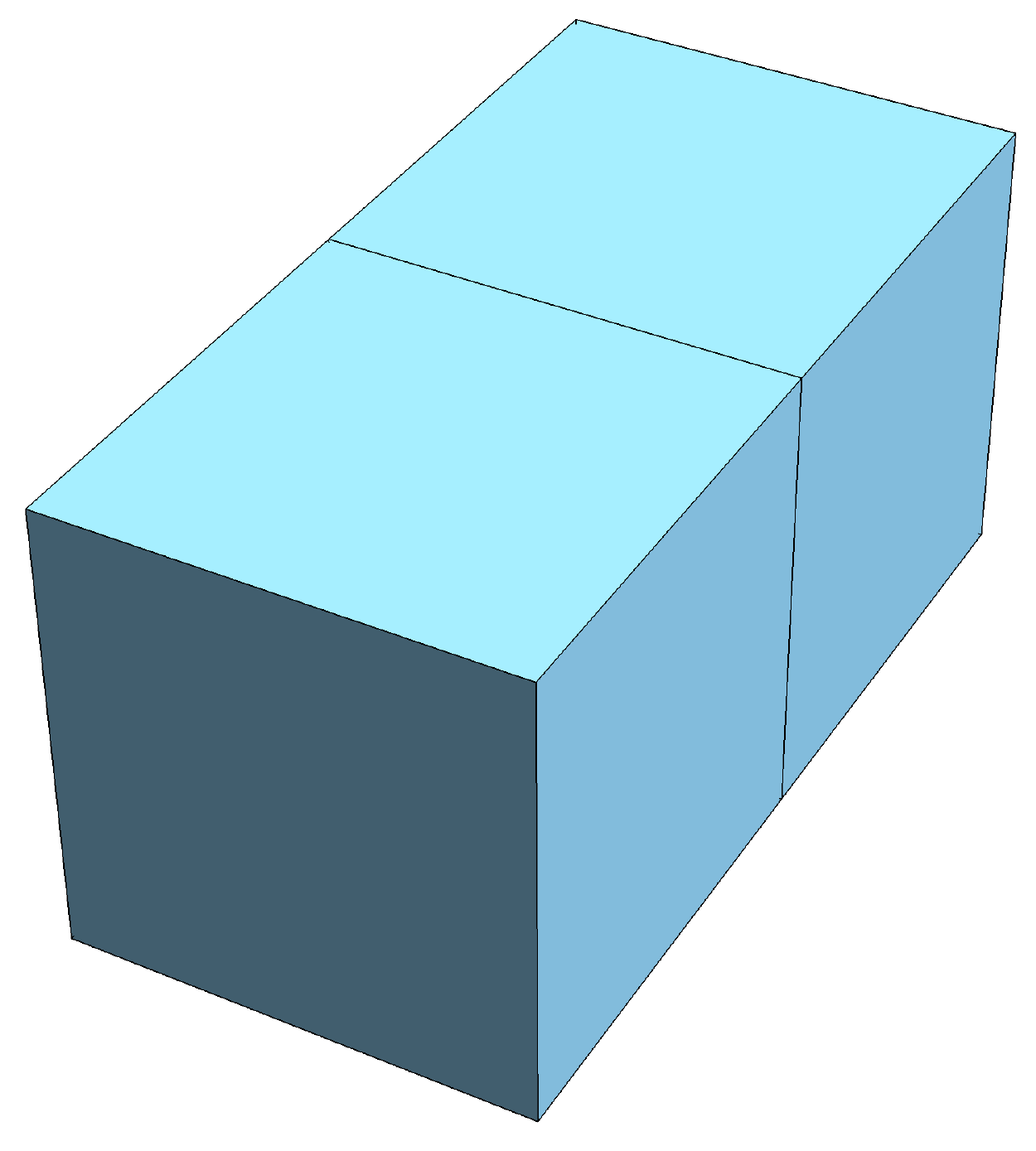}{1.5cm} && \tabimage{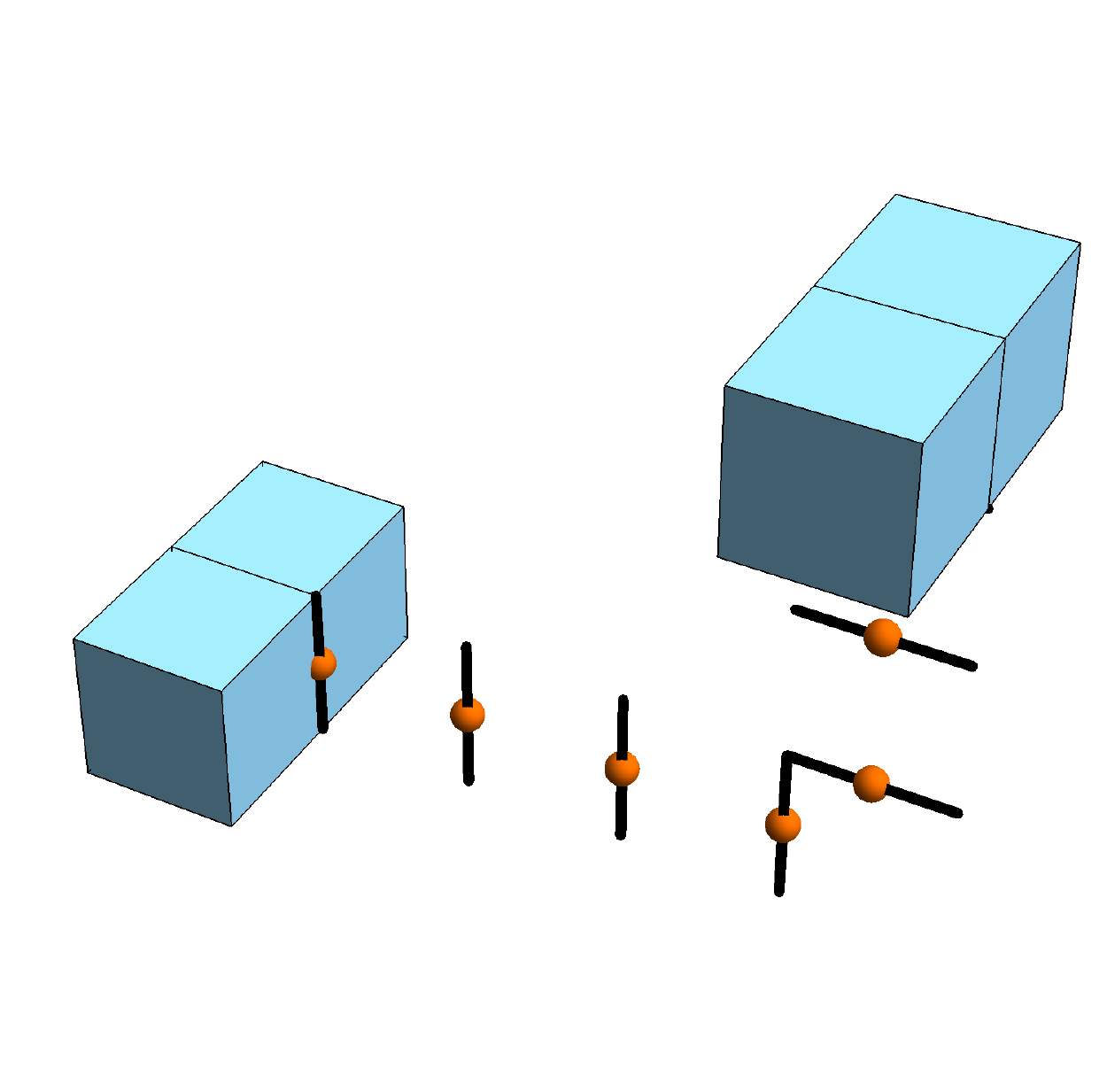}{2cm}\\
$p_i$ && Bound state of two edge-sharing $B$ terms ($p_y$ shown)&& \tabimage{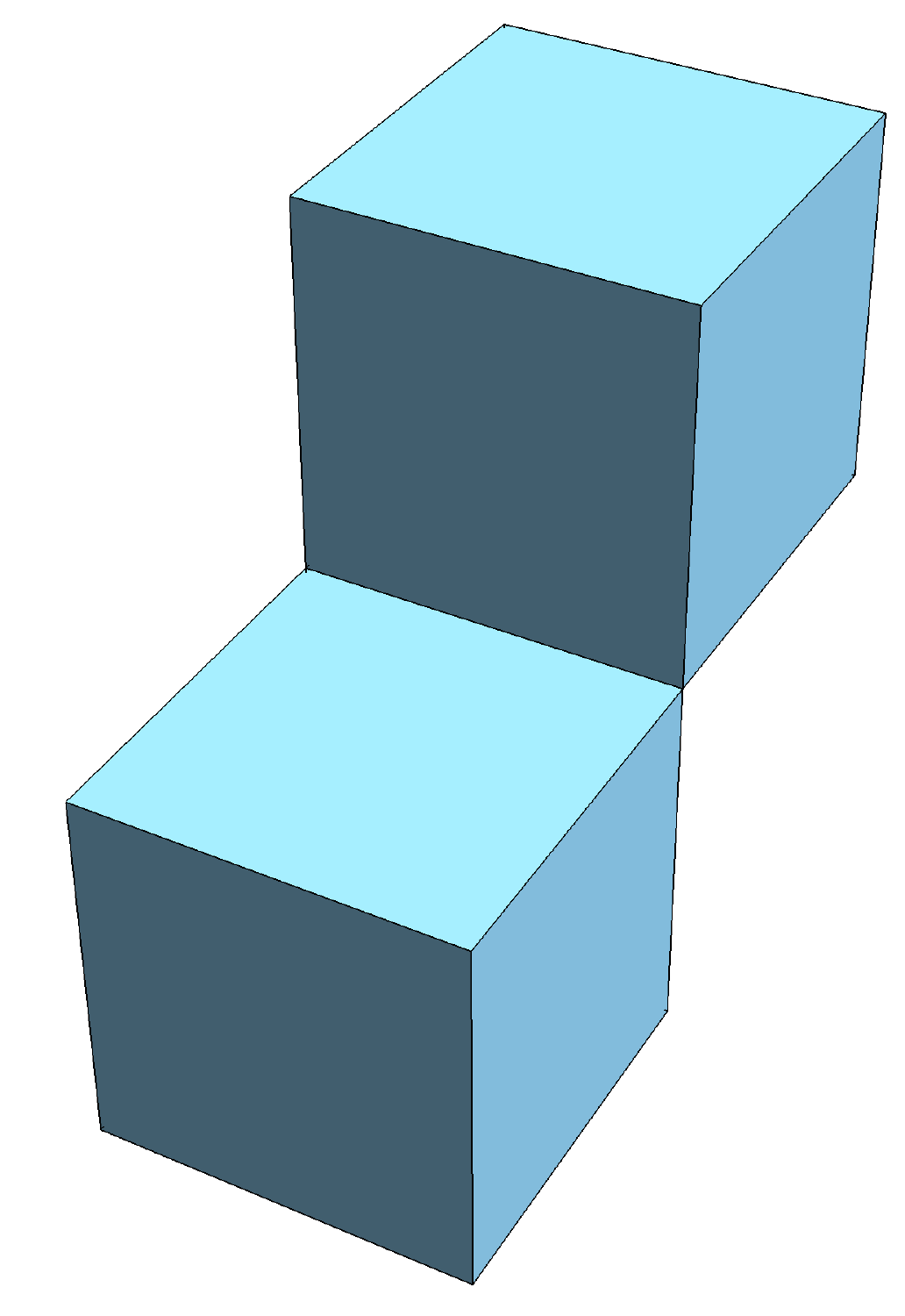}{1.5cm} && \tabimage{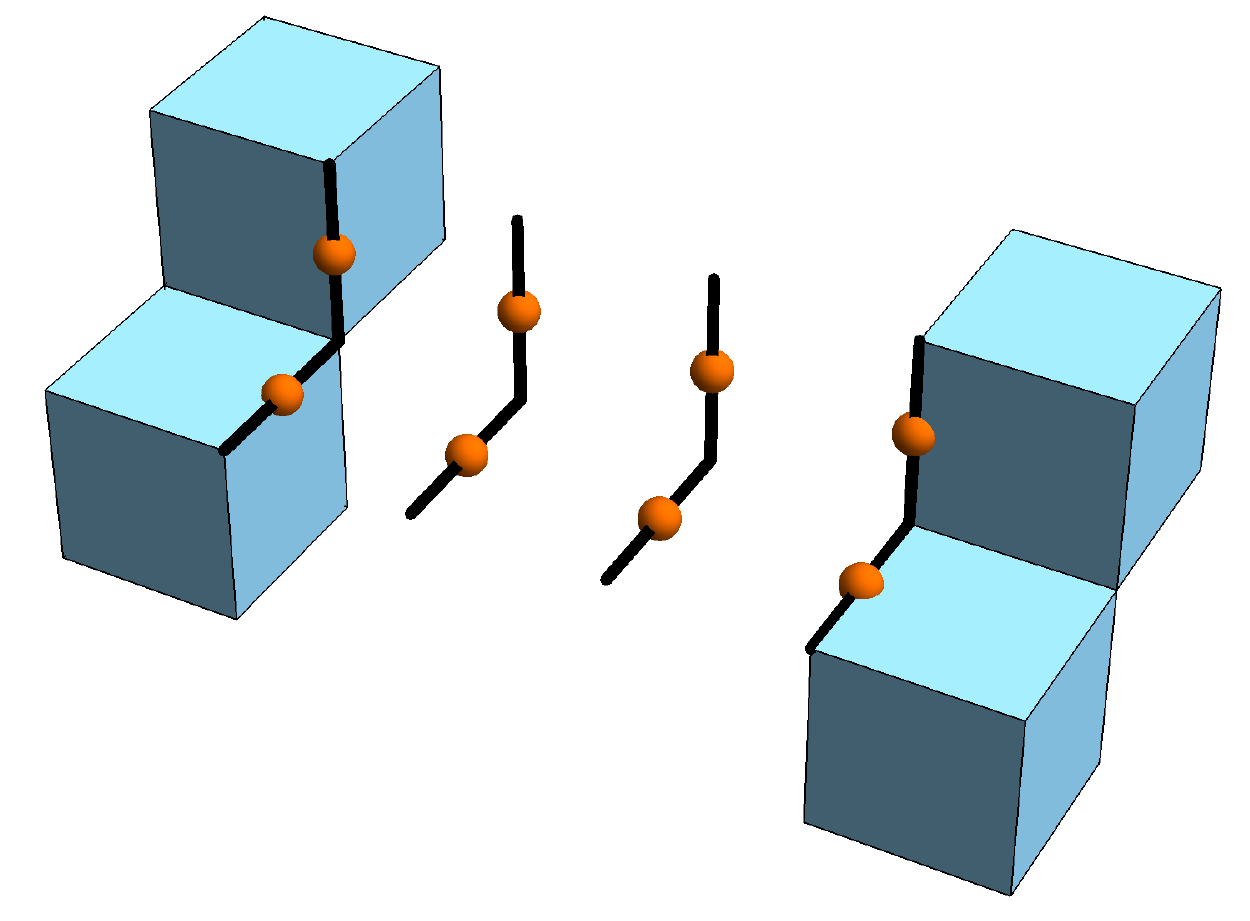}{2cm}\\
$e_i$ && $A_j$ and $A_k$ ($i\neq j \neq k$) on the same site ($e_y$ shown)&& \tabimage{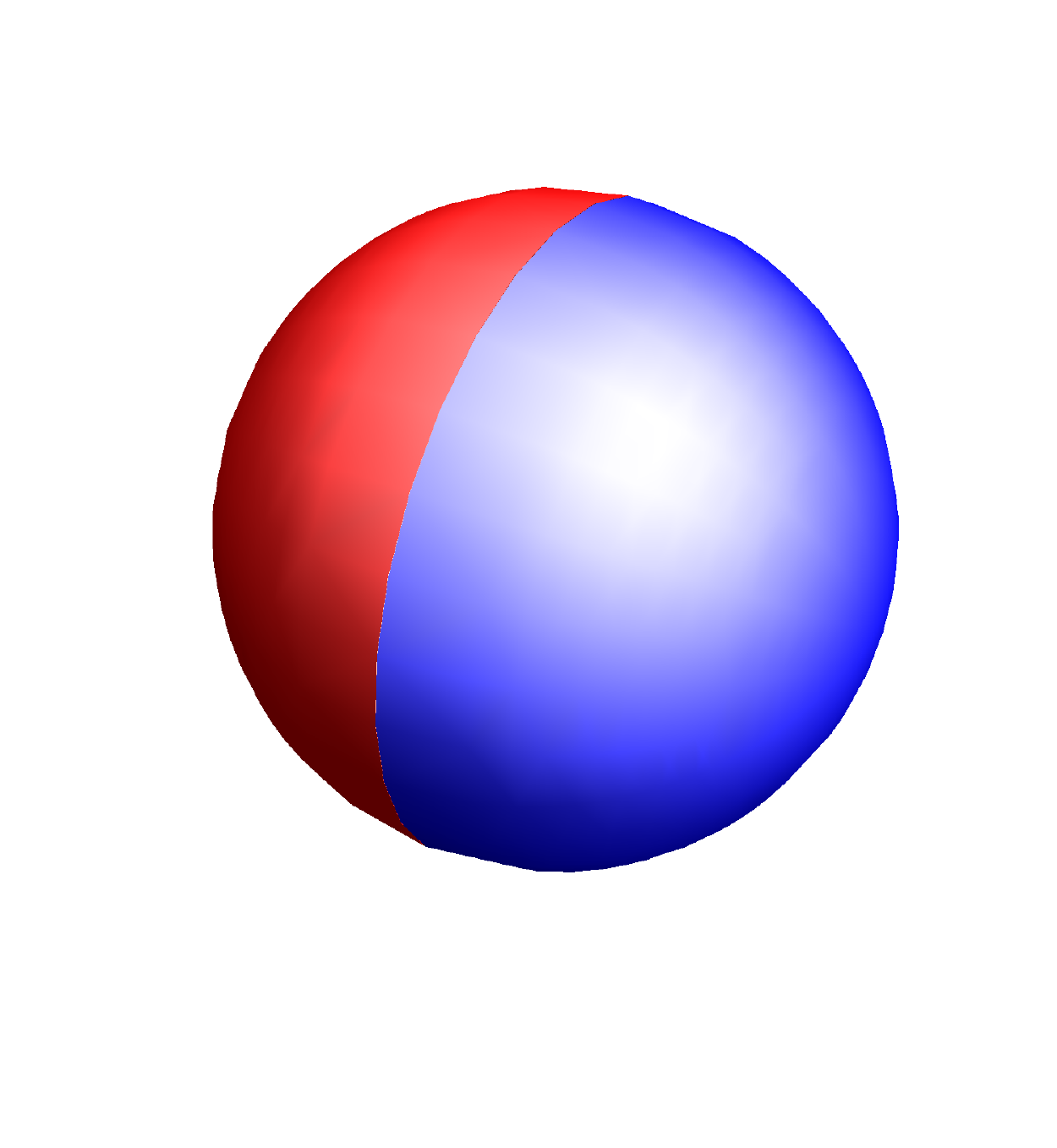}{1.5cm} && \tabimage{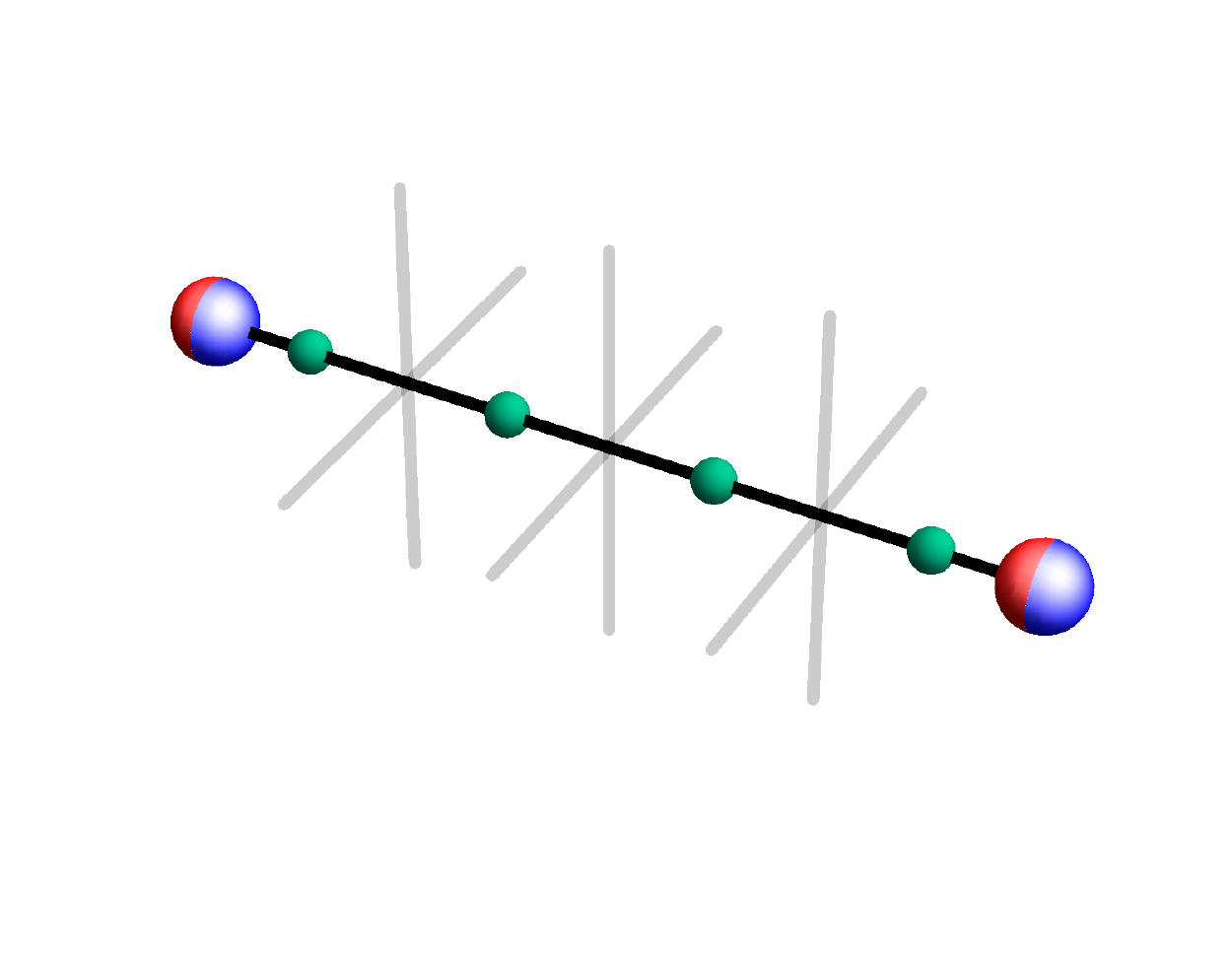}{2cm}\\
$q_{ij}$ && Bound state of two $e_i$ ($q_{yz}$ shown) && \tabimage{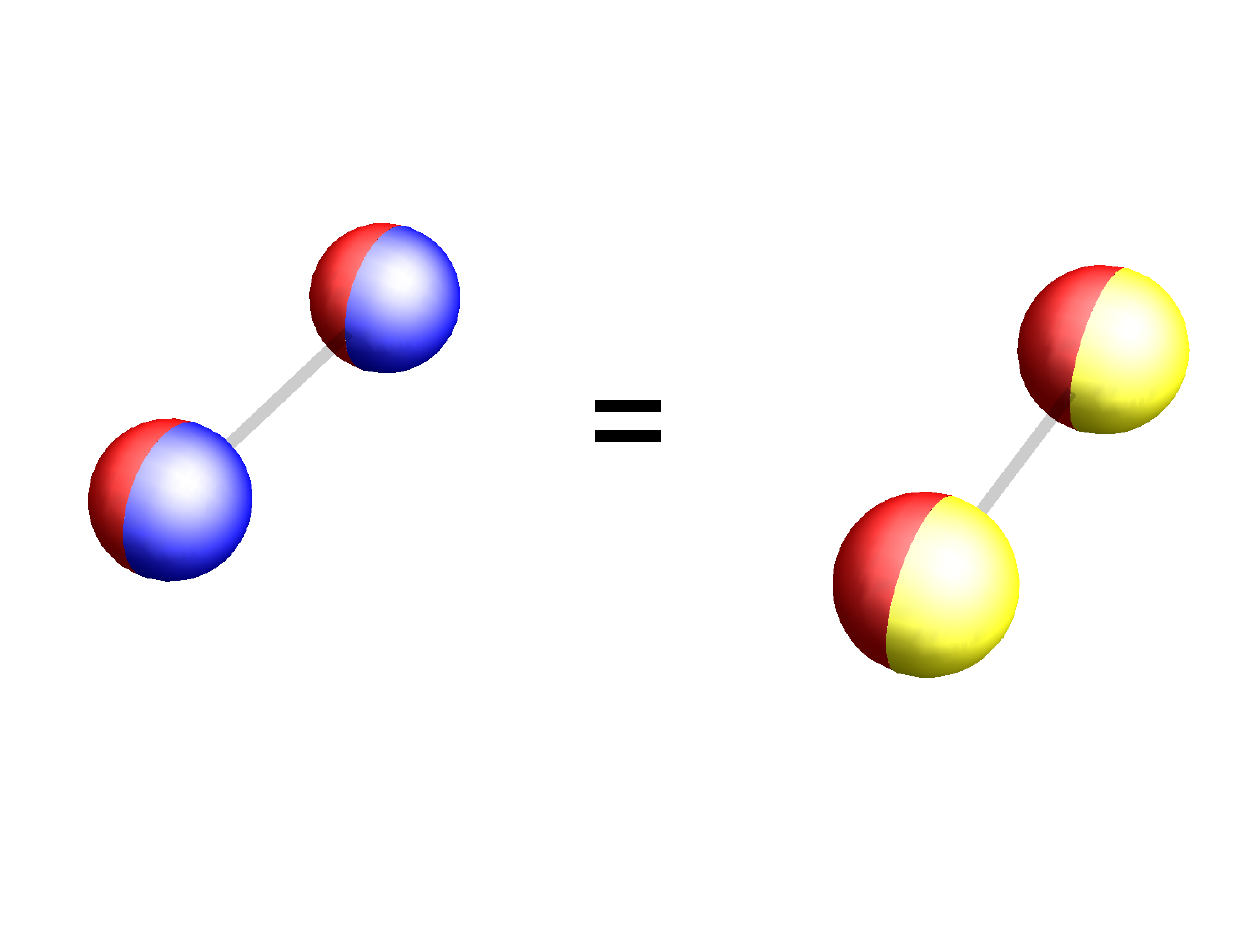}{2cm} && \tabimage{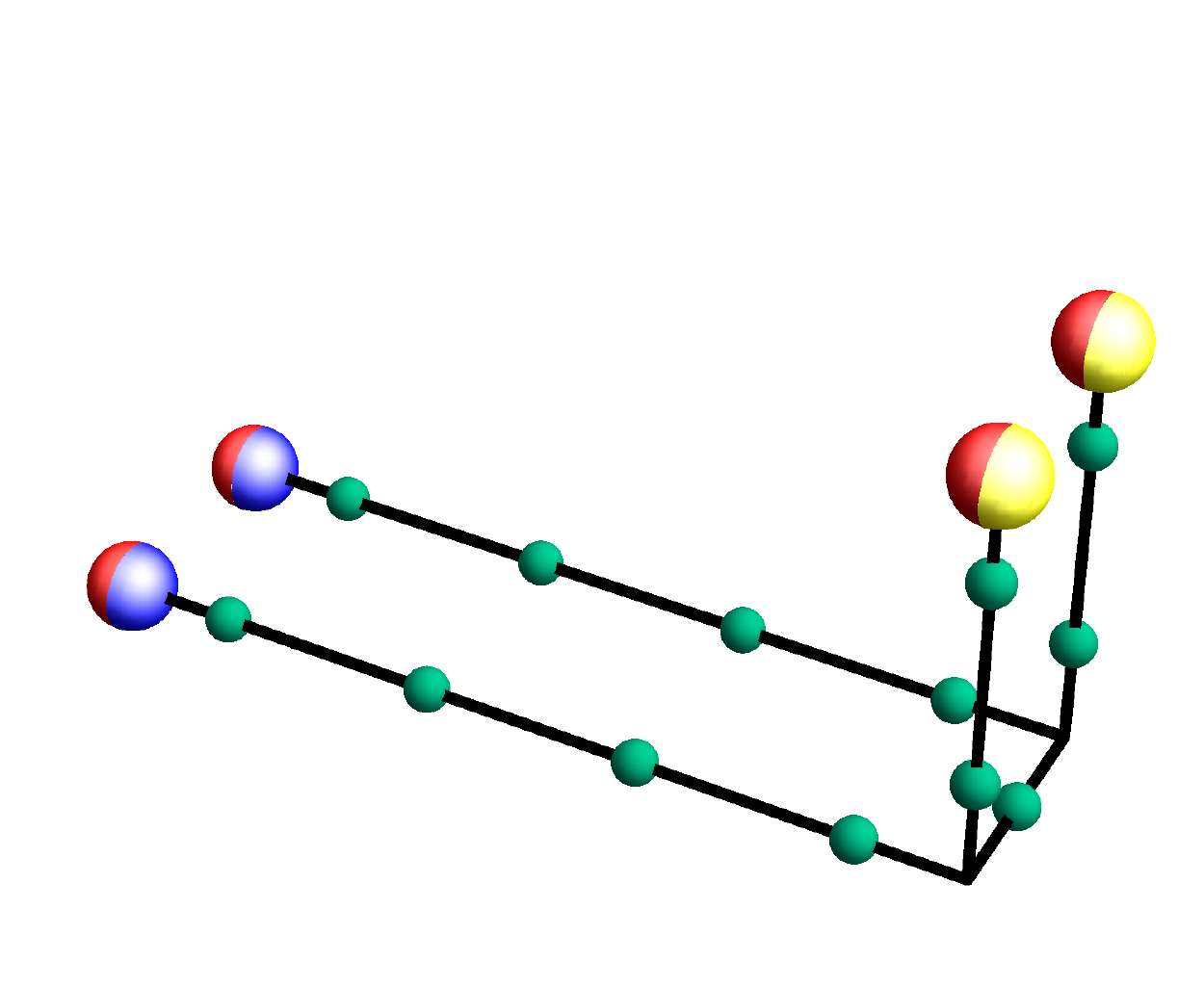}{2cm} \\
	\bottomrule[2pt]	
\end{tabular}
\caption{Excitations of the X-cube model. The subscripts $(i,j=x,y,z)$ indicate the directions in which the excitation is mobile, and $i\neq j$ is assumed. Black lines indicate lattice links, orange spheres are $X$ operators, and green spheres are $Z$ operators. Blue cubes indicate $B=-1$ and the red/blue/yellow sphere colors indicate which $A_i=-1$ according to the color scheme in Fig.~\ref{fig:XCubeTerms}.}
\label{tab:XCubeExcitations}
\end{table*}

\subsection{Bulk ``Braiding"}
\label{subsec: bulk braiding}

\begin{figure}
\centering
\includegraphics[width=.5\columnwidth]{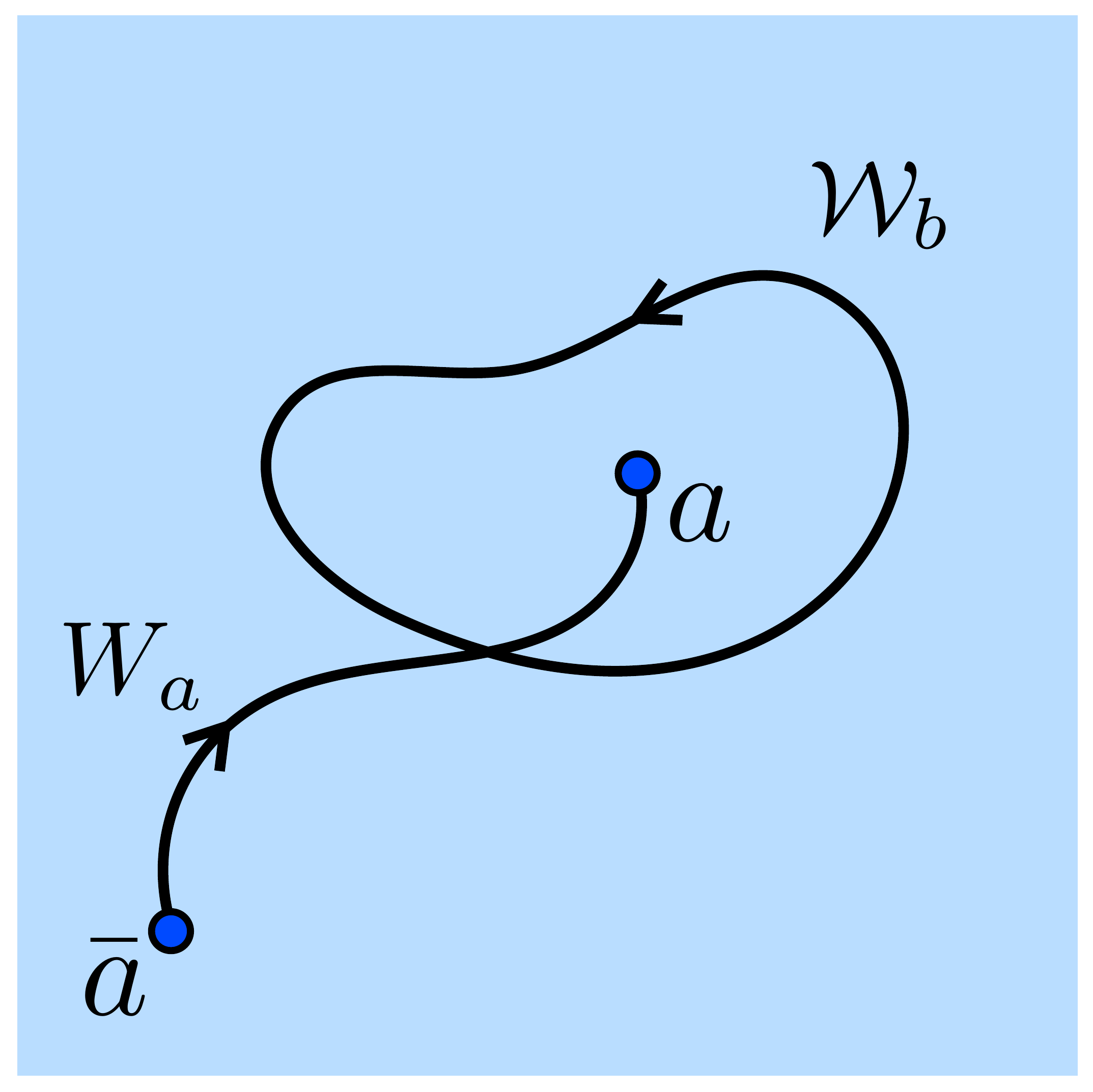}
\caption{Braiding in 2+1D topological order. The anyons $a$ and $b$ braid nontrivially if the closed Wilson loop operator $\mathcal{W}_b$ does not commute with the open Wilson line operator $W_a$, provided $W_a$ creates topological charge $a$ in the area enclosed by the support of $\mathcal{W}_b$.}
\label{fig: 2D braiding}
\end{figure}

We now turn to a discussion of braiding in the X-cube model. Several notions have been discussed informally in the previous literature~\cite{MaLayer,SlagleFieldTheory,PremCageNet,SongTwisted,ShirleyFoliatedFractional}. Although we will show that bulk braiding is generically insufficient for developing a gapped boundary criterion, we focus on a definition which is roughly equivalent to that of Refs.~\cite{MaLayer,SlagleFieldTheory,ShirleyFoliatedFractional} which will best motivate the gapped boundary construction we use. 

To define the braiding of subdimensional particles in fracton phases, we seek an analog of the braiding of pointlike excitations in Abelian topological phases in (2+1)D, which we define as follows. Roughly speaking, braiding an anyon of type $b$ around one of type $a$ can be viewed as the result of wrapping a Wilson loop (which is small compared to the system size) for the $b$ anyon around the $a$ anyon, see Fig.~\ref{fig: 2D braiding}. More precisely, let $\mathcal W_b$ be the closed Wilson loop operator associated with $b$, and let $W_a$ be the open Wilson loop operator associated with $a$ and $\bar a$.  Now suppose that $\mathcal W_b$ encircles the endpoint of $W_a$ that carries the $a$ anyon, with the other endpoint carrying $\bar a$ far away.  Then, $b$ braids nontrivially with $a$ if and only if
\begin{align}
W_a\, \mathcal W_b = e^{i\theta_{ba}}\, \mathcal W_b\, W_a
\indent\text{and}\indent
\theta_{ba}\notin 2\pi\, \mathbb Z.
\label{eq: 2D braiding def}
\end{align}
We refer to the phase $\theta_{ba}$ as the braiding phase of $b$ around $a$.

In order to discuss braiding for subdimensional particles, we need an appropriate generalization of the ``Wilson loop" notion. Indeed, while 2D particles can still be understood in terms of Wilson loops, this picture breaks down for 1D and 0D particles, which cannot move along closed paths without incurring additional energy costs.  In the known type-I fracton models, there exist such generalizations that we call ``cage operators." (Related ideas are used in Refs.~\cite{PremCageNet,ShirleyFoliatedFractional}.) They can be viewed as creating some (possibly large) set of excitations out of the vacuum, separating them, and then bringing them back together and annihilating them in a nontrivial way to produce a state with no excitations.  Some examples are shown in Fig.~\ref{fig:bulkCageOperators}. We formalize this definition in Appendix~\ref{app:bulkDefinitions}, but roughly speaking, any cage operator is associated to all particle types created in the aforementioned process of moving excitations around.

\begin{figure}
\centering
\subfloat[\label{fig:eCage}]{\includegraphics[width=0.4\columnwidth]{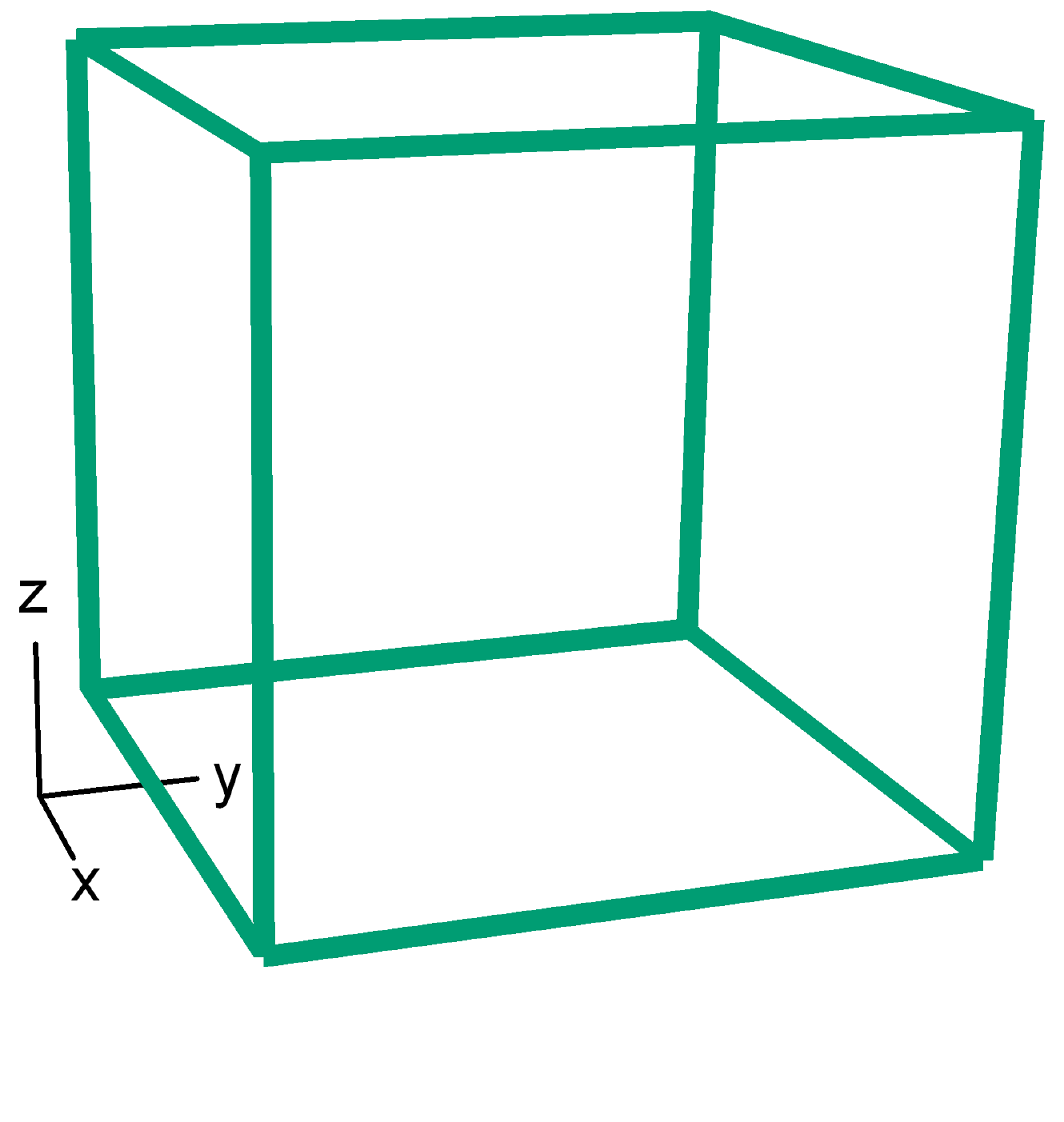}} \hfill
\subfloat[\label{fig:mxyCage}]{\includegraphics[width=0.4\columnwidth]{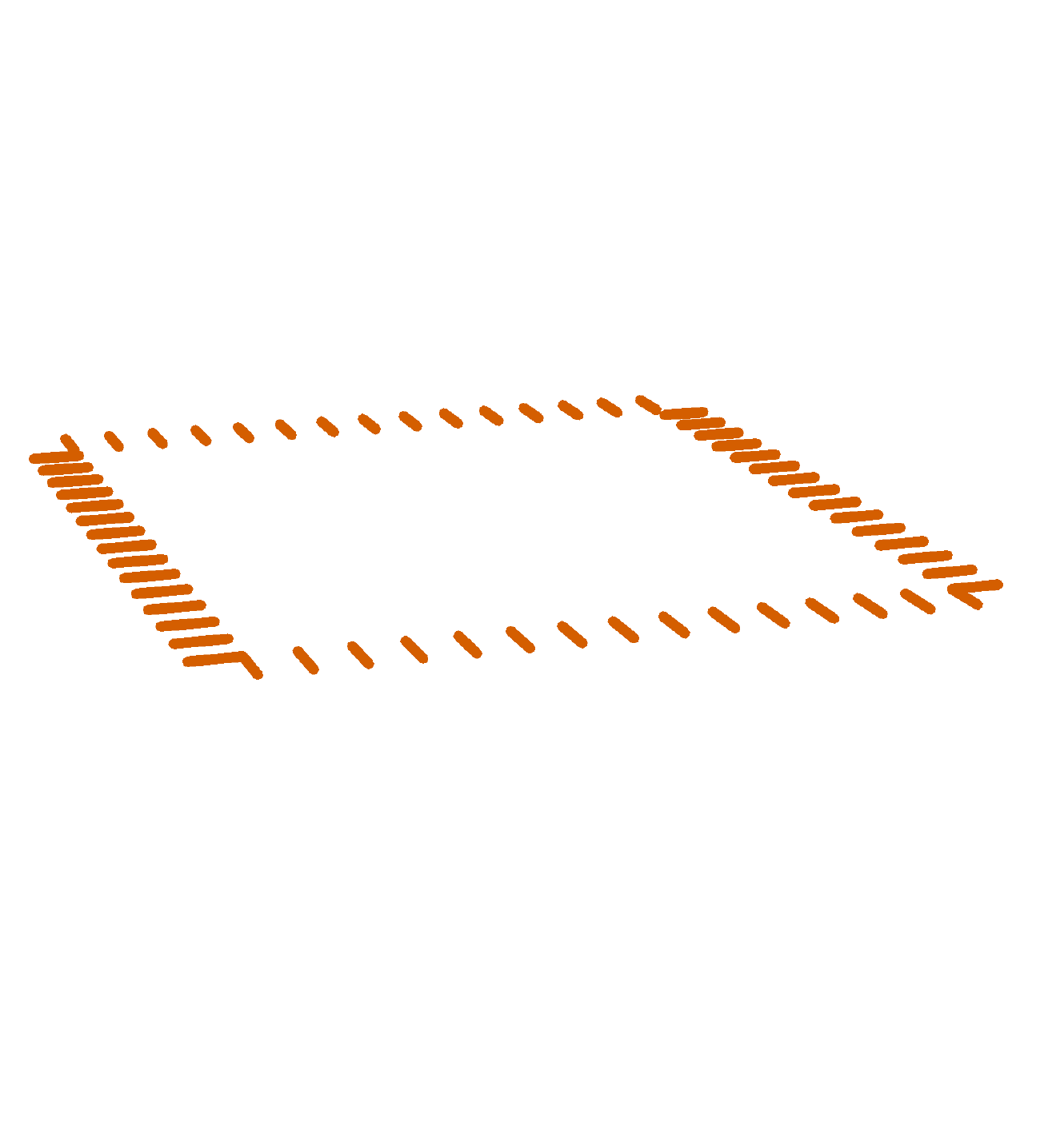}} \hfill
\subfloat[\label{fig:f0Cage1}]{\includegraphics[width=0.4\columnwidth]{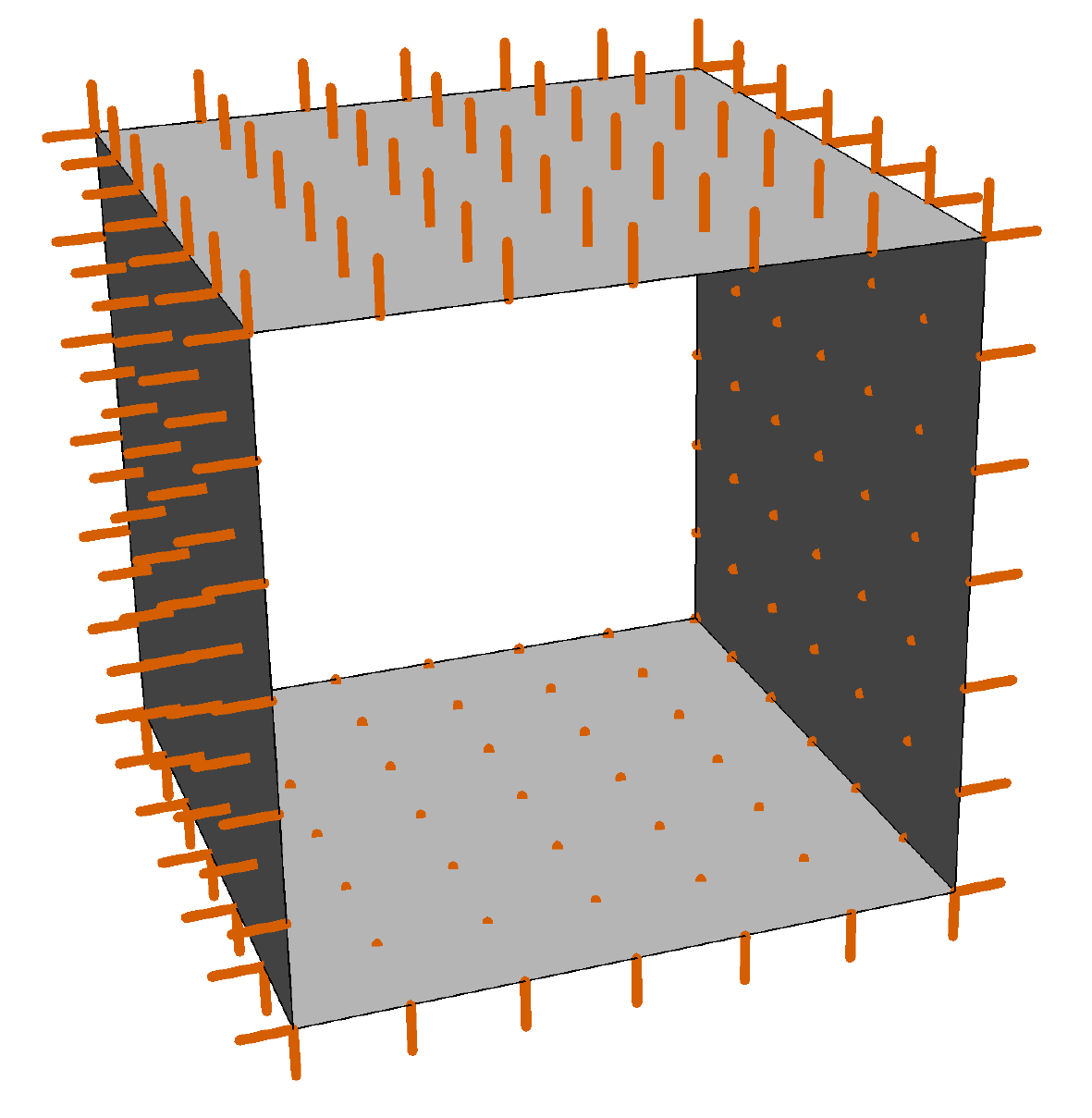}} \hfill
\subfloat[\label{fig:f0Cage2}]{\includegraphics[width=0.4\columnwidth]{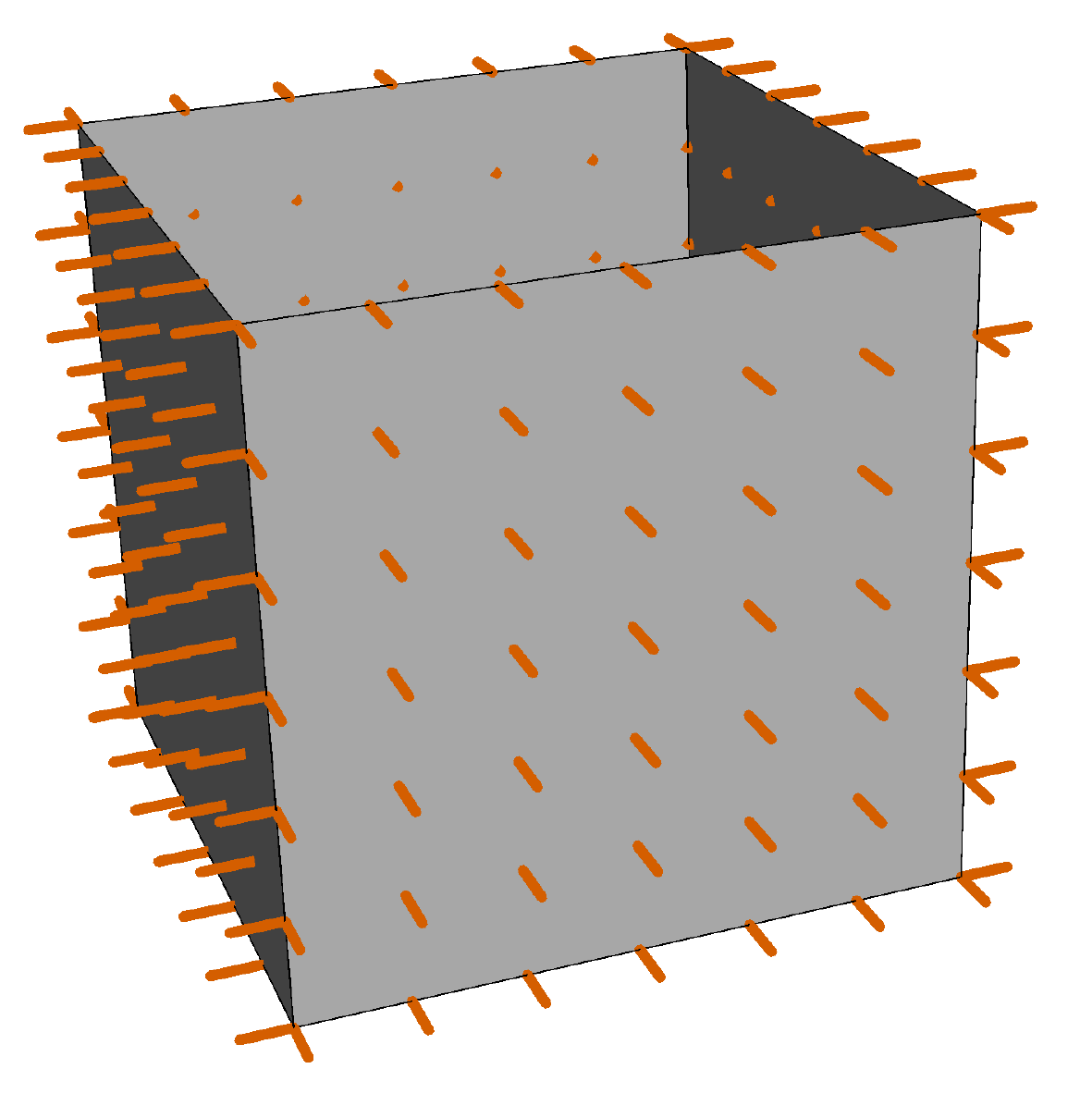}} \hfill
\caption{Cage operators for some excitations in the X-Cube model. Green (orange) links carry spins on which a $Z$ ($X$) operator acts. The grey planes are a guide to the eye. Coordinate axes in (a) are consistent throughout. (a): Cage for any $e$ particle. (b): Cage for $m_{xy}$; since $m_{xy}$ is 2D, its cage operator is just a Wilson loop. (c) and (d): Two different cage operators for $f_0$.}
\label{fig:bulkCageOperators}
\end{figure}

\begin{figure}
\centering
\subfloat[\label{fig:fractonAroundEz}]{\includegraphics[width=0.4\columnwidth]{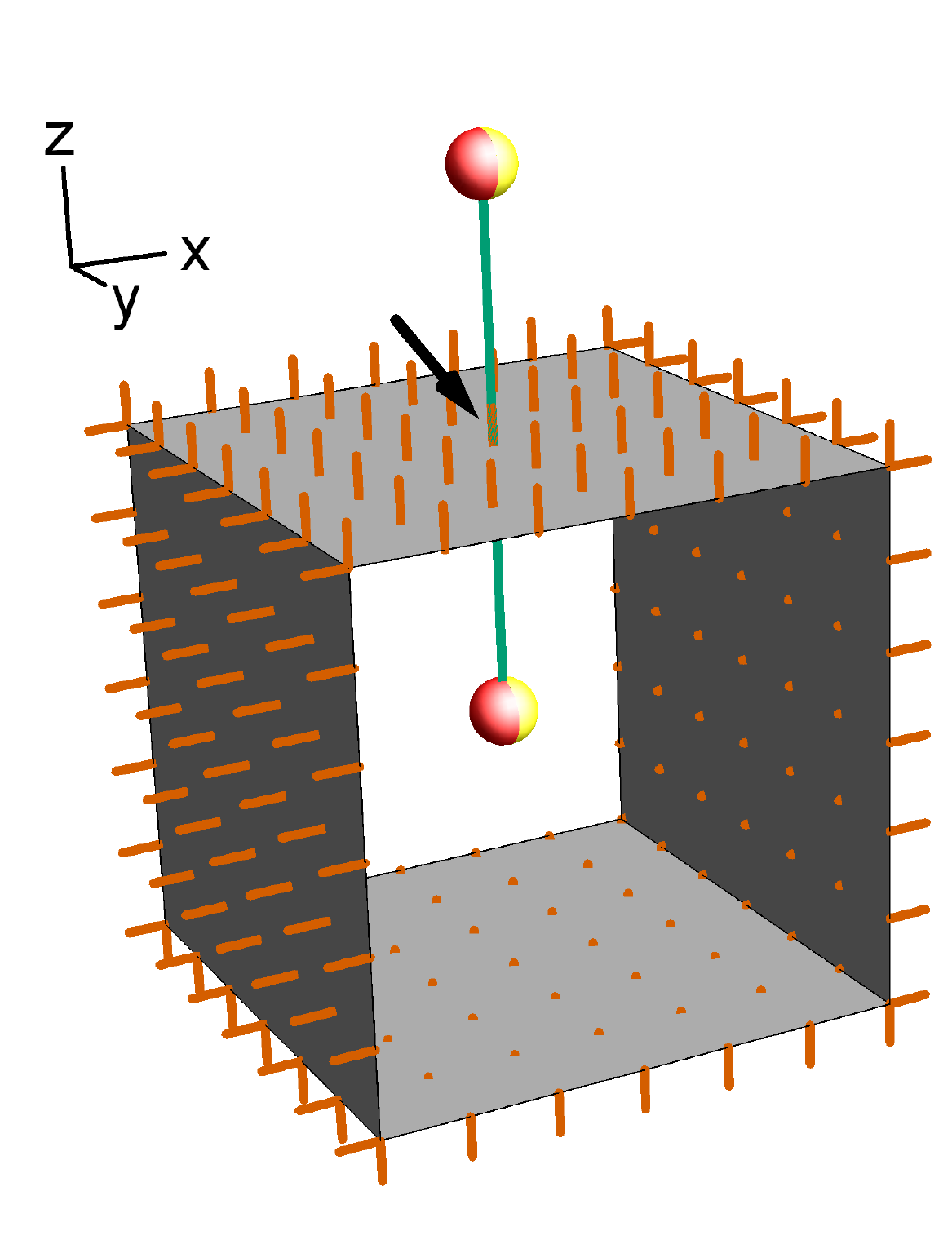}} \hfill
\subfloat[\label{fig:fractonAroundEz_trivial}]{\includegraphics[width=0.4\columnwidth]{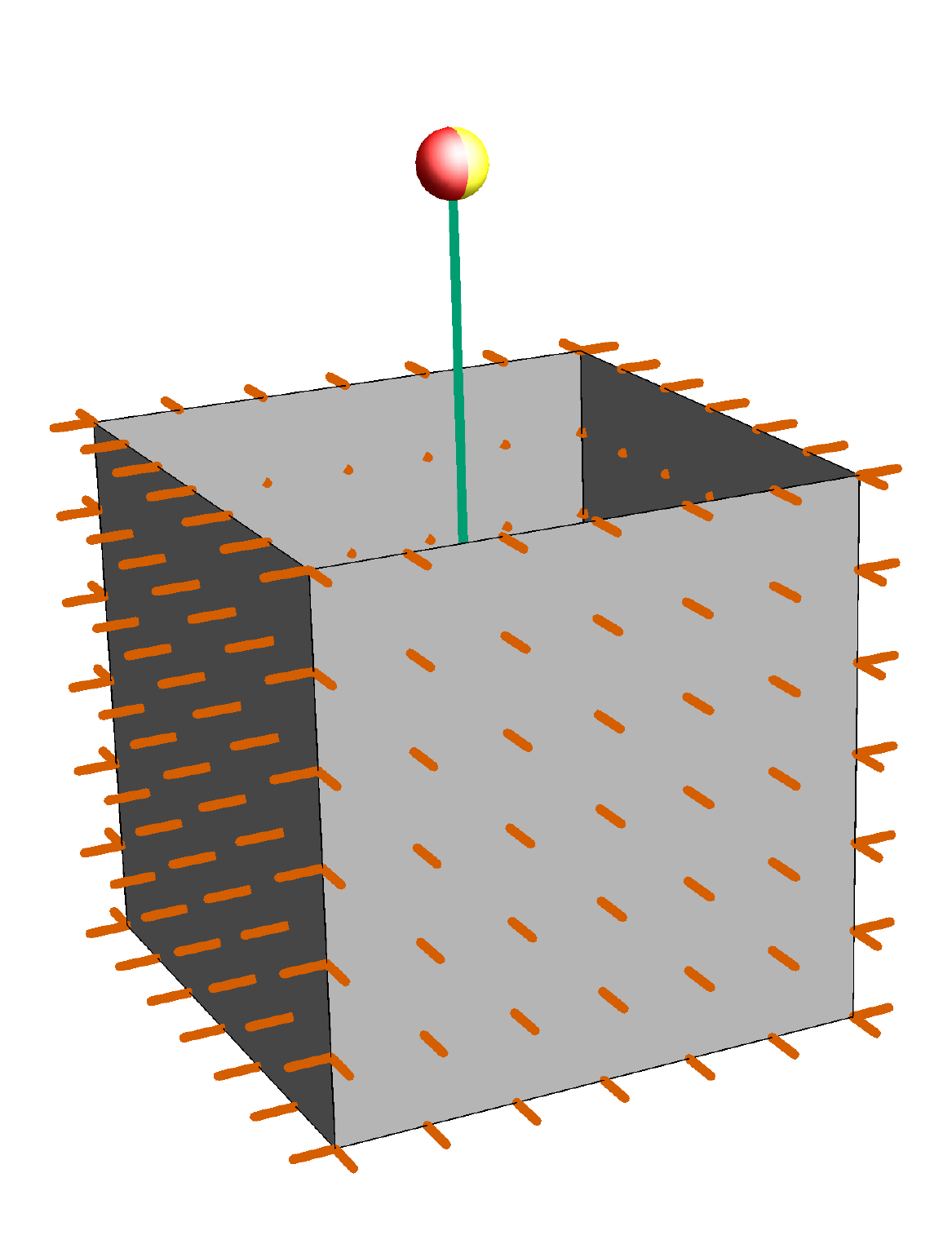}} \hfill
\caption{Two inequivalent ways to braid $f_0$ around an $e_z$. Green (orange) links carry spins on which a $Z$ ($X$) operator acts. (a) Choice of a cage operator which leads to a nontrivial braiding phase. Both the string operator creating $e_z$ excitations and the cage operator for the fracton act on the link indicated by the arrow, causing these operators to anticommute. (b) Choice of a cage operator which leads to a trivial braiding phase. Any translation of this $f_0$ cage operator commutes with  the $e_z$ string operator.}
\label{fig:f0ezBulkBraiding}
\end{figure}

A notion of braiding phase can be obtained by ``surrounding" an isolated excitation with a cage operator. Consider some operator $\mathcal{O}_a$ which creates an isolated $a$ excitation at the origin [this is the analog of the open Wilson loop operator $W_a$ in Eq.~\eqref{eq: 2D braiding def}]. We say that $b$ braids nontrivially around $a$ if, for any such $\mathcal{O}_a$, there exists a cage operator $\mathcal{C}_b$ with support far from the origin for which
\begin{align}
\mathcal{O}_a\, \mathcal{C}_b = e^{i\theta_{ba}}\, \mathcal{C}_b\, \mathcal{O}_a
\indent\text{and}\indent
\theta_{ba}\notin 2\pi\, \mathbb Z.
\label{eq: fracton bulk braiding def}
\end{align}
where the phase $\theta_{ba}$ gives the braiding statistics (here we have Abelian fractons in mind). For example, $f_0$ braids nontrivially with $e_z$ due to the process in Fig.~\ref{fig:fractonAroundEz}. 

\begin{figure}
\centering
\subfloat[\label{fig:ezAroundMxy}]{\includegraphics[width=0.33\columnwidth]{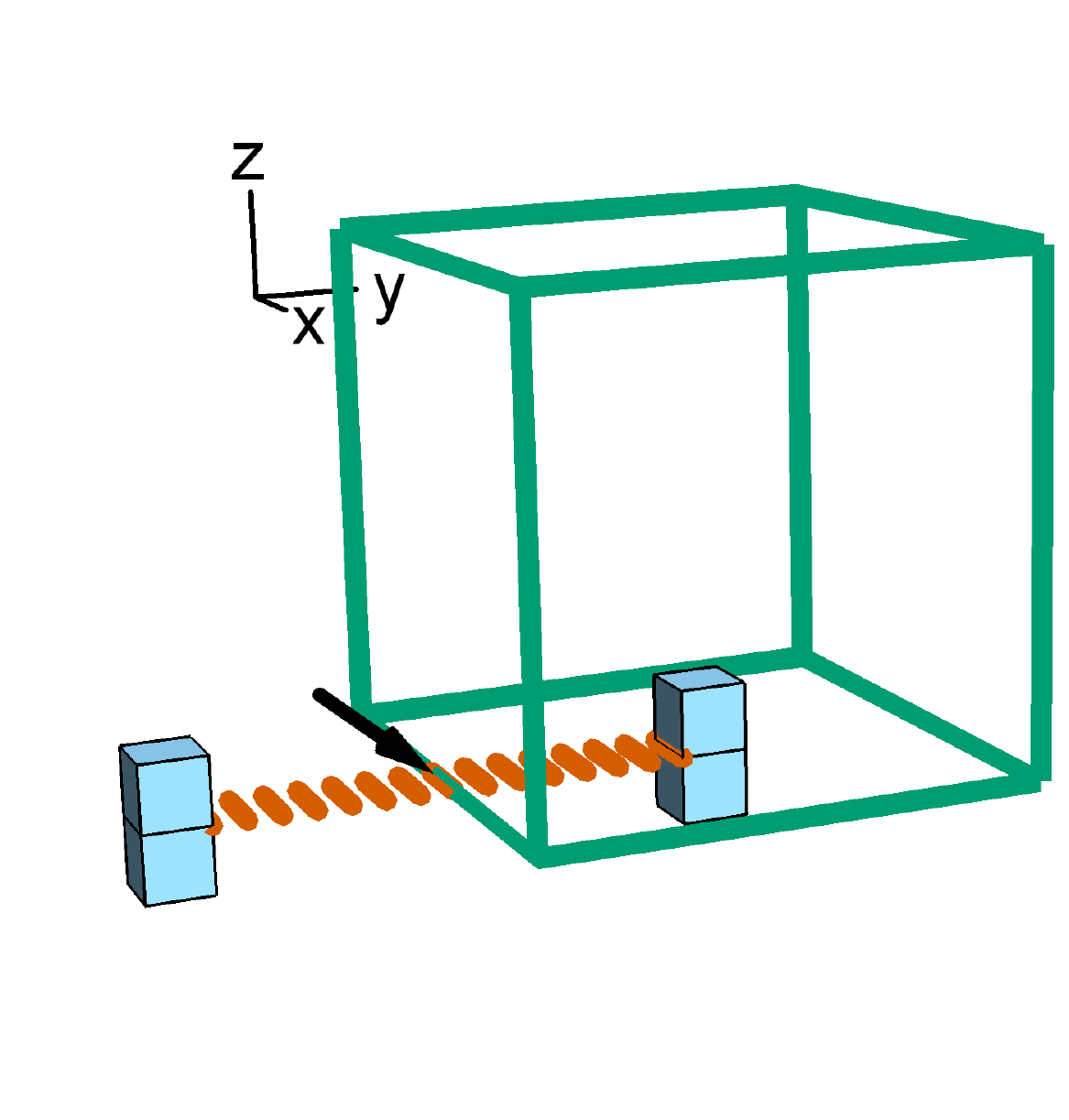}} \hfill
\subfloat[\label{fig:ezAroundMxy_trivial}]{\includegraphics[width=0.33\columnwidth]{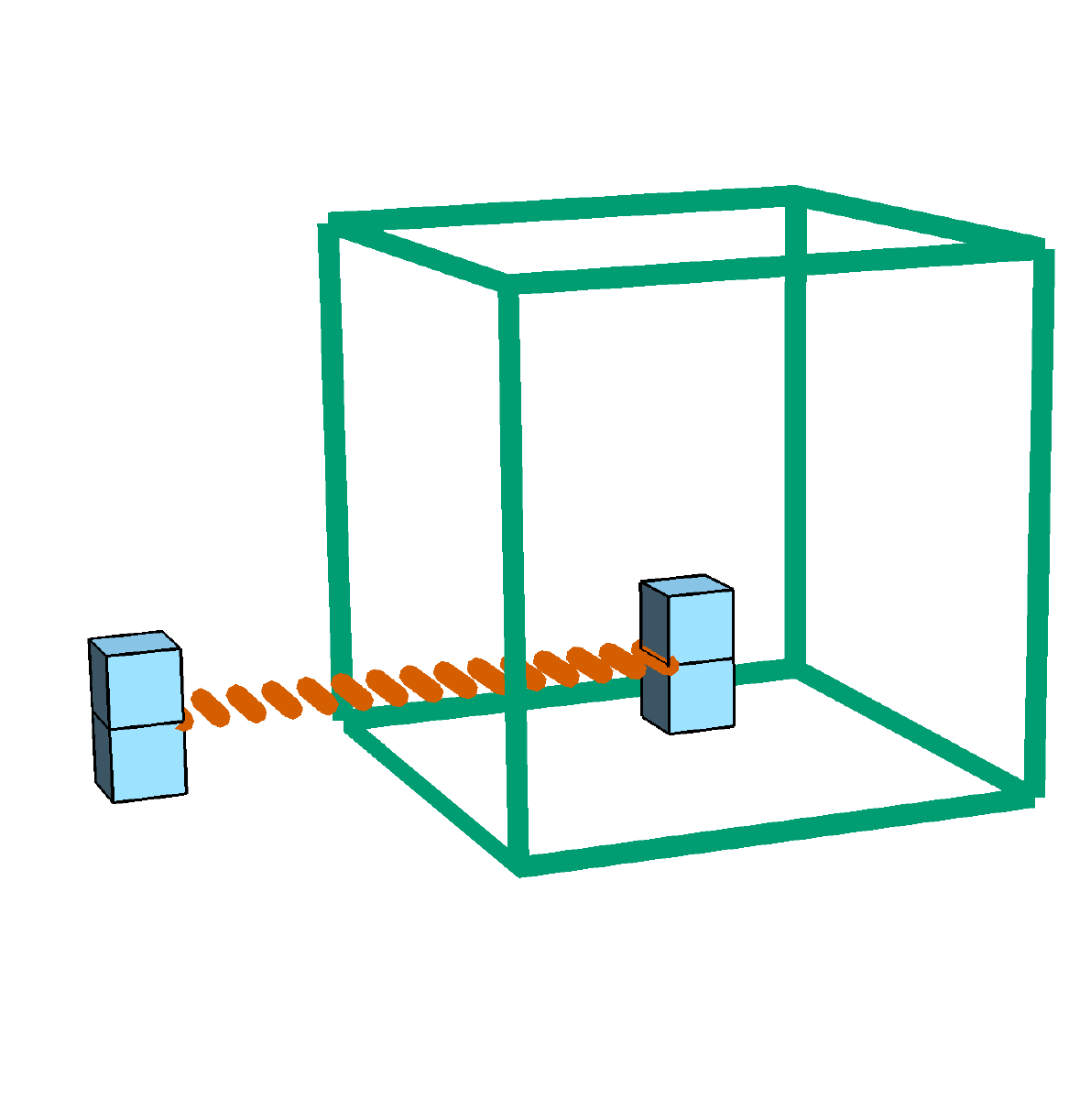}} \hfill
\subfloat[\label{fig:mxyAroundEz}]{\includegraphics[width=0.33\columnwidth]{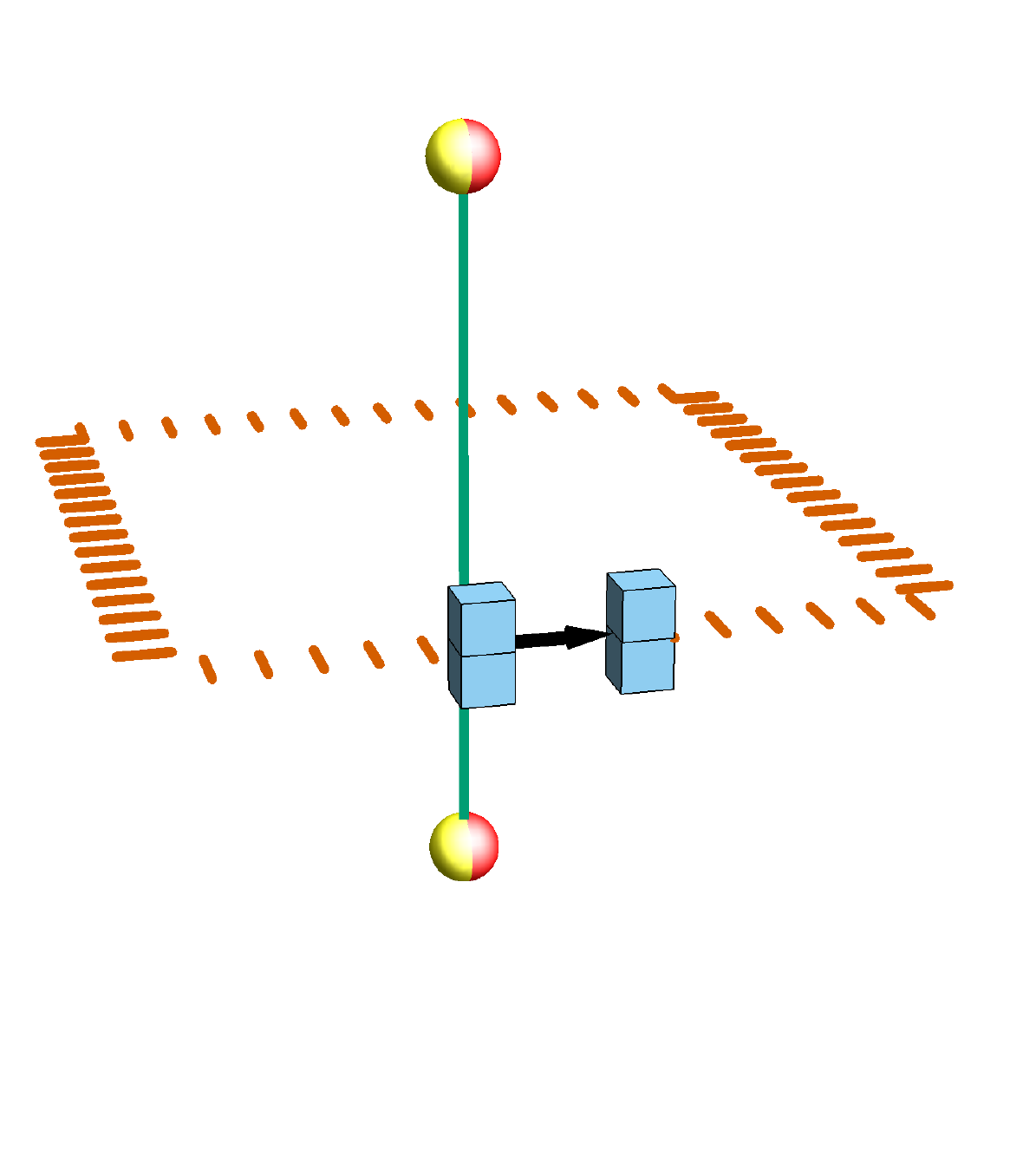}} \hfill
\caption{Several ways to braid $e_z$ and $m_{xy}$. Green (orange) links carry spins on which a $Z$ ($X$) operator acts. (a) Braiding $e_z$ around $m_{xy}$ in a way that leads to a nontrivial braiding phase; the cage and string operators shown anticommute with each other due to their overlapping support at the link indicated by the black arrow. (b) Displacing the string operator in (a) by a single lattice site in the $z$ direction causes the operators to commute, illustrating the geometric nature of bulk fracton braiding. (c) Braiding $m_{xy}$ around $e_z$; these operators commute no matter where they act relative to one another.}
\label{fig:mxyEzBulkBraiding}
\end{figure}

This definition of braiding has several properties that are dramatically different from standard topological order and which interfere with immediately generalizing the (2+1)D Lagrangian subgroup criterion to the fracton setting. First, the commutation relations of the operators $\mathcal O_a$ and $\mathcal C_b$ may change dramatically if the cage operator $\mathcal C_b$ is moved by only a small amount. For example, Fig.~\ref{fig:ezAroundMxy} shows a process where $e$ is braided around $m_{xy}$, leading to a nontrivial phase. However, moving the $e$ cage in the $z$ direction, even by a single lattice spacing, as in Fig.~\ref{fig:ezAroundMxy_trivial}, will cause the cage to commute with $\mathcal{O}_{m_{xy}}$ instead of anticommuting, even though the support of the cage never approaches the $m_{xy}$ particle. This effect does not arise in standard topological order, where the braiding of pointlike particles is independent of the precise shape of their respective Wilson loops.  That it arises in fracton phases like the X-cube model is evidence of the geometric (rather than purely topological) nature of subdimensional particles in (3+1)D.

Second, this definition of bulk braiding is nonreciprocal, that is, $\theta_{ab}$ need not equal $-\theta_{ba}$. (Importantly, although braiding is not reciprocal, it is unambiguous---$\theta_{ba}$ depends only on the positions of the excitations created by $\mathcal{O}_a$ relative to the cage $\mathcal{C}_b$, not on $\mathcal{O}_a$ itself.) To see why, consider braiding $e_z$ and $m_{xy}$, with cage operators shown in Figs.~\ref{fig:eCage} and \ref{fig:mxyCage}.  Any Wilson loop for $m_{xy}$ will commute with an operator creating an isolated $e_z$, as shown in Fig.~\ref{fig:mxyAroundEz}; in other words, $m_{xy}$ braids trivially around $e_z$. However, given an $m_{xy}$ at the origin, there is always an $e$ cage that anticommutes with it (see Fig.~\ref{fig:ezAroundMxy}), that is, $e_z$ braids nontrivially around $m_{xy}$.

Third, we note that the assignment of cage operators to particle types is not unique, unlike the assignment of Wilson loop operators to pointlike particles in standard Abelian topological phases.  For example, in the X-cube model there are three inequivalent cage operators $\mathcal C_{f_0}$ for the fracton that cannot be smoothly deformed into one another by applying a sequence of local operators.  One of these is shown in Fig.~\ref{fig:f0Cage1}, and another is shown in Fig.~\ref{fig:f0Cage2}.  (The third inequivalent cage operator can be obtained from the one in Fig.~\ref{fig:f0Cage2} by a $\pi/2$ rotation around the $z$-axis.). The former cage operator anticommutes with an $e_z$ string operator, as shown in Fig.~\ref{fig:fractonAroundEz} while the latter cage operator commutes with the same string, as shown in Fig.~\ref{fig:fractonAroundEz_trivial}; yet, they are both valid cage operators for the fracton $f_0$.

These ambiguities suggest that, unlike in Abelian toplogical phases in (2+1)D, the bulk braiding properties of subdimensional particles in fracton phases are not sufficient to uniquely determine the possible gapped boundaries (at least when the above definition of bulk braiding, which also appears elsewhere in the literature, is used). The precise set of ambiguities depends on a choice of definition of braiding. For example, the definition of braiding that we have used is essentially equivalent to that in Ref.~\cite{ShirleyFoliatedFractional}, but Ref.~\cite{ShirleyFoliatedFractional} associates braiding processes to operators rather than particles, and we need the latter in order to study gapped boundaries. On the other hand, the definition considered in Refs.~\cite{PremCageNet,SongTwisted} has the same issues of geometry dependence, but the nature of nonreciprocity is different because processes like braiding $e_z$ around $m_{xy}$ are simply undefined (for example, the process corresponding to Fig.~\ref{fig:ezAroundMxy} is associated to $q$ particles instead of to $e_z$ particles). It also suffers from some ambiguities about the association of cages with particle types, although they are different from the ambiguities in the definition we used.

We will see that the notion of boundary braiding, which we define in Sec.~\ref{sec: boundary braiding}, does not suffer from any of these ambiguities.  It appears that defining braiding with respect to a boundary with a particular geometry is sufficient to determine a unique (and reciprocal) notion of braiding for all particle types on that boundary, and in turn the set of possible gapped boundaries (modulo anyon condensation, as we explain).

\subsection{Bulk ``Fusion"}

\begin{figure}
\centering
\includegraphics[width=0.5\columnwidth]{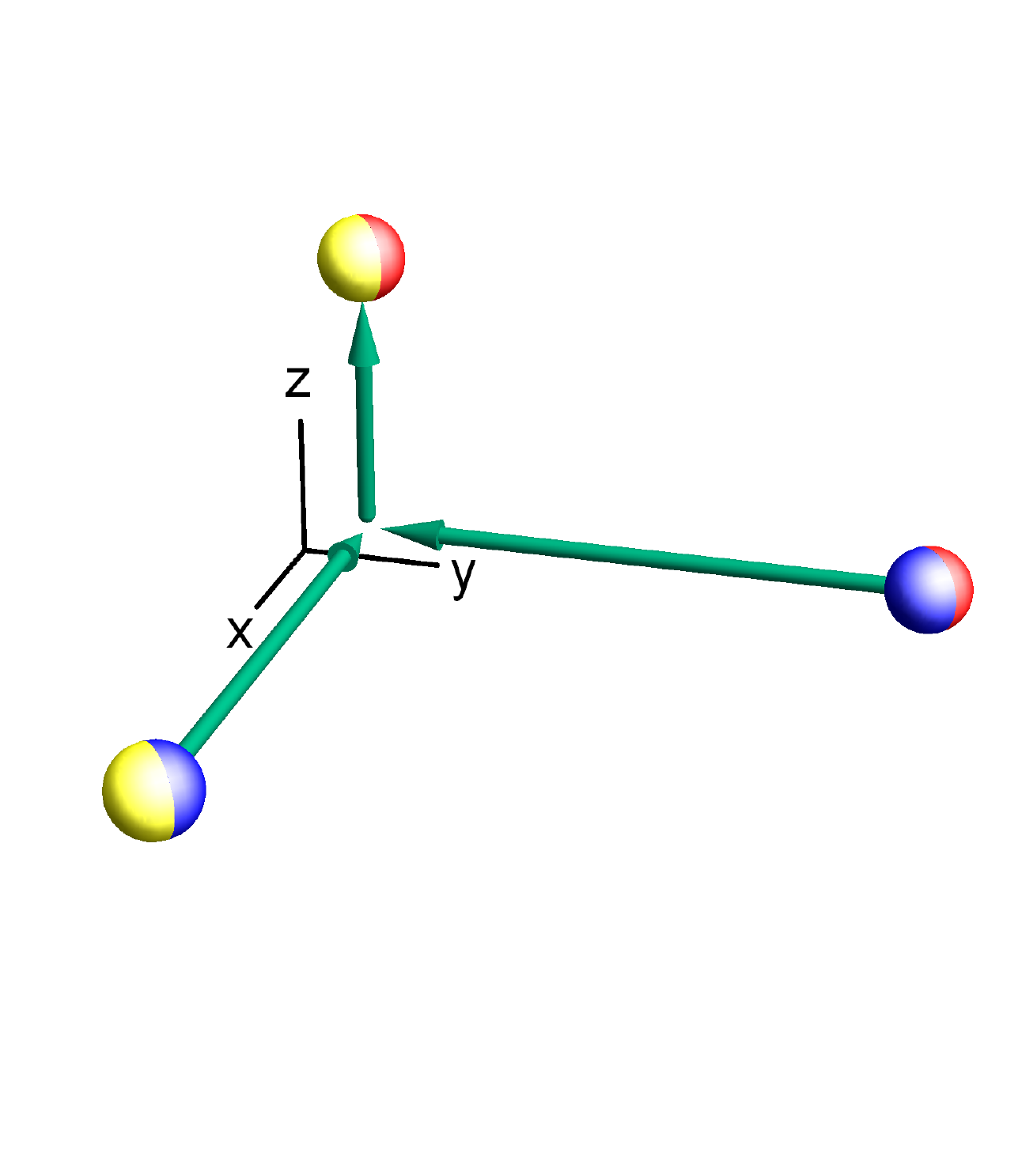}
\caption{``Fusion" process for $e_x \times e_y \sim e_z$.}
\label{fig:eFusion}
\end{figure}

Our subsequent discussion of the gapped boundaries of the X-cube model will also make use of a loose notion of ``fusion rule." Since $A_i^2 = B^2=1$, two $+$ or $\cube$ excitations of the same type cancel out when they lie on top of one another.  Thus, in a sense, these excitations ``fuse" with themselves to the identity.  More importantly, however, there is a sense in which
\begin{align}
e_i \times e_j \sim e_k\enspace
\text{for}\enspace
i,j,k\in\{x,y,z\}\enspace
\text{and}\enspace
 i\neq j \neq k.
 \label{eq: e-e fusion}
\end{align}
as shown in Fig.~\ref{fig:eFusion}, one can bring together an appropriately positioned $e_x$ and $e_y$ to produce an $e_z$.  It is also straightforward to see that
\begin{align}
m_{ij}\times m_{ik} \sim p_i
\label{eq: m-m fusion}
\end{align}
in a similar way. 

We must distinguish the notion of ``fusion" in Eqs.~\eqref{eq: e-e fusion} and \eqref{eq: m-m fusion} from the notion of fusion in topological order. In standard topological order, the notion of fusion refers to the idea that nearby excitations (i.e., a few correlation lengths away from one another) exhibit the properties of their fusion product when viewed from far away (i.e., many correlation lengths). In the case of subdimensional particles in fracton phases, this is not the case: an $e_x$ and an $e_y$ excitation do not become an $e_z$ until they ``collide" at a vertex of the cubic lattice.
Thus, like braiding, fusion depends on the local lattice geometry.  Another example of this dependence on lattice geometry is the fact that two neighboring $f_0$ excitations are not equivalent to the identity, but are instead a separate excitation $m_{ij}$ or $p_i$.  (Similarly, neighboring $e_x$ excitations may form $q_{xz}$ or $q_{zy}$ bound states.) Which excitation is formed by the neighboring fractons depends on the lattice geometry; the fusion product is not specified until we know the relative positions of the $f_0$ excitation in the cubic lattice.

\section{Gapped Boundaries of the X-cube Model}
\label{sec:XCubeBoundaries}

We now investigate some of the possible gapped boundaries of the X-cube model.  We focus first on the simplest boundaries, which arise when the cubic lattice is terminated on a crystallographic plane, before moving on to the boundaries that arise from a more complicated lattice termination.

\subsection{(001) Boundary}
\label{subsec: (001) Boundary}

\begin{figure}
\centering
\subfloat[\label{fig:ee001Boundary}]{\includegraphics[width=0.9\columnwidth]{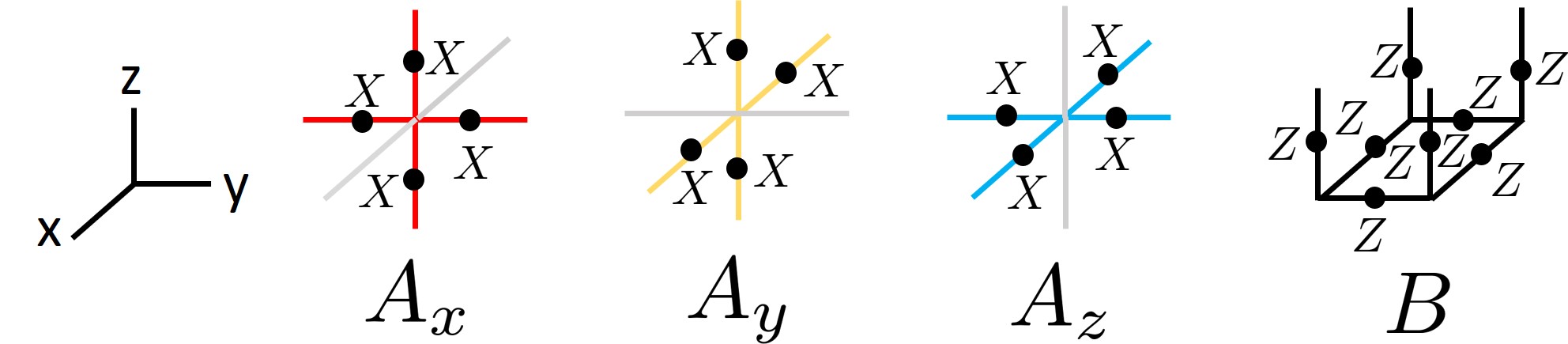}} \hfill
\subfloat[\label{fig:mm001Boundary}]{\includegraphics[width=0.8\columnwidth]{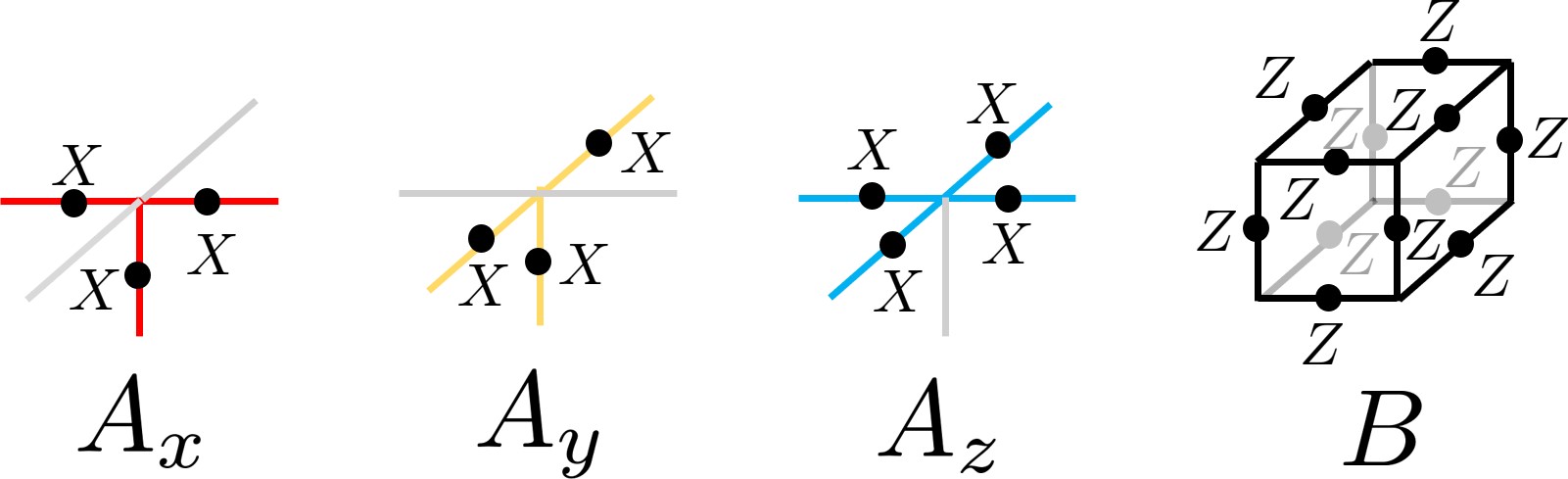}} \hfill
\subfloat[\label{fig:me001Boundary}]{\includegraphics[width=0.4\columnwidth]{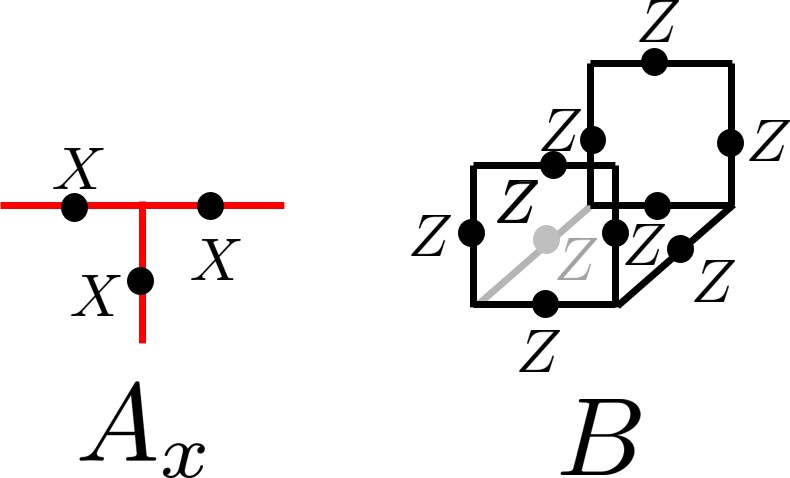}}\hfill
\subfloat[\label{fig:em001Boundary}]{\includegraphics[width=0.4\columnwidth]{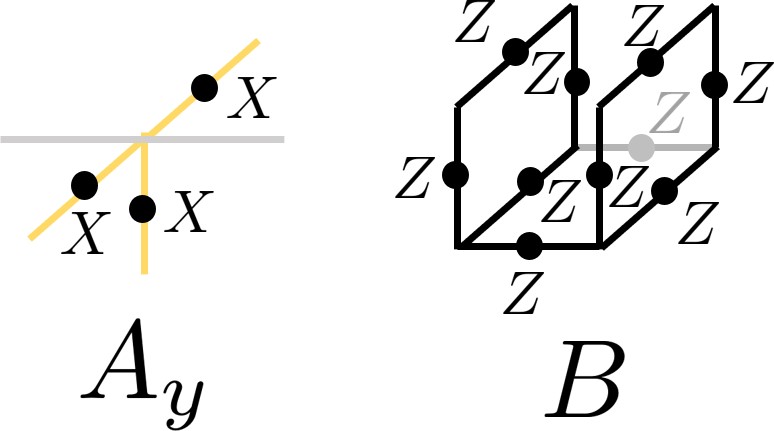}}\hfill
\caption{Boundary Hamiltonians for each gapped boundary condition for the $(001)$ surface of the X-cube model. (a) $(ee)$ boundary. (b) $(mm)$ boundary. (c) $(me)$ boundary. There are no $A_y$ or $A_z$ operators for the topmost layer of sites, but sites in the next layer down have all three bulk $A_i$ terms. (d) $(em)$ boundary.}
\label{fig:001Boundary}
\end{figure}

We begin by considering a semi-infinite system which occupies $\mathbb{R}^2 \times (-\infty, 0)$. What types of gapped boundary conditions are allowed? For simplicity we assume translation invariance (but no rotational invariance).

Boundary Hamiltonians corresponding to four possibilities are shown in Fig.~\ref{fig:001Boundary}. These boundary conditions can be simply written down, but to explain why one might expect them to describe all possible distinct gapped boundaries, it is useful to consider the isotropic layer construction of Refs.~\cite{VijayLayer,MaLayer}. Their construction consists of coupling interpenetrating layers of the (2+1)D toric code~\cite{KitaevQC03}, which undergoes a phase transition to the X-cube phase as a function of a certain coupling. The interpenetrating layers belong to three stacks of toric-code layers oriented perpendicular to each coordinate axis. (Because we are using the layer construction simply as motivation and do not derive any of our results from it, we refer the reader to Refs.~\cite{VijayLayer,MaLayer} for details of the layer construction.)

In the weak-coupling limit, the set of gapped boundaries of the layer construction is inherited from the allowed gapped boundaries of each toric code layer, which in turn comes from the set of Lagrangian subgroups of the quasiparticles in a layer. A gapped boundary condition for each layer is obtained by condensing either the $e$ (``rough boundary") or the $m$ (``smooth boundary") anyon in that layer.  There are therefore exactly four possible boundary conditions that obey translation symmetry in the plane of the boundary. We denote these boundaries as $(ee),(mm), (em),$ and $(me)$, where $(ab)$ means that the $a$ (resp. $b$) toric code quasiparticle is condensed in the layers oriented perpendicular to the $x$ (resp. $y$) direction. In the strong-coupling limit, one can show that they lead to the Hamiltonians in Fig.~\ref{fig:001Boundary}. The $(ee)$ boundary is ``rough" in that it has spins on ``dangling" $z$-oriented links, whereas the $(mm)$ boundary is ``smooth" in that it has no dangling $z$-oriented links. The topmost layer of the $(em)$ (resp. $(me)$) boundaries contain spins only on the $x$- (resp. $y$-) oriented links. It is not too hard to convince oneself (or to run a computerized check for small system size in a slab geometry) that all of these boundary conditions are gapped, in that no local operator commutes with both them and the bulk terms.

\begin{figure}
\centering
\includegraphics[width=.75\columnwidth]{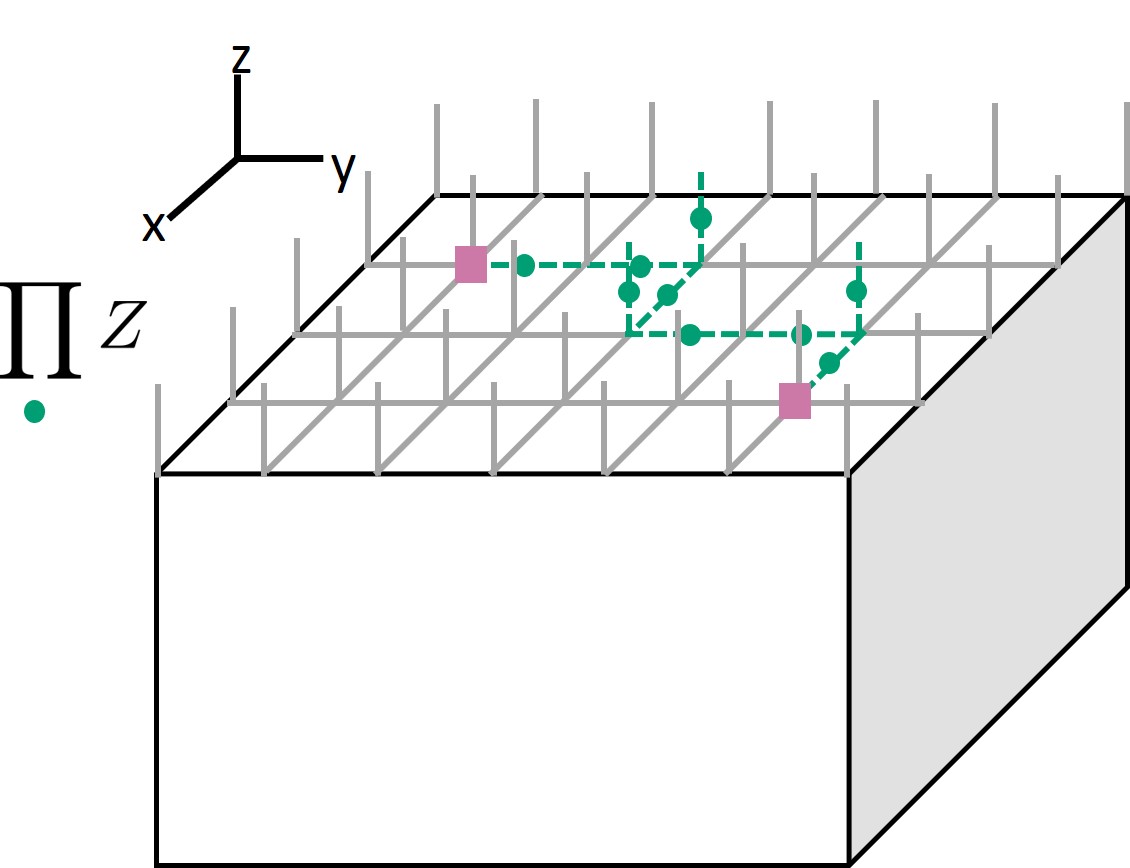}
\caption{String operator (green dashed line) consisting of $Z$ operators (green circles) that creates isolated $e_x=e_y$ excitations (purple squares) at its ends on an $(ee)$ boundary of the X-cube model. Acting with $Z$ on a vertically oriented bond fuses an $e_z$ onto the original excitation, transmuting it between $e_x$ and $e_y$. Such operators allow $e_x=e_y$ excitations to move in two dimensions on the surface.
}
\label{fig:ee2DString}
\end{figure}

We now examine the physical properties of such boundaries. We first consider the $(ee)$ boundary. Here it is obvious that a $Z$ operator acting on any ``dangling" surface bond creates a single $e_z$ excitation---that is, this excitation is condensed on the boundary. (Note that $q_{xz}$ and $q_{yz}$ are also condensed since they are bound states of $e_z$ excitations. Nothing particularly interesting happens to the $f_0$ and $m_{ij}$ sector.) Interestingly, this condensation implies that on the surface, $e_x$ can be turned into $e_y$ (and vice versa) by the action of a local operator, thanks to the fusion rule~\eqref{eq: e-e fusion}. In other words, the 1D particles $e_x$ and $e_y$ become a single 2D particle on the boundary due to the ability to capture $e_z$ particles from the condensate ``for free." An example string operator for this 2D particle is shown in Fig.~\ref{fig:ee2DString}.

\begin{figure}[b!]
\centering
\includegraphics[width=0.75\columnwidth]{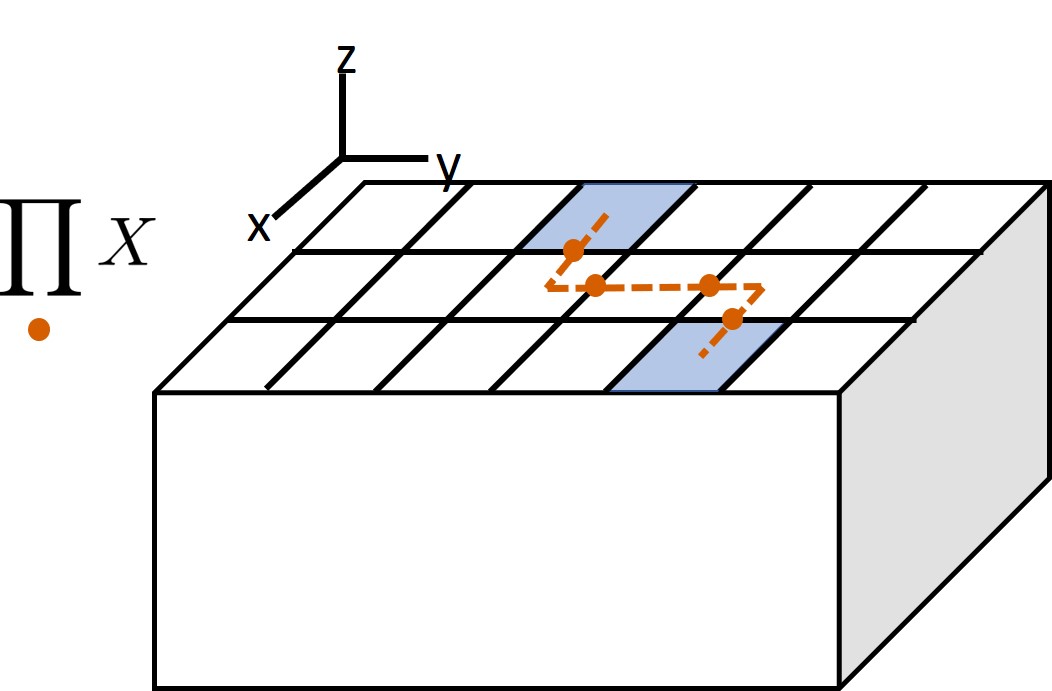}
\caption{String operator (orange dashed line) consisting of $X$ operators (orange circles) that creates isolated $f_0$ excitations (blue shading) at its ends on an $(mm)$ boundary of the X-cube model. Such operators allow $f_0$ excitations to move in two dimensions on the surface.}
\label{fig:mmFractonString}
\end{figure}

\begin{table*}
\centering
\renewcommand{\arraystretch}{1.3}
\begin{tabular}{@{}p{2cm}lp{2cm}ll}
\toprule[2pt]
\textbf{Boundary Type}&&\textbf{Condensed Excitations}&&\textbf{Surface Mobility Changes} \\ \hline
$(ee)$ && $e_z, q_{xz}, q_{yz}$ && $e_x = e_y$ mobile in $x,y$\\
$(mm)$ && $m_{xz}$, $m_{yz}$, $p_z$ && $f_0$ mobile in $x,y$\\
$(me)$ && $m_{yz}, q_{xz}$ && $f_0$ mobile in $x$, bulk $e_z$ converts to boundary $e_y$\\
$(em)$ && $m_{xz},q_{yz}$ && $f_0$ mobile in $y$, bulk $e_z$ converts to boundary $e_x$
\\	\bottomrule[2pt]	
\end{tabular}
\caption{Types of gapped $(001)$ boundaries of the X-cube model and properties of excitations on them.}
\label{tab:001BoundaryTypes}
\end{table*}

Next we consider the $(mm)$ boundary. Here we see that on the surface, applying an $X$ operator on a link in the plane of the boundary can create either an $m_{xz}$ or a $m_{yz}$ excitation; thus, these two excitations are condensed at the boundary. [Note that, due to the $m$-$m$ fusion rule \eqref{eq: m-m fusion}, the $p_z$ excitation is also condensed at the boundary.] Since the $m_{iz}$ excitations are themselves bound states of two $f_0$ excitations, this means that $f_0$, which is immobile in the bulk, becomes a 2D particle at the surface, see Fig.~\ref{fig:mmFractonString}.

\begin{figure}
\centering
\subfloat[\label{fig:qCondensed}]{\includegraphics[width=0.45\columnwidth]{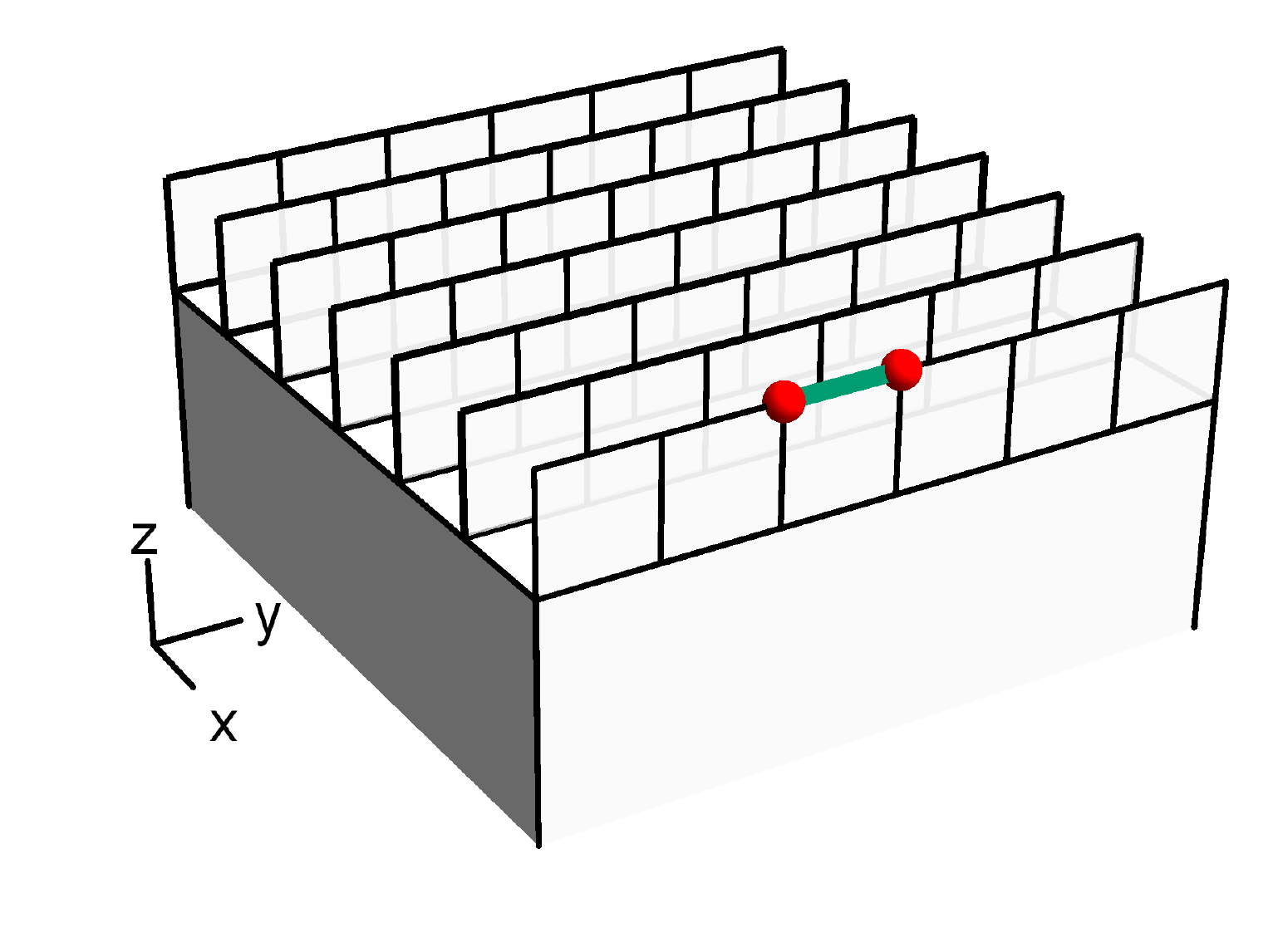}} \hfill
\subfloat[\label{fig:ezWithQCondensed}]{\includegraphics[width=0.45\columnwidth]{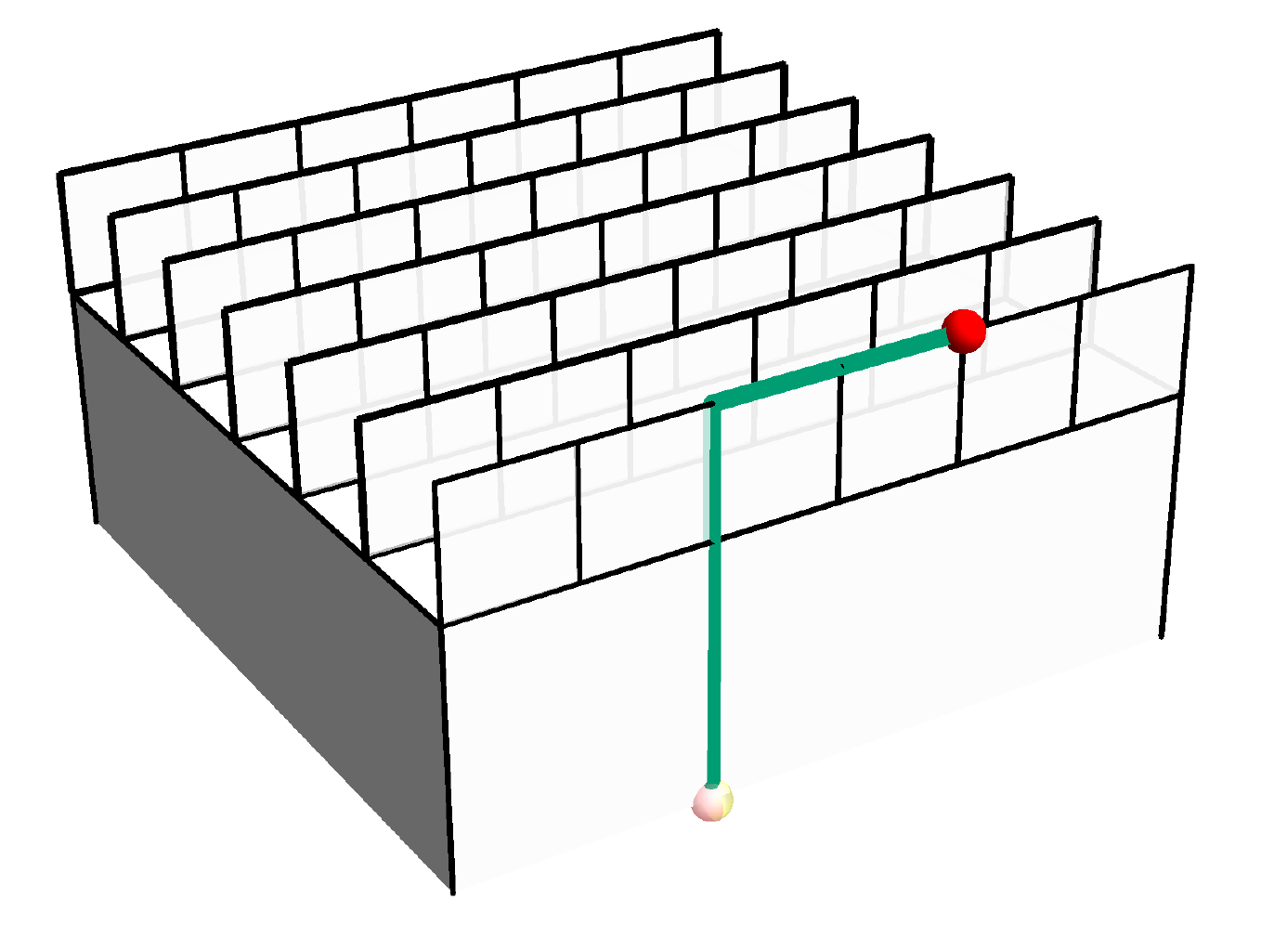}} \hfill
\caption{Condensation of $q_{xz}$ on $(001)$ surface of the X-cube model with $(me)$ boundary conditions. Green links mean $Z$ acts on this link. (a) Operator that creates a single $q_{xz}$ (its directionality can be checked by moving it into the bulk). (b) Operator that takes a bulk $e_z$ excitation (translucent), moves it to the boundary, and moves it in the $y$ direction by fusing it with $q_{xz}$ excitations.}
\end{figure}

Finally, we consider the $(me)$ boundary [the $(em)$ boundary is the same up to a $\pi/2$ rotation]. Here we can act on a horizontal bond in the plane of the surface with either a $Z$ or an $X$ operator. In the former case a single $q_{xz}$ is created, as in Fig.~\ref{fig:qCondensed}, while in the latter case a single $m_{yz}$ is created.  Thus, $q_{xz}$ and $m_{yz}$ are condensed at the boundary. Since $m_{yz}$ is condensed, the fracton $f_0$ becomes a 1D particle that is mobile in the $x$ direction at the boundary.  Furthermore, the condensation of $q_{xz}$ implies that a bulk $e_z$ excitation can be converted into an $e_y$ at the boundary by fusing with the condensed $q_{xz}$ particle, see Fig.~\ref{fig:ezWithQCondensed}.

What has happened? For each boundary condition, some set of excitations that are mobile in the $z$ direction condenses. Via the ``fusion rules," this condensation results in some excitations becoming more mobile on the surface than they are in the bulk. We summarize the results of this section in Table \ref{tab:001BoundaryTypes}. It is not obvious why these sets of excitations can condense to form a gapped boundary or if these are the only possible such sets. This table motivates the rest of our paper, wherein we conjecture a criterion relating the existence of a gapped boundary to a set of particles condensed at that boundary.

The above examples show that 2D excitations that do not exist in the bulk can appear at a gapped boundary.  We will find it useful to think of this ``new" particle content in terms of (2+1)D topological order at the surface of the (3+1)D fractonic bulk. Similar ideas have been discussed in the study of surface topological order that can arise at gapped boundaries of (3+1)D symmetry-protected topological phases~\cite{vonKeyserlingkSurfaceAnyons,BondersonTPfaffianSurface,FidkowskiSO3Level6Surface,JianLayerConstruction,MetlitskiTPfaffianSurface}.

\begin{figure}
\includegraphics[width=0.75\columnwidth]{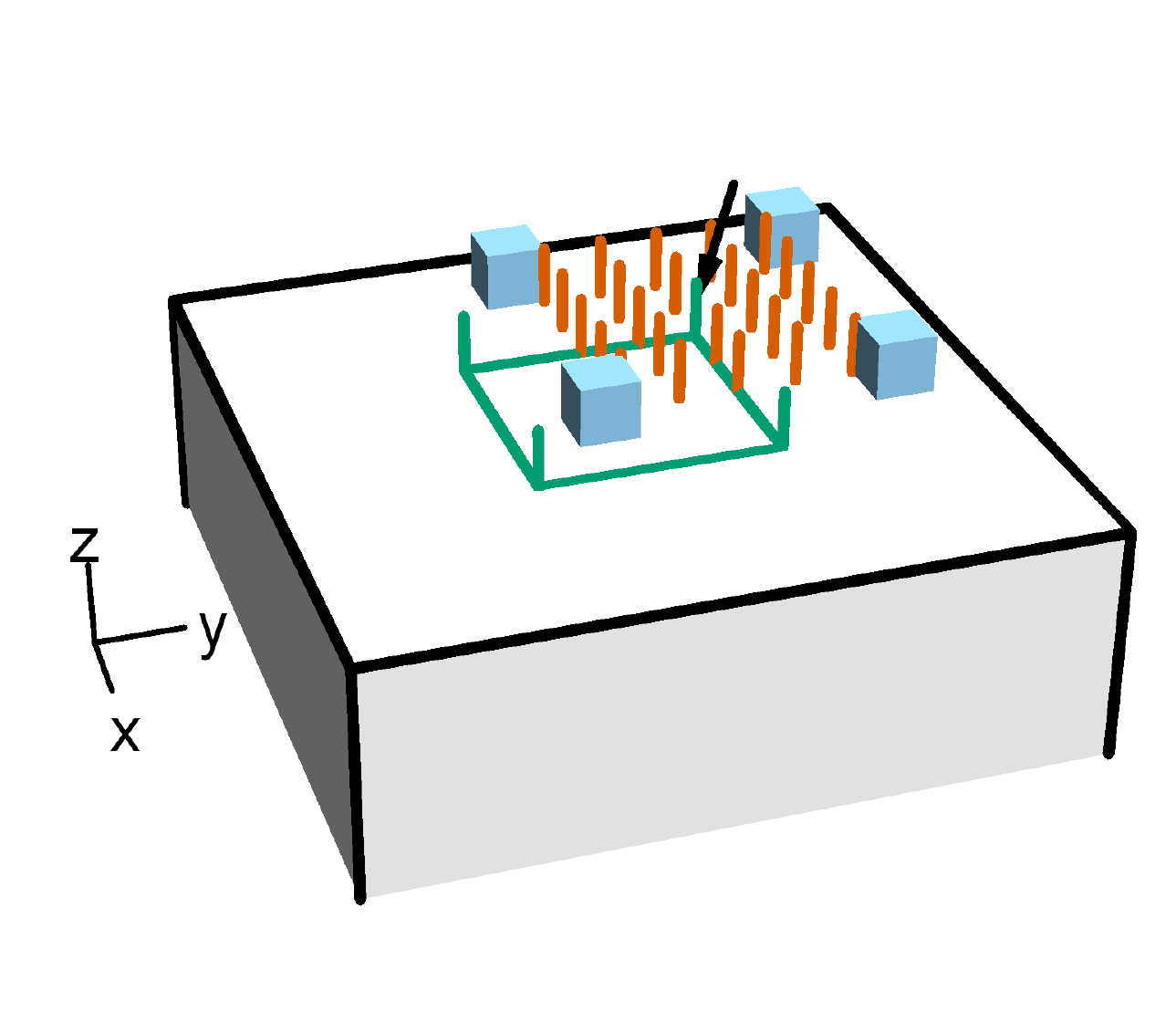}
\caption{Surface $e$ excitation encircling a surface $f_0$ excitation on the $(ee)$ boundary of the X-Cube model. Orange links are part of the membrane $X$ operator creating the $f_0$ excitations (blue cubes). Green links are part of the surface Wilson loop operator for the surface $e$ excitation. The arrow indicates the only link where both the membrane and Wilson loop have support, causing the operators to anticommute.}
\label{fig:eeSurfaceEAroundf0}
\end{figure}

We first consider the $(ee)$ boundary. In this case, the surface has two deconfined 2D particles: $m_{xy}$ and $e_x=e_y \equiv e$. These excitations are fully deconfined in the plane of the surface and cannot be moved into the bulk, and it is straightforward to check that these excitations indeed obey the $\mathbb{Z}_2$ toric code fusion rules and statistics. Thus, it is reasonable to view them as pointlike particles belonging to a $\mathbb Z_2$ topologically ordered layer at the surface. However, the surface can of course harbor other excitations besides $e$ and $m_{xy}$. For example the $m_{xz}$ and $m_{yz}$ excitations can be brought to the surface from the bulk and, within the surface, are one-dimensional objects. However, one can check that both $e$ and $m_{xy}$ braid trivially with them, so they do not constitute nontrivial quasiparticles from the point of view of the surface. On the other hand, one can show that the surface $e$ particle picks up a phase $\pi$ when it encircles an isolated $f_0$ excitation, see Fig.~\ref{fig:eeSurfaceEAroundf0}.  Note that this phase does not depend on the spatial orientation of the membrane operator that supports the fracton.  Since $f_0$ is immobile in isolation, and can only move in coordination with other fractons far away by crossing a large energy barrier, we view it as a ``topological defect" that is extrinsic to the (2+1)D topological order at the surface.  Hence, we should think of this surface as having $\mathbb{Z}_2$ topological order, but in the presence of certain topological defects that are associated with fractons. 

The $(mm)$ boundary is somewhat similar. It also has two deconfined 2D particles: $q_{xy}$ and $f_0$. These also obey $\mathbb Z_2$ toric-code braiding and fusion rules, and this time $f_0$ braids nontrivially with the 1D $e_x$ and $e_y$ excitations. Note that two $e_x$ or two $e_y$ can be created on the surface exactly as they are created in the bulk, or one $e_x$ and one $e_y$ can be brought to the surface by bringing an $e_z$ there from the bulk and splitting it according to the fusion rule~\eqref{eq: e-e fusion}.  Thus, the $(mm)$ boundary can also be viewed as a layer with $\mathbb Z_2$ topological order that can host a set of unusual ``defects," the bulk $e_x$ and $e_y$ excitations, that are mobile only in the $x$- and $y$-directions.

The $(me)$ boundary only harbors 2D excitations that are also 2D in the bulk, namely $m_{xy}$ and $q_{xy}$.  These particles also obey $\mathbb Z_2$ toric-code fusion and braiding rules.  However, because this surface does not harbor any ``new" 2D particles, we should not view these excitations as belonging to a topological order at the surface that was induced by the condensation of the $m_{yz}$ and $q_{xz}$ excitations.  Instead, the condensation of the $m_{yz}$ and $q_{xz}$ excitations converts the $f_0$ into a 1D particle at the surface.  Because this ``new" 1D particle is outside of the scope of (2+1)D topological order (as are all subdimensional excitations), it does not seem appropriate to view this boundary as exhibiting surface topological order \textit{per se}.

\subsection{(110) Boundary}
\label{subsec:XCube110}

From Table \ref{tab:001BoundaryTypes}, we note that we have only found gapped boundaries where excitations with mobility in the $z$-direction were condensed. One thus might conjecture that gapped surfaces are related only to the condensation of particles which are mobile in the direction normal to the surface. However, we will now show that mobility is too strong of a condition by considering boundaries normal to the $(110)$ direction. In particular, we exhibit three boundary conditions where \textit{immobile} bulk excitations are condensed. We focus on these three boundary conditions, which will play an important role in our later understanding, and will not provide an exhaustive list of all possible $(110)$ gapped boundaries.

\begin{figure}
\centering
\includegraphics[width=0.7\columnwidth]{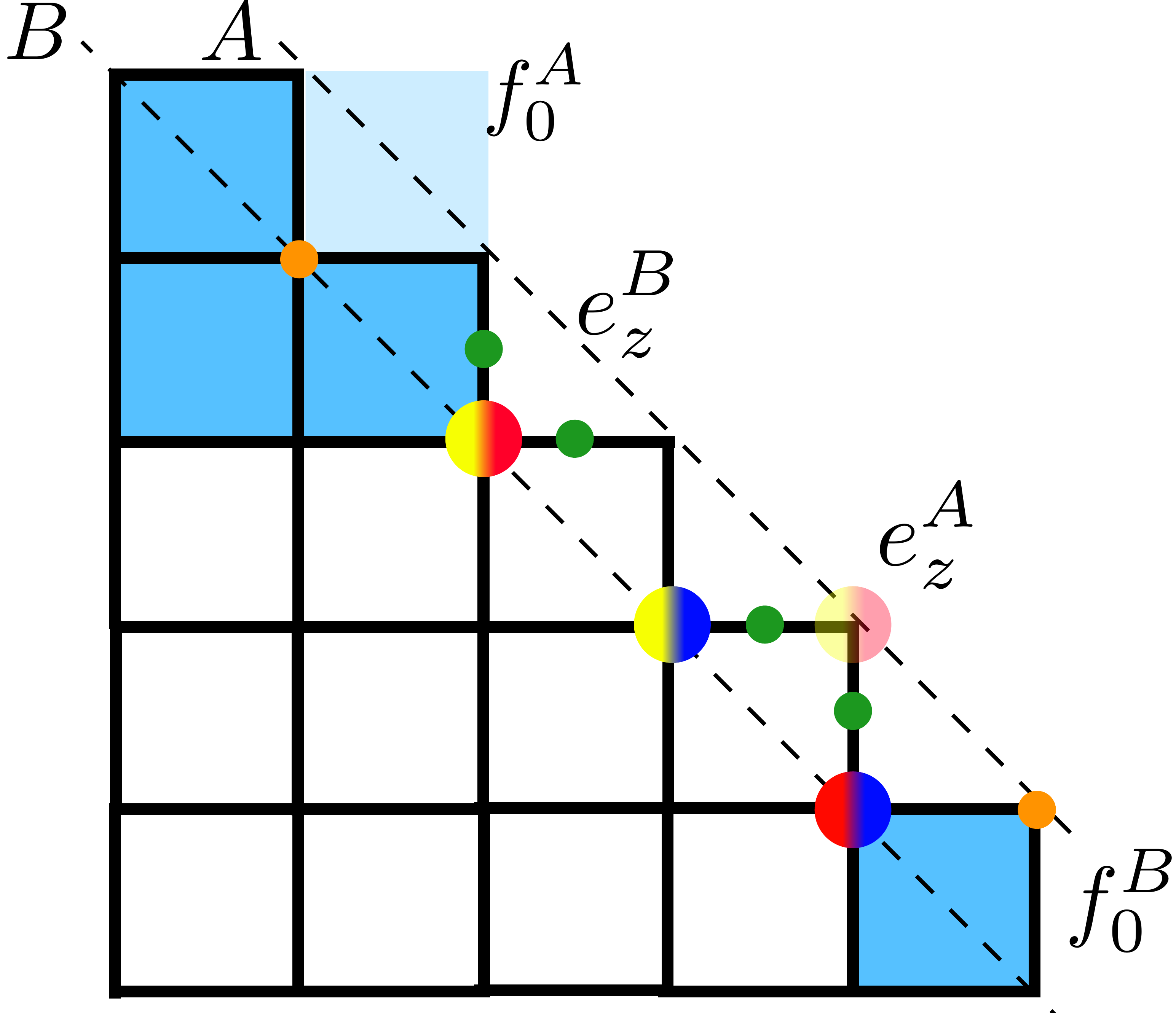}
\caption{Top view of the cubic lattice with $(110)$ boundary, with boundary excitations depicted in Fig.~\ref{fig:110Boundary}.  The vertices that live on the boundary can be divided into two sublattices, $A$ and $B$ (dashed lines).  The $f^A_0$ excitation is a bound state of three fractons (blue squares) that would be equivalent modulo local operators to a single fracton in the bulk (transparent blue square).  The $f^B_0$ excitation is equivalent to a bulk $f_0$.  The $e^A_z$ excitation is a bound state of an $e_x$ and an $e_y$ excitation (blue/yellow and blue/red circles) that would be equivalent modulo local operators to an $e_z$ in the bulk (transparent red/yellow circle).  The $e^B_z$ excitation is equivalent to a bulk $e_z$.}
\label{fig:110Boundary_2Dproj}
\end{figure}

\begin{figure}
\centering
\subfloat[\label{fig:110Boundary_f0}]{\includegraphics[width=0.5\columnwidth]{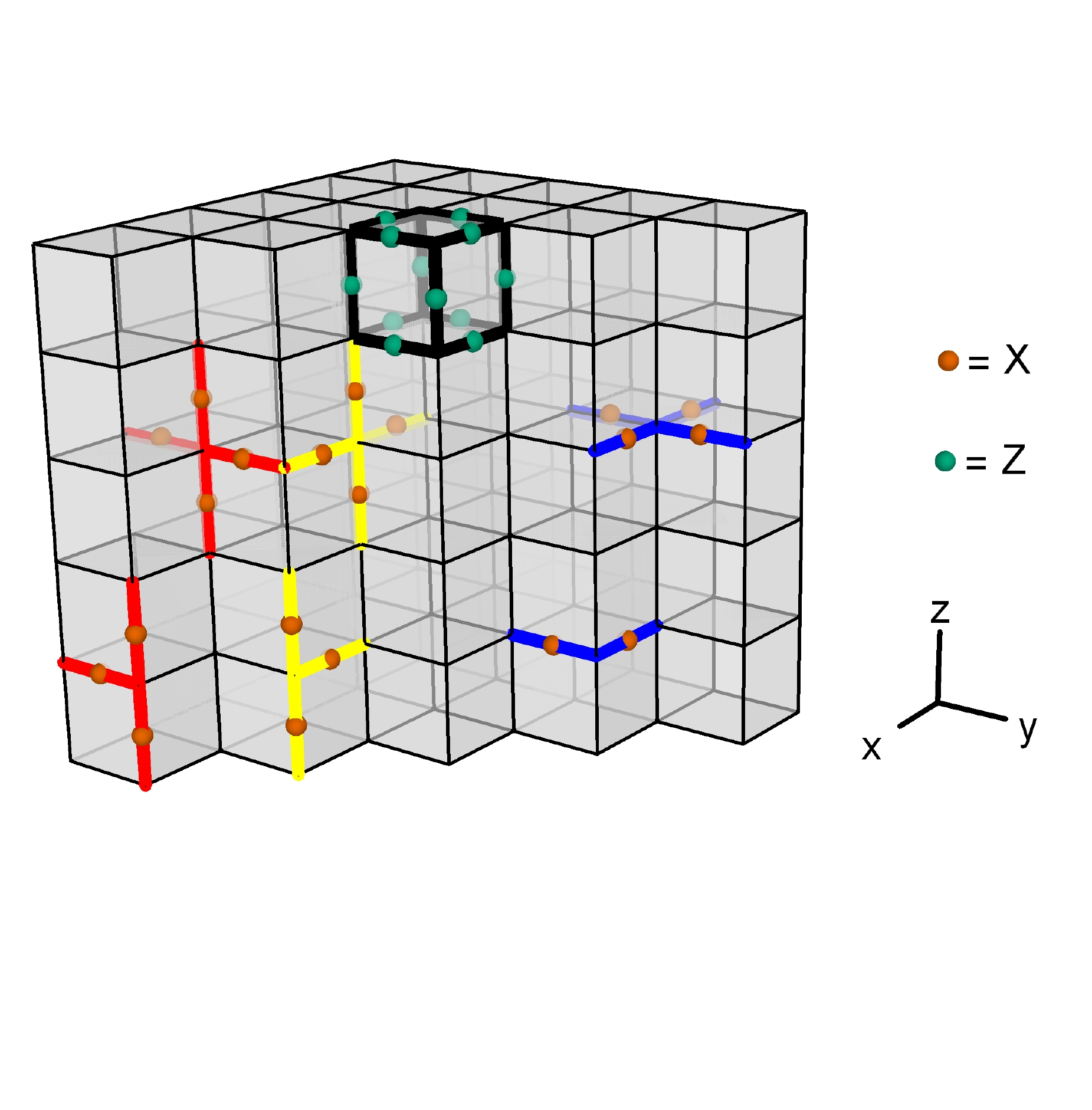}} \hfill
\subfloat[\label{fig:110Boundary_f0Excitations}]{\includegraphics[width=0.45\columnwidth]{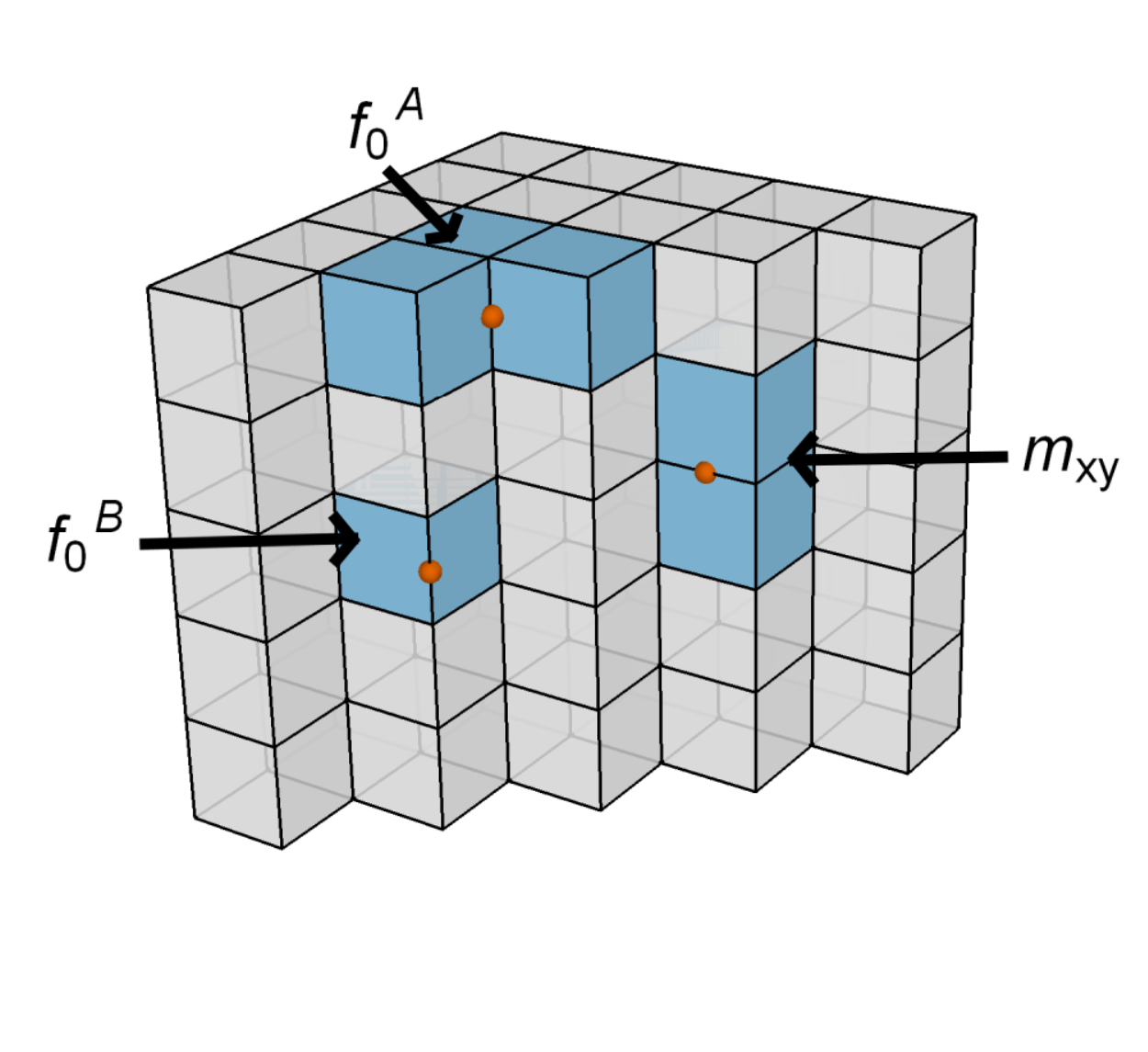}} \hfill
\subfloat[\label{fig:110Boundary_f0ez}]{\includegraphics[width=0.5\columnwidth]{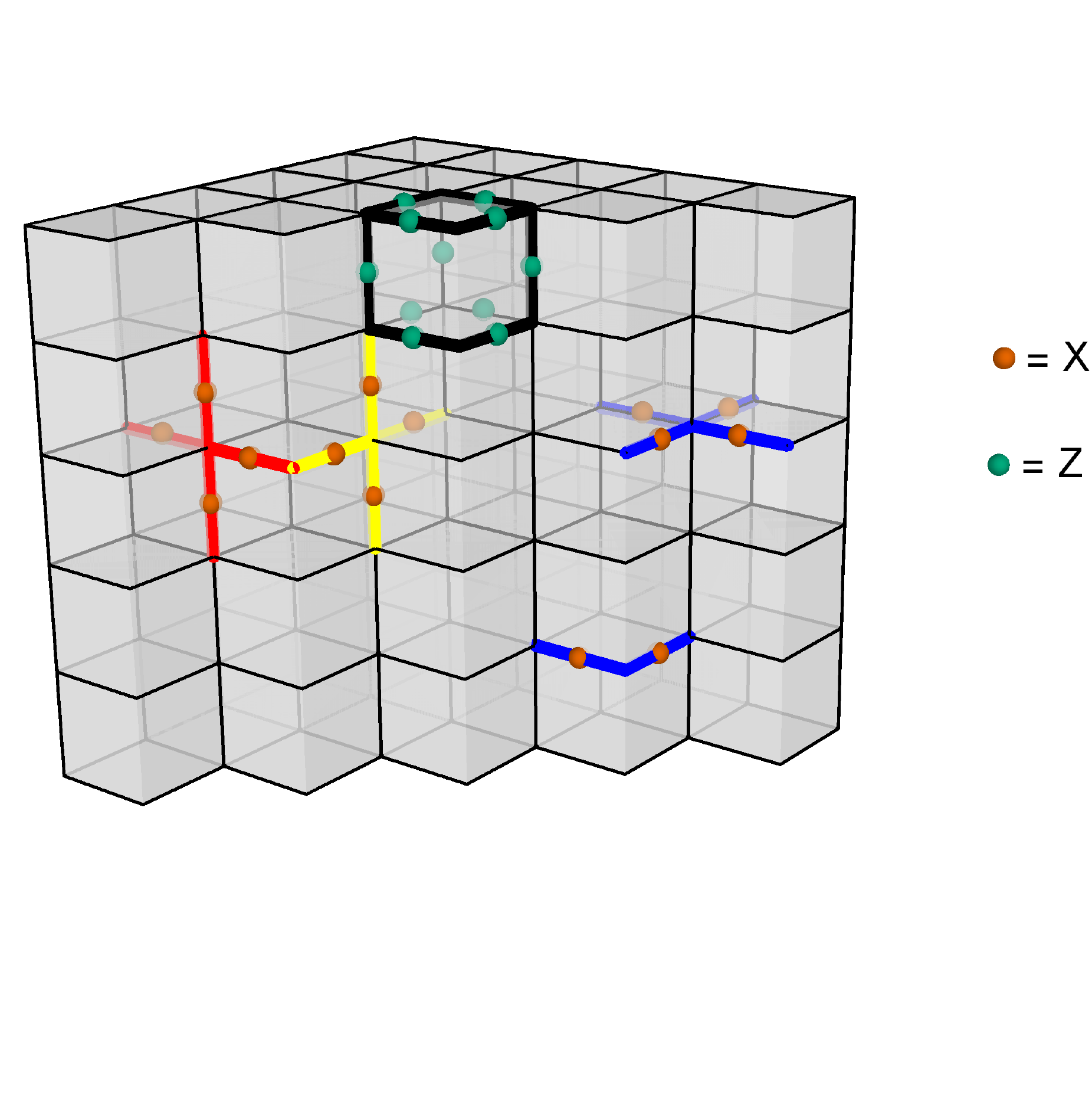}} \hfill
\subfloat[\label{fig:110Boundary_f0ezExcitations}]{\includegraphics[width=0.45\columnwidth]{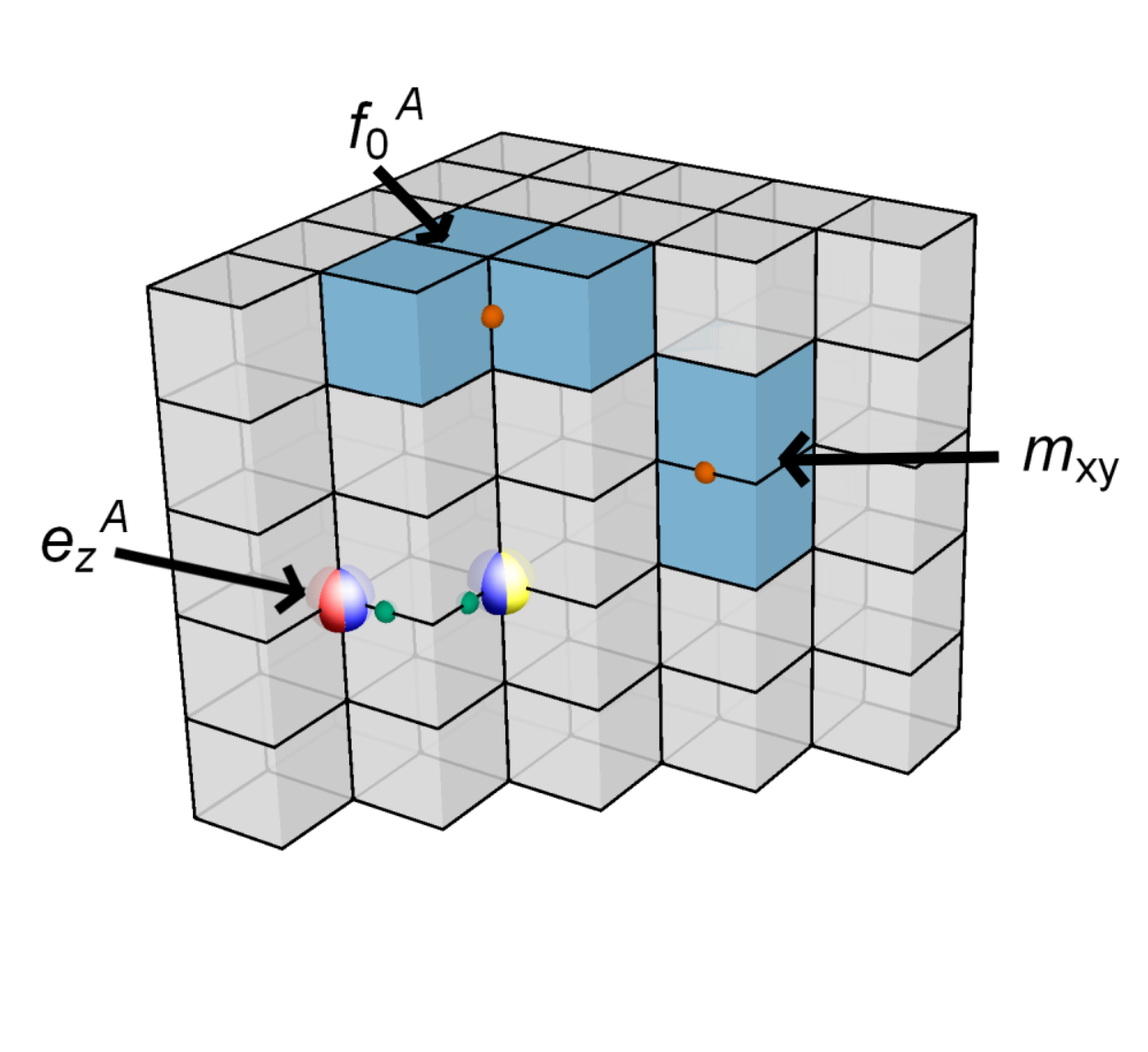}} \hfill
\subfloat[\label{fig:110Boundary_f0BezB}]{\includegraphics[width=0.45\columnwidth]{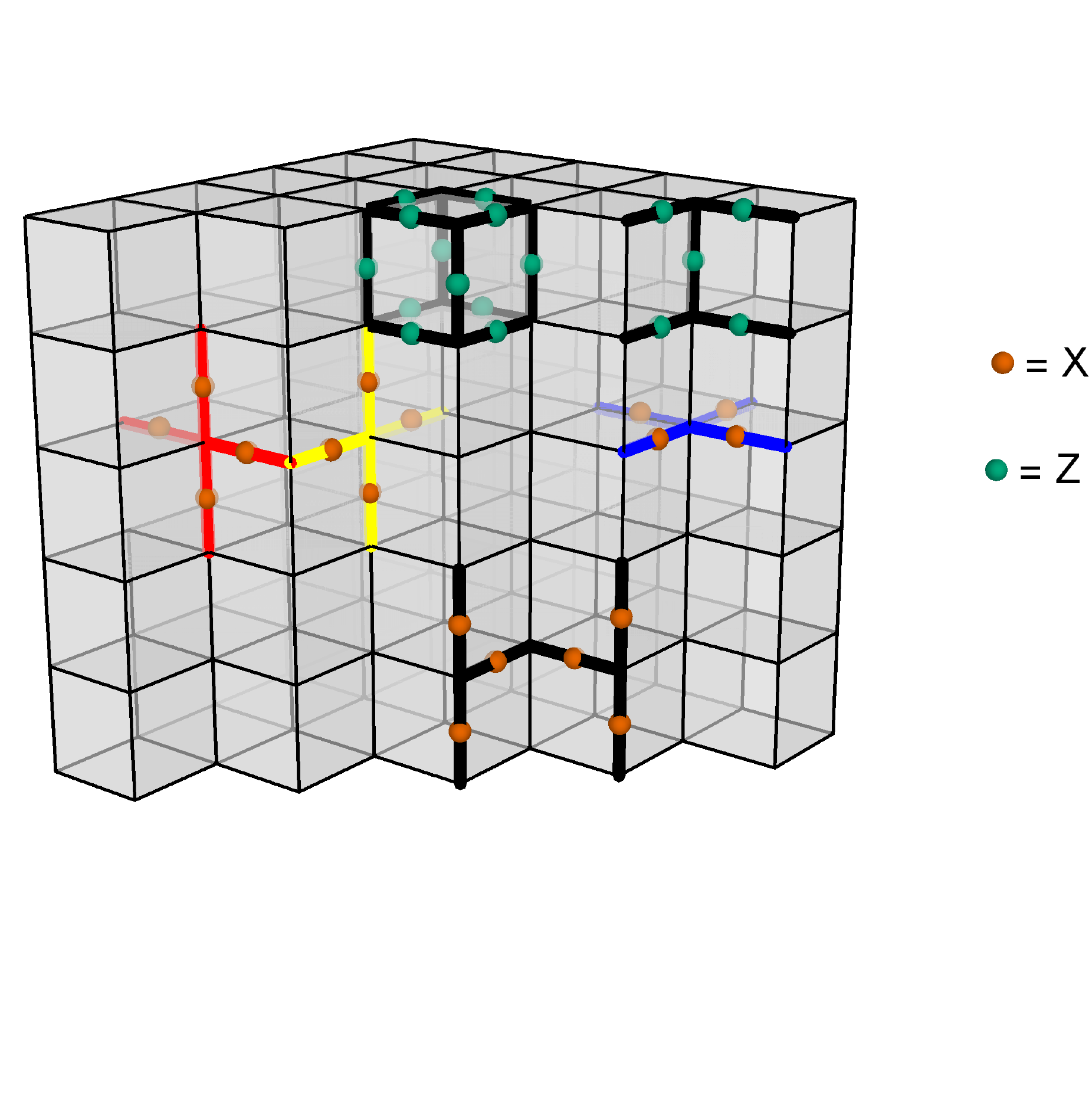}} \hfill
\subfloat[\label{fig:110Boundary_f0BezBExcitations}]{\includegraphics[width=0.45\columnwidth]{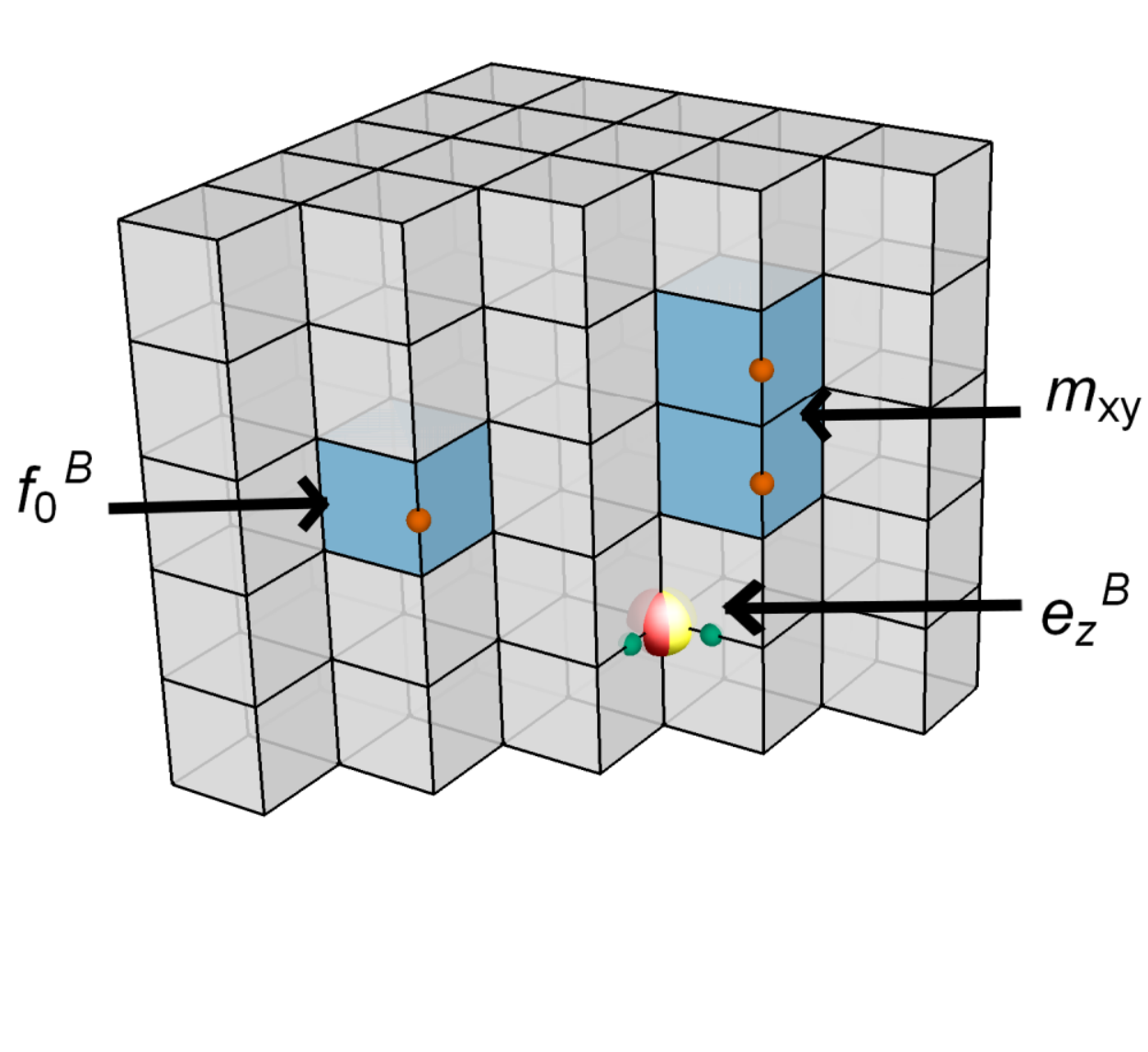}} \hfill
\caption{(a), (c), and (e): Terms in the boundary Hamiltonian for three gapped boundaries at the (110) surface of the X-cube model. Green (orange) spheres represent $Z$ ($X$) operators. For boundary condition (c), there are no spins on the outermost $z$-oriented links of the lattice. (b), (d), and (f): Action of local operators that create condensed excitations for the boundary conditions (a), (c), and (e) respectively. Operators are represented in the same way as in (a), (c), and (e). Excitations are illustrated using the visual representations in Table \ref{tab:XCubeExcitations}.}
\label{fig:110Boundary}
\end{figure}

We first discuss the relevant excitations at the $(110)$ boundary.  Viewed from the top, the boundary can be divided into two sublattices, $A$ and $B$, as shown in Fig.~\ref{fig:110Boundary_2Dproj}. Up to translations that preserve the boundary, there are \textit{two} inequivalent boundary $f_0$ excitations, which we label $f_0^A$ and $f_0^B$ as shown in Fig.~\ref{fig:110Boundary_2Dproj}.  $f^A_0$ can also be viewed as a three-fracton bound state, see Fig.~\ref{fig:110Boundary_2Dproj}. In the bulk, such excitations would be related by translation symmetry, but the boundary breaks that translation symmetry and renders them inequivalent. Further, because $f_0$ is immobile, there is no local operator that relates $f_0^A$ to $f_0^B$. This is the sense in which we are justified in treating these as different excitations. Similarly, there are two inequivalent $e_z$ excitations on the boundary, which we call $e^A_z$ and $e^B_z$, see Fig.~\ref{fig:110Boundary_2Dproj}.

The three boundary Hamiltonians we consider are shown in Figs.~\ref{fig:110Boundary_f0}, \ref{fig:110Boundary_f0ez}, and \ref{fig:110Boundary_f0BezB}, and the operators that create condensed excitations are shown in Figs.~\ref{fig:110Boundary_f0Excitations}, \ref{fig:110Boundary_f0ezExcitations}, and \ref{fig:110Boundary_f0BezBExcitations}.  The Hamiltonian in Fig.~\ref{fig:110Boundary_f0} condenses $f^A_0$, $f^B_0$, see Fig.~\ref{fig:110Boundary_f0Excitations}, but does not condense any particles in the $e$ sector.  The Hamiltonian in Fig.~\ref{fig:110Boundary_f0ez} condenses $f^A_0$ and $e^A_z$, see Fig.~\ref{fig:110Boundary_f0ezExcitations}, while the Hamiltonian in Fig.~\ref{fig:110Boundary_f0BezB} condenses $f^B_0$ and $e^B_z$.  All three boundaries host a condensed $m_{xy}$ excitation. 

In addition to the fact that neither $f_0$ nor $e_z$ excitations are mobile in the $(110)$ direction, the fact that all three boundaries are gapped is remarkable. As discussed in Sec.~\ref{subsec: bulk braiding}, there is a sense in which $e_z$ and $f_0$ braid nontrivially in the bulk.  Nevertheless, the boundary Hamiltonians shown in Figs.~ \ref{fig:110Boundary_f0ez} and \ref{fig:110Boundary_f0BezB} host condensed particles in both the fracton and $e$ sectors.  These examples rule out any naive generalization of a Lagrangian subgroup criterion based on bulk braiding, as such a generalization would indicate that these two boundaries would be gapless. As we will see in the next section, the missing ingredient that must be taken into account is the geometry of the boundary.

\section{Boundary Braiding of Fractons}
\label{sec: boundary braiding}

The above discussion suggests that purely bulk braiding properties are insufficient to determine the set of possible gapped boundaries for all surface terminations. Our conjecture about gapped boundaries will therefore require a boundary-based definition of braiding for subdimensional excitations, which we now discuss.

\subsection{Preliminary Definitions}
\label{subsec: Preliminary Definitions}

In the arguments we discussed in Sec.~\ref{sec:2DReview} in the context of (2+1)D topological order, \textit{open} Wilson lines that terminate at the boundary play a key role. We presently define the fractonic analogue of such an object. We assume the system has a boundary throughout.

\textit{Definition}: Given a fractonic system with a boundary, a \textit{boundary half-cage (BHC) operator} $W_a$ for an excitation $a$ at that boundary is an operator that:
\begin{itemize}
\item Creates no excitations in the bulk,
\item Creates isolated $a$ excitations on the boundary,
\item Has support on the boundary only close to the $a$ excitations, and
\item Has support on a finite region of space (i.e., one that does not grow with system size).
\end{itemize} 
The term ``close" used above should be roughly interpreted as ``within a few correlation lengths."

\begin{figure}
\centering
\subfloat[\label{fig:ezHalfCage001}]{\includegraphics[width=0.4\columnwidth]{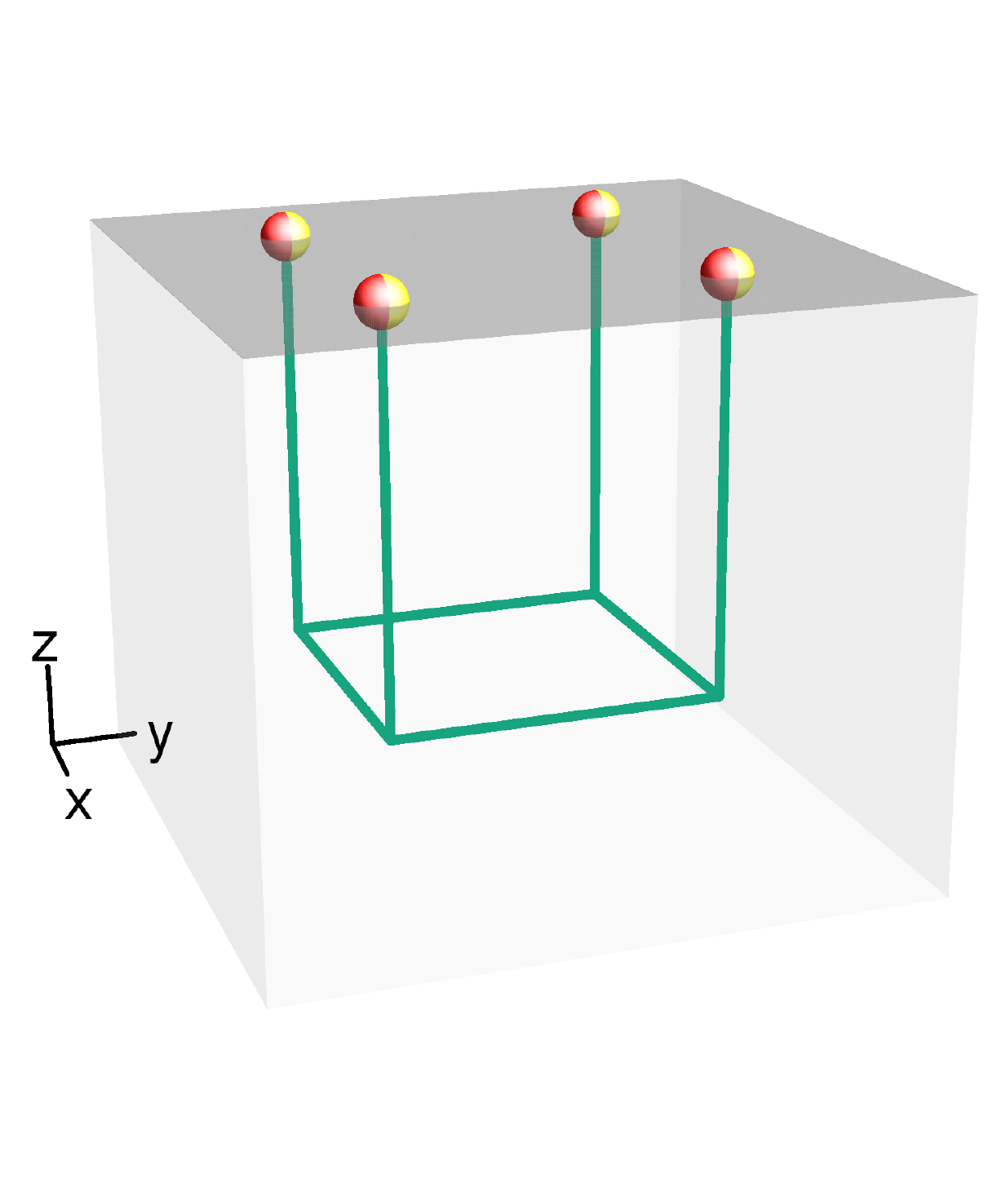}} \hfill
\subfloat[\label{fig:myzHalfCage001}]{\includegraphics[width=0.4\columnwidth]{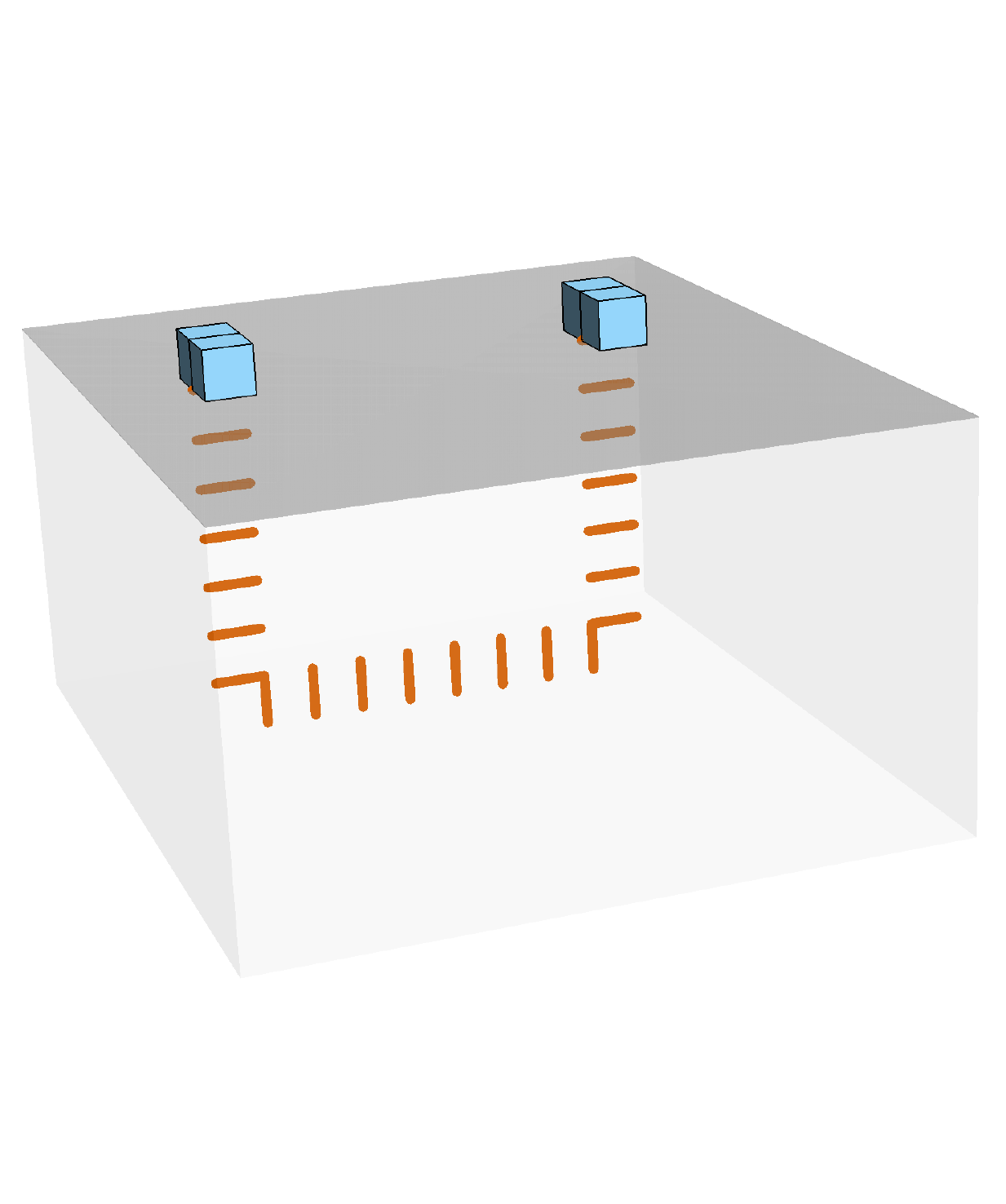}} \hfill
\subfloat[\label{fig:ezAHalfCage}]{\includegraphics[width=0.45\columnwidth]{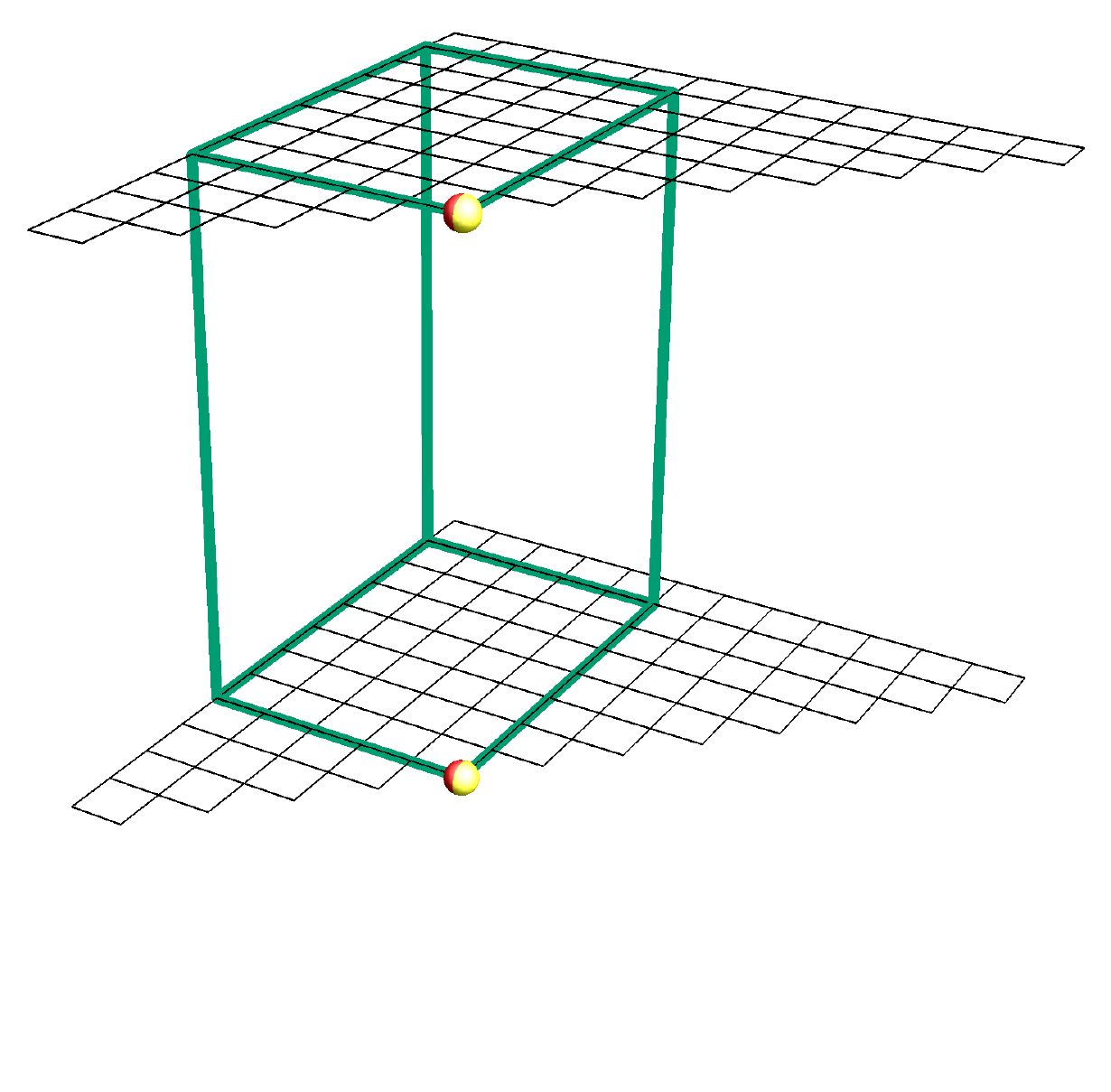}} \hfill
\subfloat[\label{fig:ezBHalfCage}]{\includegraphics[width=0.45\columnwidth]{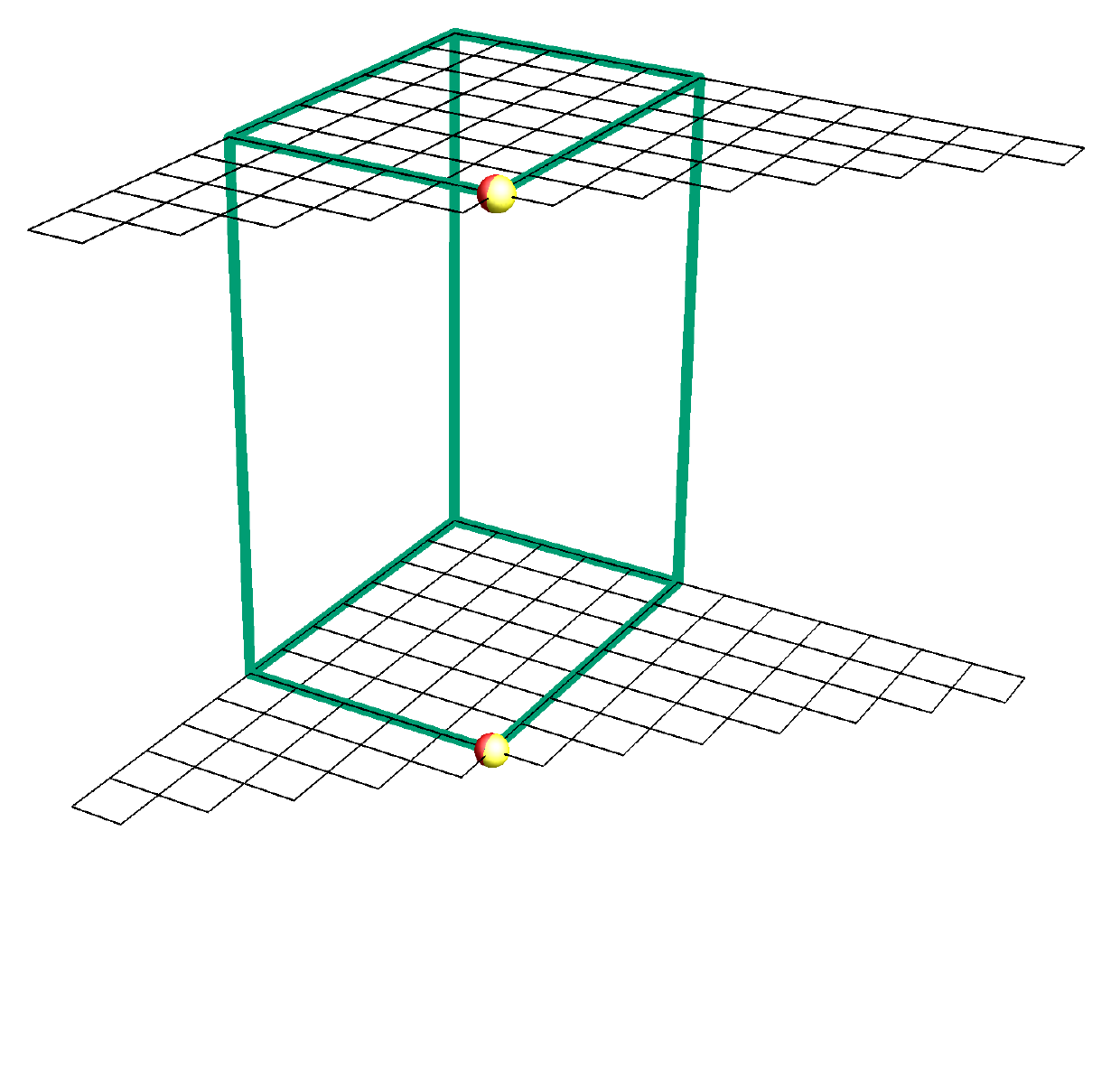}} \hfill
\subfloat[\label{fig:f0HalfCage}]{\includegraphics[width=0.45\columnwidth]{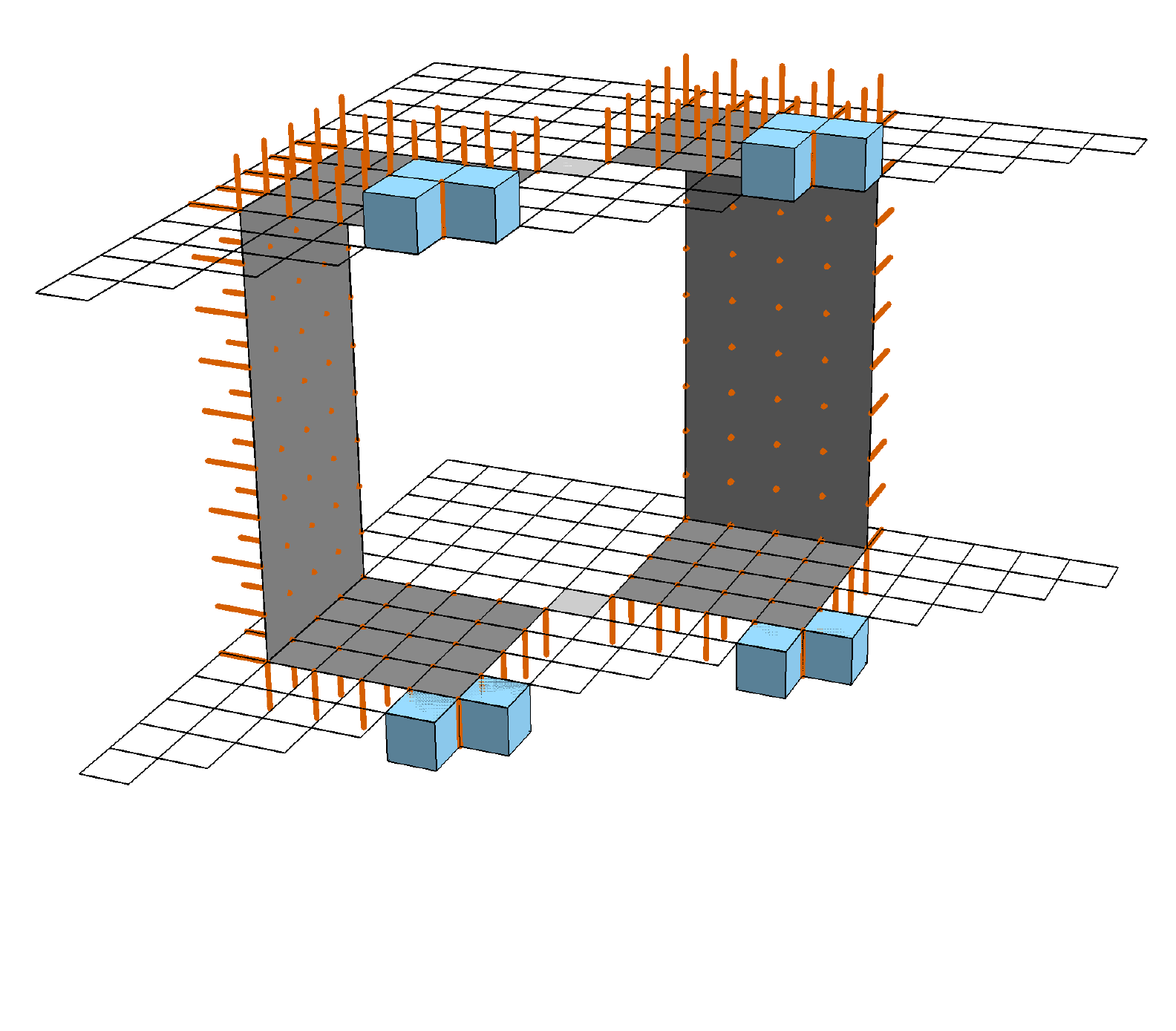}} \hfill
\caption{Some examples of boundary half-cages in the X-Cube model.(a) and (b): Boundary half-cage for $e_z$ and $m_{yz}$, respectively, on the (001) surface. (c) and (d): Boundary half-cage for $e_z^A$ and $e_z^B$, respectively, on the (110) surface. The operators differ by a single lattice site. (e): Boundary half-cage for $f_0^A$ on the (110) surface.}
\label{fig:boundaryHalfCages}
\end{figure}

\begin{figure}
\centering
\includegraphics[width=0.45\columnwidth]{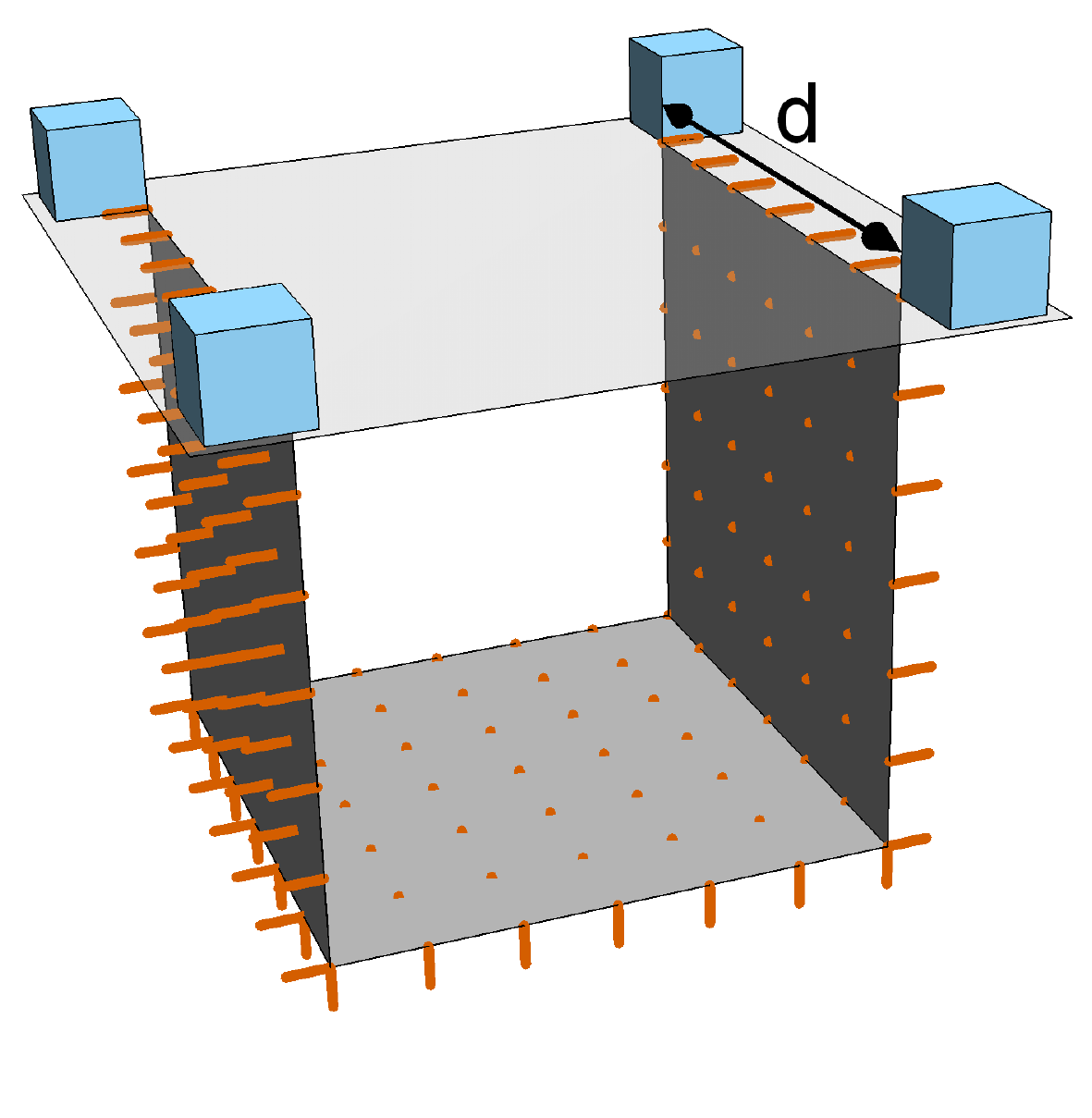}
\caption{Operator that creates isolated $f_0$ excitations on the (001) surface (light grey plane). This operator is not a boundary half-cage because it has support on the boundary at points arbitrarily far away from the $f_0$ excitations.}
\label{fig:badf0Cage001}
\end{figure}

Some examples of BHC operators for particular boundaries of the X-Cube model are shown in Fig.~\ref{fig:boundaryHalfCages}. Crucially, the existence of such operators is strongly dependent on the boundary itself. For example, no BHC exists for the $f_0$ excitations on the $(001)$ boundary of the X-cube model: one can verify that any attempt at building such an operator violates the third criterion above, since pairs of faraway fractons on the boundary are connected by strings of operators, see Fig.~\ref{fig:badf0Cage001}.  However, a BHC operator for $f_0$ does exist on the $(110)$ boundary, see Fig.~\ref{fig:f0HalfCage}.

For practical purposes, it is desirable to have a rule for when BHCs exist. It appears that the mobility of a particle in the direction normal to the boundary is sufficient but not necessary for the existence of a BHC.  For example, a BHC operator for $f_0$ exists on the $(110)$ boundary of the X-cube model even though $f_0$ is immobile. In all examples we have studied, the existence of BHCs is tied to a weaker form of mobility, which we call ``cascading:" given a particular boundary termination, any excitation for which a BHC exists must be able to \textit{cascade} in the direction normal to the boundary. To avoid postponing the discussion of boundary braiding, we define cascading in Appendix \ref{app:bulkDefinitions}.

BHCs can now be used to define boundary braiding:

\textit{Definition}: Suppose BHCs exist for two excitations $a$ and $b$. Then we say $a$ and $b$ \textit{braid nontrivially on the boundary} if, for any BHC $W_a$ for $a$, there exists a BHC $W_b$ for $b$ such that $[W_a,W_b] \neq 0$ and all excitations created by $W_a$ are well-separated from all excitations created by $W_b$. The braiding is Abelian if 
\begin{align}
\label{eq: abelian boundary braiding}
W_aW_b = e^{i\theta_{ab}}W_b W_a
\end{align}
for some braiding phase $\theta_{ab}$.

Applied in the context of (2+1)D topological order, this definition reduces to the definition of particle exchange at the boundary that was given in Sec.~\ref{sec:2DReview}.  Indeed, open Wilson loops that terminate on the boundary satisfy our definition of BHCs.

In the bulk, cage operators like the ones defined in Sec.~\ref{subsec: bulk braiding} can be split into two bulk half-cage operators (see Appendix~\ref{app:bulkDefinitions}) and nontrivial braiding can be defined making use of half-cages in place of the operators $\mathcal O_a$ used in Eq.~\eqref{eq: fracton bulk braiding def}. However, as discussed in Sec.~\ref{subsec: bulk braiding}, the nonreciprocity of bulk braiding makes the triviality or nontriviality of braiding ambiguous. It appears that there is no such ambiguity for boundary braiding, which takes into account the geometry of the boundary through the use of BHCs.

To illustrate these concepts and to elaborate on the highly geometry-dependent nature of boundary braiding, we give some examples of boundary braiding in the X-cube model.

\subsection{Boundary Braiding in the X-cube Model}

\begin{figure}
\centering
\includegraphics[width=0.45\columnwidth]{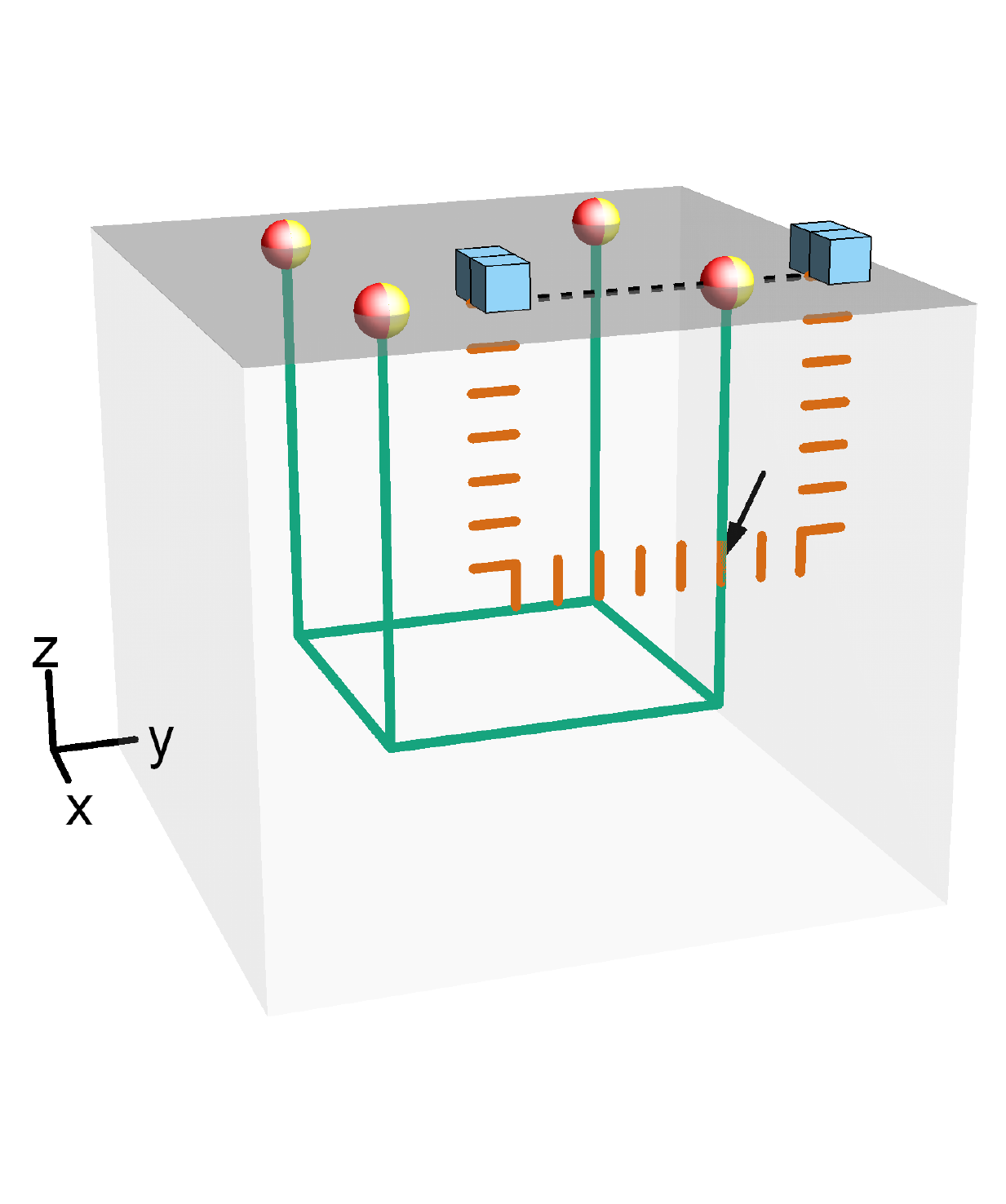}
\caption{Boundary braiding of $e_z$ and $m_{yz}$ on the $(001)$ surface of the X-cube model. The boundary half-cage operators anticommute because both act on the link indicated by the black arrow. The dashed line is a guide to the eye - placing the $m_{yz}$ excitations at a different position in $x$ may not yield nontrivial braiding.}
\label{fig:emBoundaryBraid001}
\end{figure}

The simplest example of nontrivial boundary braiding is $e_z$ and $m_{yz}$ on the $(001)$ surface of the X-cube model. Their BHC operators are shown in Figs.~\ref{fig:ezHalfCage001} and \ref{fig:myzHalfCage001}. If the operators are arranged geometrically as in Fig.~\ref{fig:emBoundaryBraid001}, then they anticommute, so these excitations braid nontrivially. We will see that this is the reason both cannot condense simultaneously on the $(001)$ boundary. Much as we saw for bulk braiding, boundary braiding is geometry dependent; translating the $m_{yz}$ BHC operator by even a single lattice site in the $x$ direction causes them to commute. 

On the other hand, $f_0$ simply does not have a BHC on the $(001)$ boundary; for example, the operator shown in Fig.~\ref{fig:badf0Cage001} has boundary support far from the $f_0$ excitations. Therefore, boundary braiding in the sense defined here does not exist for $f_0$ excitations on this boundary. As alluded to in Sec.~\ref{subsec: Preliminary Definitions}, the nonexistence of a BHC for $f_0$ on the $(001)$ boundary is related to the fact that $f_0$ does not cascade in the $\hat{\bv{z}}$-direction (see Appendix \ref{app:bulkDefinitions} for a definition of cascading).  However, if the separation $d$ in Fig.~\ref{fig:badf0Cage001} is taken to be a single lattice spacing, then we obtain a BHC for $m_{xz}$. Indeed, $m_{xz}$ is mobile in the $\hat{\bv{z}}$-direction (therefore, in particular, it cascades in that direction), so it is reasonable to expect that it has a BHC on the $(001)$ boundary.  We saw in Sec.~\ref{subsec: (001) Boundary} that, while the $(mm)$ boundary in the $(001)$ direction hosts a condensed $m_{xz}$ excitation, there does not seem to be a boundary Hamiltonian that condenses $f_0$ on the $(001)$ boundary.  The conjecture that we formulate in Sec.~\ref{sec: Boundary Lagrangian Subgroup Conjecture} would attribute this fact to the nonexistence of BHCs for $f_0$ on the $(001)$ boundary.

\begin{figure}
\centering
\subfloat[\label{fig:f0AezABraid}]{\includegraphics[width=0.45\columnwidth]{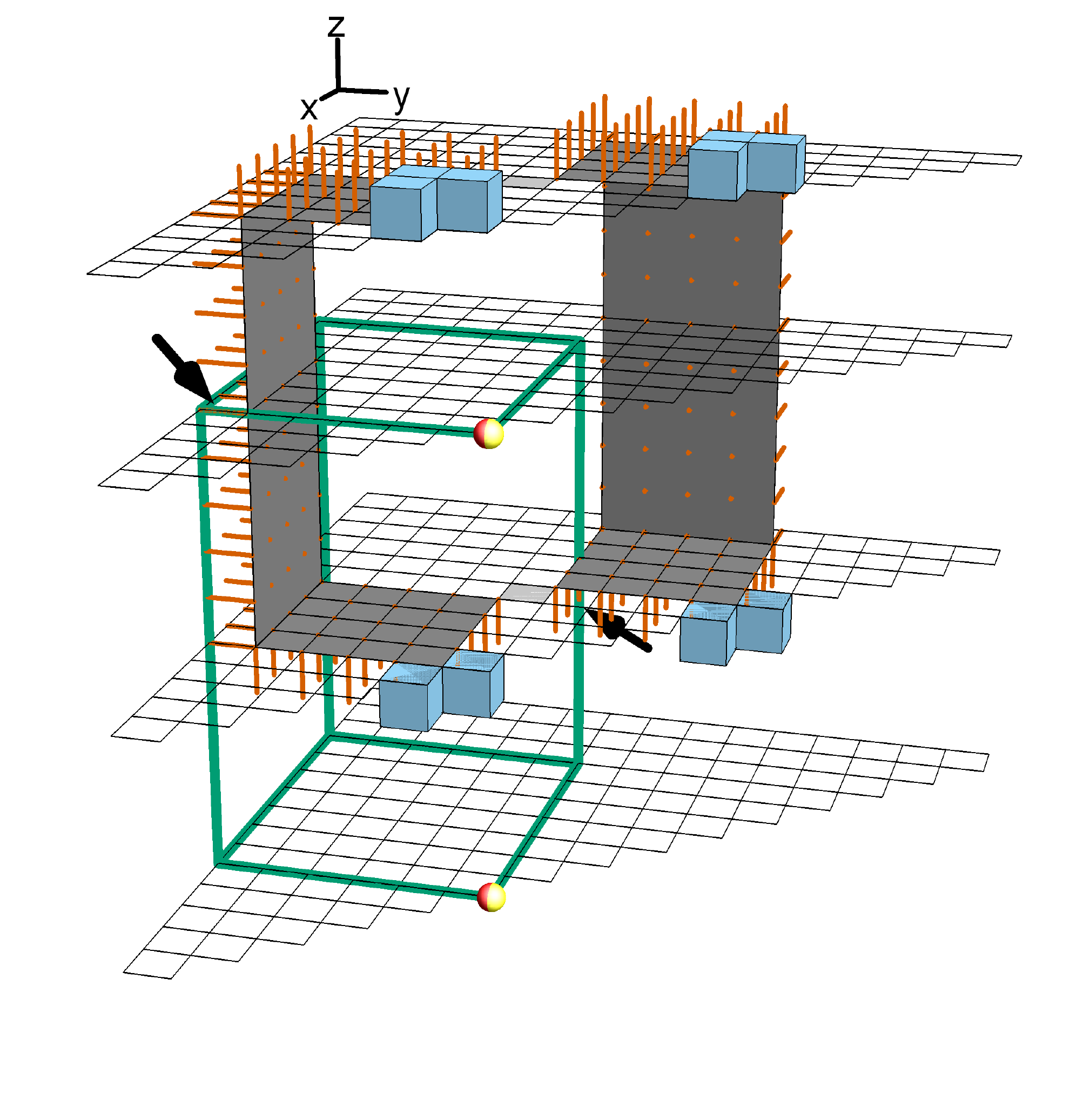}} \hfill
\subfloat[\label{fig:f0AezBBraid}]{\includegraphics[width=0.45\columnwidth]{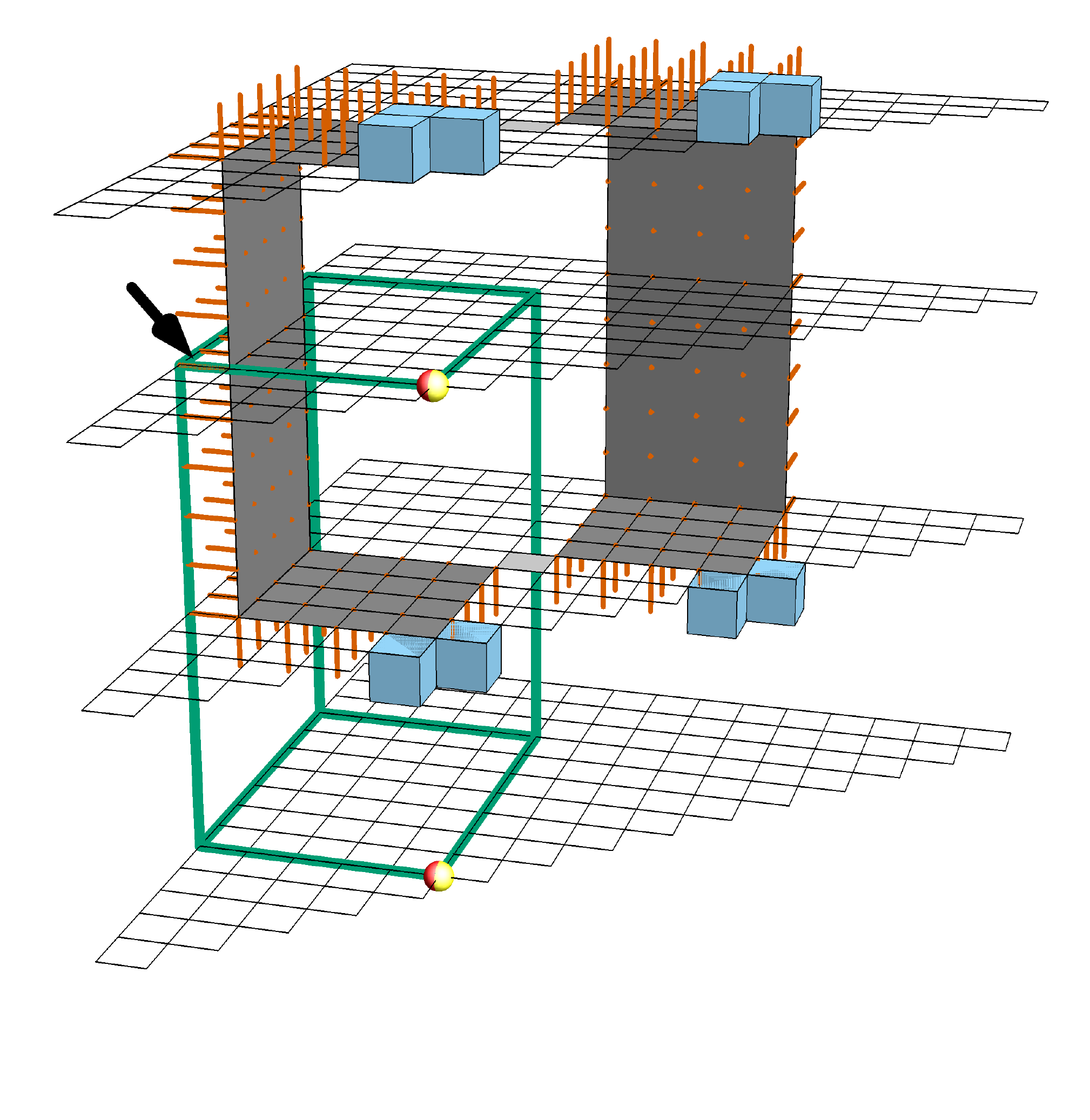}} \hfill
\subfloat[\label{fig:f0BezABraid}]{\includegraphics[width=0.45\columnwidth]{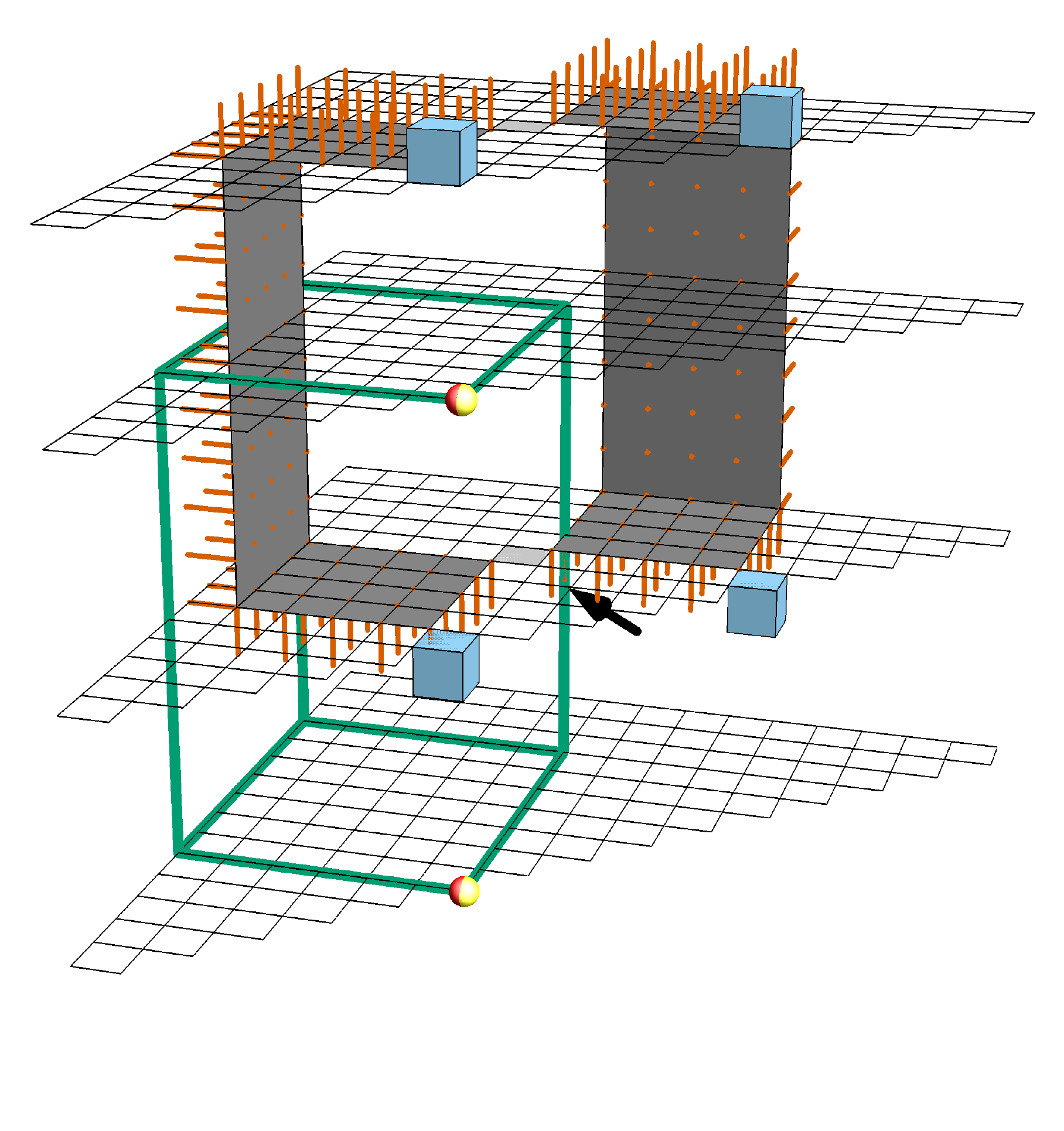}} \hfill
\subfloat[\label{fig:f0BezBBraid}]{\includegraphics[width=0.45\columnwidth]{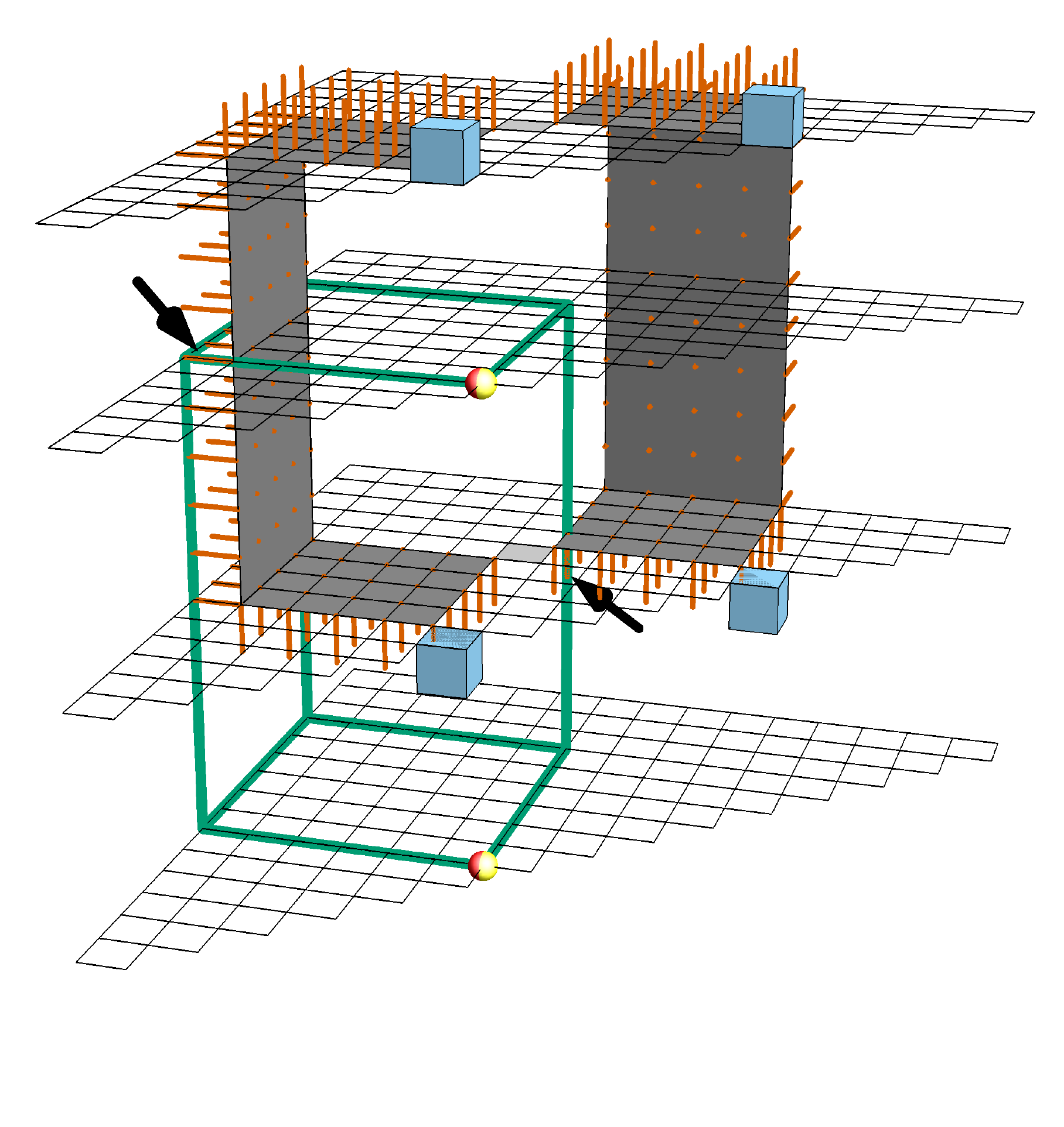}} \hfill
\caption{Boundary braiding on the (110) surface of the X-Cube model for: (a) $f_0^A$ and $e_z^A$, which is trivial, (b) $f_0^A$ and $e_z^B$, which is nontrivial, (c) $f_0^B$ and $e_z^A$, which is nontrivial, and (d) $f_0^B$ and $e_z^B$, which is trivial. Black arrows point to the links where the boundary half-cages have overlapping support.}
\label{fig:f0ez110Braiding}
\end{figure}

We now consider boundary braiding on the $(110)$ surface, starting with the BHCs. The BHC operators for the two inequivalent fractons $f^A_0$ and $f^B_0$ are displaced from each other by a single lattice site; the one for $f_0^A$ is shown in Fig.~\ref{fig:f0HalfCage}. The same is true for the inequivalent $e^{A}_z$ and $e^B_z$ excitations on the boundary, whose BHC operators are shown in Fig.~\ref{fig:ezAHalfCage} and \ref{fig:ezBHalfCage}.
 
The boundary braiding behavior of $f_0$ on the $(110)$ surface differs substantially from the naive expectations based on bulk braiding. (Recall from Sec.~\ref{subsec: bulk braiding} that there is a sense in which $f_0$ and $e_z$ braid nontrivially in the bulk.) Remarkably, $f_0^A$ particles braid \textit{trivially} with $e_z^A$, but \textit{nontrivially} with $e_z^B$, at the $(110)$ surface.  [Likewise, $f^B_0$ and $e^B_z$ braid trivially at the $(110)$ boundary, while $f^B_0$ and $e^A_z$ braid nontrivially.] An example of commuting BHCs for $f_0^A$ and $e_z^A$ is shown in Fig.~\ref{fig:f0AezABraid}, while an example of anticommuting BHCs for $f_0^A$ and $e_z^B$ is shown in Fig.~\ref{fig:f0AezBBraid}. The fact that the BHCs for $e_z^A$ and $e_z^B$ are slightly displaced with respect to one another constitutes a crucial difference due to the rigidity of the half-cages---which arises from the restricted mobility of the excitations---as we can see in Fig.~\ref{fig:f0ez110Braiding}.  

As we will prove shortly, excitations with nontrivial boundary braiding cannot condense simultaneously on the same boundary. This is the reason that there are two distinct $(110)$ boundaries with condensed $f_0$ and $e_z$ excitations; one boundary, whose Hamiltonian is given in Fig.~\ref{fig:110Boundary_f0ez}, hosts condensed $f_0^A$ and $e^A_z$ excitations, while the other (with Hamiltonian given in Fig.~\ref{fig:110Boundary_f0BezB}) hosts condensed $f_0^B$ and $e_z^B$ excitations.  It is also the reason why the boundary that hosts condensed $f^A_0$ and $f^B_0$ excitations (with Hamiltonian given in Fig.~\ref{fig:110Boundary_f0}) does not also host any condensed particles in the $e$ sector.

\section{Boundary Lagrangian Subgroup Conjecture}
\label{sec: Boundary Lagrangian Subgroup Conjecture}

Armed with these definitions, we can now state the conjecture. Given the notion of boundary braiding discussed in Sec.~\ref{sec: boundary braiding}, this conjecture is a natural extension of the result reviewed in Sec.~\ref{sec:2DReview} for gapped boundaries of Abelian topological phases in (2+1)D:

\textit{Consider an Abelian type-I fracton phase with a specific boundary termination. Let $S$ be the set of excitations for which BHCs exist. Then the boundary is gapped if and only if a ``boundary Lagrangian subgroup" (BLS) $B \subset S$ is condensed at the boundary. A BLS $B$ has the following properties:
\begin{enumerate}
\item $B$ is closed under fusion.
\item All excitations in $B$ exhibit trivial boundary braiding with one another.
\item All uncondensed excitations in $S$ exhibit nontrivial boundary braiding with an element of $B$.
\end{enumerate}
Furthermore, distinct gapped boundaries correspond to distinct BLSs modulo the condensation of excitations on the boundary for which no BHC exists.}

The definition of a BLS looks similar to the definition of a Lagrangian subgroup in (2+1)D topological order.  However, standard Lagrangian subgroups are determined using bulk braiding data, whereas BLSs are formulated in terms of BHCs and the notion of boundary braiding. These are equivalent in (2+1)D topological order due to the bulk-boundary correspondence, but the distinction is important for fractons. As we saw in the previous subsection, BHCs may exist for an excitation in one boundary geometry but not in another. This is the sense in which bulk braiding information is insufficient to determine the allowed gapped boundaries; although the BHCs may be moved into the bulk (assuming translation invariance away from the boundary), many other ``half-cages" can generally exist. The geometry of the boundary is a crucial input that determines which half-cages give us the braiding data that is required to understand the boundary.

Because our conjecture is formulated in terms of BLSs, it does not restrict the possibility of condensing particles for which no BHC exists.  Indeed, such condensation processes can occur, and we discuss examples of them at the microscopic level in Sec.~\ref{sec: Surface Condensation}. In general, these condensation processes drive surface phase transitions wherein the surface gap closes and reopens; the new surface phase that results from such a transition can result in a distinct gapped boundary that is associated with the same BLS (we discuss such an example in Sec.~\ref{sec: Surface Condensation}).  This is why the conjecture states that BLSs are in one-to-one correspondence with gapped boundaries modulo such condensation processes.  However, it is also possible to use such condensation processes to drive transitions between gapped boundaries with \textit{different} BLSs.  In all examples we have considered, it appears that condensing a 2D particle at the surface does not change the BLS, while condensing a 1D particle does.

Our task in the remainder of this section is to justify this conjecture, although we can only partially prove it since there is not yet a fully general framework for describing even Abelian type-I fracton topological phases.  We treat our examples as sufficient evidence to justify assuming that some nonempty set of excitations for which BHCs exist is condensed at any gapped boundary. We are not aware of any counterexamples. Given this assumption, justifying the conjecture means that we should justify the presence of each of the conditions (1)--(3) in the definition of a BLS.

As in (2+1)D, condition (1) is a self-evident property of condensation---if local operators can create excitations $a$ and $b$ at the surface, then the product of such operators is a local operator which creates the fusion of $a$ and $b$.

In the following subsections, we will prove that condition (2) is necessary to obtain a gapped boundary and give some intuition and simple arguments about why condition (3) should be necessary. We will give some rough discussion about the sufficiency of our criterion at the end of the section. 

\subsection{Necessity of Condition (2) (Trivial Braiding)}

The argument that this condition is necessary is a modification of Levin's argument for Abelian topological order in $(2+1)$D which we reviewed in Sec.~\ref{sec:2DReview}.

Suppose that the excitations $a$ and $b$ are both condensed at a gapped boundary, which we take without loss of generality to be at $z=0$.

Consider any BHCs $W_a$ and $W_b$ for $a$ and $b$, respectively, such that all excitations created by both operators are well-separated. Since $a$ is condensed on the boundary, there is a product $U_a$ of local unitary operators, localized near the boundary $a$ excitations, that annihilate the $a$ excitations created by $W_a$. Since the support of $W_b$ on the boundary is far from the support of $W_a$ on the boundary, we have $[U_a,W_b]=0$. Since $W_a$ creates no bulk excitations and the boundary is assumed to be gapped, we have for any ground state $\ket{G}$ that
\begin{align}
U_aW_a\ket{G} = W_aU_a\ket{G} = \ket{G},
\end{align}
since $U_a$ is local and $W_a$ has bounded support on the boundary. Likewise, there exists a product $U_b$ of local operators such that $[U_b,W_a]=0$ and
\begin{align}
U_bW_b\ket{G}=W_bU_b\ket{G}=\ket{G}.
\end{align}
We have absorbed some phases into the definitions of the operators. Since the boundary excitations created by $W_a$ and $W_b$ are far apart and $U_{a}$ and $U_{b}$ are local operators, we also have $[U_a,U_b]=0$.

It follows immediately that
\begin{equation}
U_aW_aU_bW_b\ket{G} = U_bW_bU_aW_a\ket{G} = \ket{G},
\end{equation}
but applying the definition of boundary braiding for Abelian subdimensional particles given in Eq.~\eqref{eq: abelian boundary braiding} gives
\begin{equation}
U_aW_a U_bW_b\ket{G} = e^{i\phi}\, U_aU_bW_bW_a\ket{G} = e^{i\phi}\ket{G}.
\end{equation}
Therefore $\phi=0$ (mod $2\pi$) for all $W_a$ and $W_b$, i.e., $a$ and $b$ braid trivially at the boundary.

\subsection{Necessity of Condition (3) (Maximality of the BLS)}

Condition (3) is essentially a braiding nondegeneracy condition and has an intuitive justification that is analogous to the $(2+1)$D case. There are some difficulties in trying to generalize the semiformal proof given in Ref.~\cite{LevinGappedBoundaries} of the necessity of this condition in the case of $(2+1)$D Abelian topological order; in particular, the proof relies heavily on very general knowledge of the operators that create isolated excitations [Wilson loops/string operators in $(2+1)$D], which is not yet well-understood for fractons. We therefore only give an argument, not a proof.

The starting point of the argument is analogous to the corresponding argument in \cite{LevinGappedBoundaries}. If there is an uncondensed excitation $a\in S$ at the surface, it cannot be annihilated by a local operator with support on a finite patch of the surface surrounding $a$. In particular, it cannot be annihilated by deleting from the boundary a finite region containing $a$, so $a$ is not detectable in any such finite region. Since $a$ thus cannot be detected locally, it is reasonable to think that it is possible to detect $a$ nonlocally, for example by an interference experiment involving braiding.  [In $(2+1)$D topological order, this fact is guaranteed if the braided fusion category that describes the topological order is modular~\cite{KitaevHoneycomb}.]

For a fracton system, there are two ways to have well-defined braiding at the surface. The first way is to wrap a Wilson loop around $a$ that corresponds to a 2D particle $b$ which is mobile only in the plane of the surface, similar to Fig.~\ref{fig: 2D braiding}. (Note that such a Wilson loop always exists for 2D particles; on the other hand, braiding a 1D excitation mobile in the plane of the surface around $a$ is not well-defined because such a particle cannot form a closed path.) The second way is to enclose $a$ with a BHC operator $W_c$, for some condensed particle $c$,  and annihilate the $c$ excitations locally on the boundary. The first case cannot occur.  To see why, suppose that the $a$ excitation in question has been created by a BHC $W_a$. By definition, the support of $W_a$ on the surface is localized near the excitations it creates, so every surface Wilson loop $W_b$ whose support is far from an $a$ excitation commutes with $W_a$. Since any Wilson loop $W_b$ on the boundary can be locally deformed into a Wilson loop that does not intersect with the boundary support of $W_a$, $b$ braids trivially around $a$ on the surface. The only remaining way to detect $a$ nonlocally is through boundary braiding as defined in Sec.~\ref{sec: boundary braiding}, so condition (3) must hold.

\subsection{Sufficiency of the BLS Criterion}

Any proof of the ``sufficiency" part of the conjecture would presumably require us to have a general characterization of Abelian fractons, perhaps one analogous to the $K$-matrix formalism for Abelian anyons, so we will not attempt a proof here. Our primary justification for conjecturing sufficiency of our criterion is simply that we have studied examples (including the checkerboard model\cite{VijayFractons}, Sierpinski prism model\cite{YoshidaFractal}, and $\mathbb{Z}_N$ generalizations of some of these models) and have found in all cases that picking a BLS is always sufficient to determine a gapped boundary.

\section{Surface Condensation}
\label{sec: Surface Condensation}

It is clear on physical grounds that distinct boundary Lagrangian subgroups lead to physically distinct gapped boundaries, as the action of local operators can create different particles in each case. However, the definition of a BLS does not restrict the condensation of excitations for which no BHC exists. In this section, we investigate such condensation.

As we saw in the X-cube case, the gapped boundaries obtained after condensing particles in a BLS  can be thought of as supporting (2+1)D topological order, possibly with defects that include immobile or 1D excitations. What happens if we further condense excitations on this surface?

Condensing a 2D excitation that is mobile in the plane of the boundary may be treated as usual in topological order---condensing an excitation confines every 2D particle that braids nontrivially with it, in the sense of Fig.~\ref{fig: 2D braiding}. This can either drive the system to a new (gapped) surface topological order or trivialize the 2D topological order on the top surface.

\begin{figure}
\centering
\includegraphics[width=8cm]{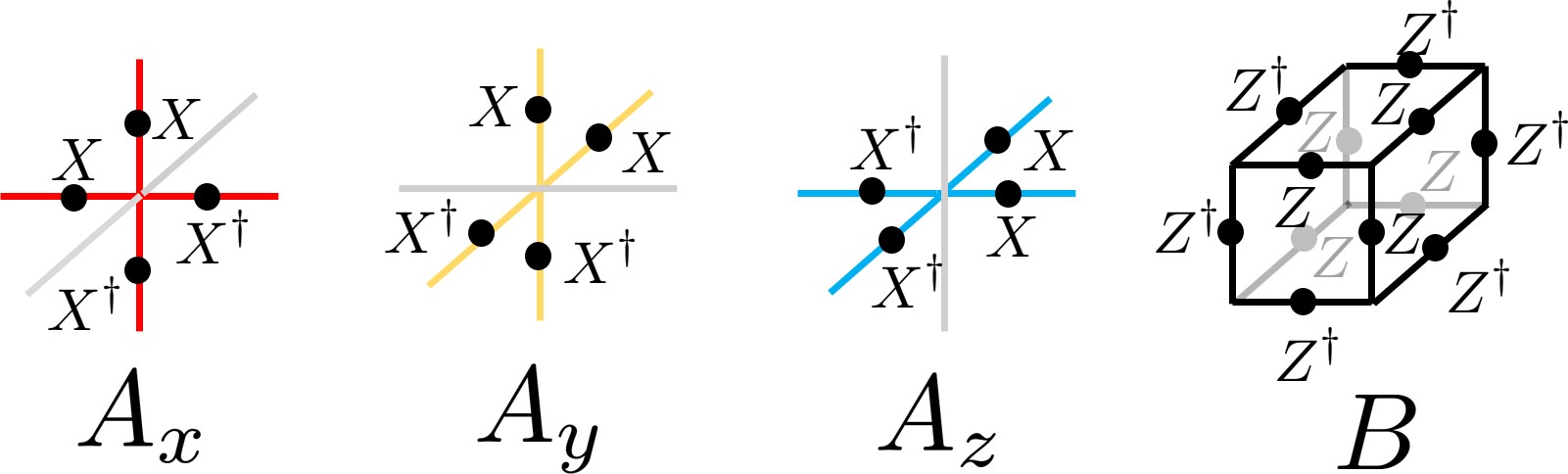}
\caption{Terms in the bulk Hamiltonian of the $\mathbb{Z}_N$ X-cube model. $N$-state systems live on the links and $X$ and $Z$ are generalized Pauli matrices.}
\label{fig:XCubeZNBulk}
\end{figure}

\begin{figure}
\centering
\subfloat[\label{fig:XCubeZ4ECondensed}]{\includegraphics[width=\columnwidth]{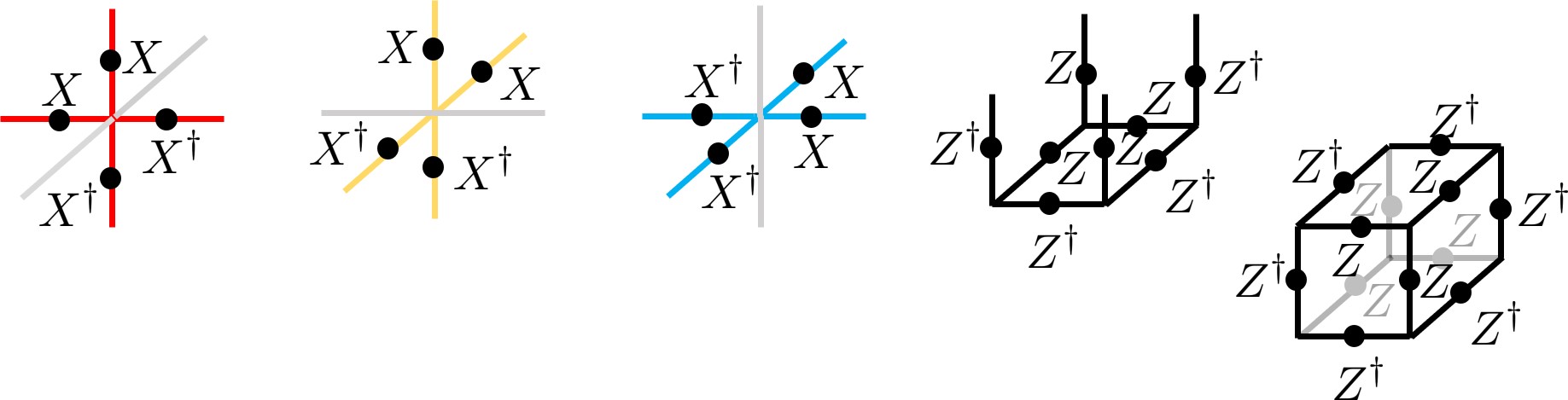}} \hfill
\subfloat[\label{fig:XCubeZ4WithTC}]{\includegraphics[width=\columnwidth]{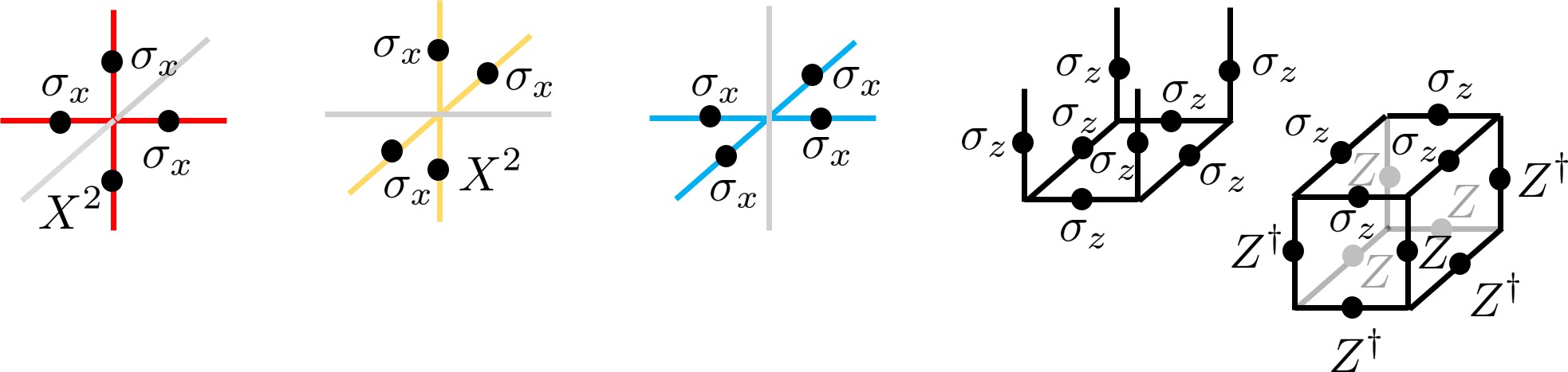}}
\caption{Terms in boundary Hamiltonian of the $\mathbb{Z}_4$ X-cube model with (a) $e_z^{1,2,3}$ condensed and (b) with the in-plane $e^2$ excitation condensed on the surface. The bottom face of the open cube operator and the top face of the closed one live in the same layer of the system. All spins are $\mathbb{Z}_4$ in (a) and all spins with $\sigma_{x,z}$ operators are effective $\mathbb{Z}_2$ degrees of freedom in (b).}
\label{fig:Z4Boundaries}
\end{figure}

As an example, consider the $(001)$ boundary of the $\mathbb{Z}_4$ X-cube model with $e^{1,2,3}_z$ condensed (here the superscripts indicate the $\mathbb{Z}_4$ charge). The bulk Hamiltonian (for general $\mathbb{Z}_N$) is a small modification, shown in Fig.~\ref{fig:XCubeZNBulk}, of the $\mathbb{Z}_2$ case given in Eq. \eqref{eqn:XCubeH}. Here $X$ and $Z$ are generalized Pauli matrices obeying $XZ = e^{2\pi i/N}ZX$. The boundary Hamiltonian for this boundary condition is shown in Fig. \ref{fig:XCubeZ4ECondensed}.

This results in a gapped boundary that is analogous to the $(ee)$ boundary of the $\mathbb Z_2$ X-cube model studied in Sec.~\ref{subsec: (001) Boundary}.  In particular, the condensation of $e^{1,2,3}_z$ allows one to identify the $e_x^{1,2,3}$ and $e^{1,2,3}_y$ excitations on the surface via fusion with the condensate.  As a result, there are two types of 2D excitations on the surface, $e_x^{1,2,3}\sim e_y^{1,2,3}\equiv e^{1,2,3}$ and $m_{xy}^{1,2,3}$, which obey the braiding and fusion rules of the $\mathbb{Z}_4$ toric code in 2D. As in the $\mathbb Z_2$ case, there are additional subdimensional surface excitations that can be viewed as defects.

Consider first condensing the in-plane excitation $e^1$, which necessarily condenses $e^{2,3}$ as well owing to the fusion rules $e^1\times e^2\sim e^3$ and $e^3\times e^1\sim1$. This is done physically by adding a term $-h\sum (Z + Z^{\dagger})$ to the Hamiltonian on all of the surface links and taking $h$ large. It is easy to check that at lowest order in perturbation theory, the effective Hamiltonian in the low-energy subspace, which has all surface spins in the $Z=+1$ state, is exactly the same as the original Hamiltonian, but with the boundary at $z=-1$ instead of $z=0$. That is, the surface layer trivialized; this is to be expected because condensing $e^1$ in the $\mathbb{Z}_4$ toric code trivializes the topological order.

However, there is a nontrivial anyon condensation procedure available: condensing only $e^2$. Following the usual rules for anyon condensation, such a procedure should turn the surface topological order into $\mathbb{Z}_2$ toric code order. We show this explicitly.

This condensation procedure is implemented by adding  a term $-h\sum Z^2$ acting on all surface links and taking $h$ large, since these $Z^2$ operators create $e^2$ excitations. In this limit, every surface spin is pinned to the $Z = Z^{\dagger}= \pm 1$ subspace, so the surface spins are now effectively spin-1/2 particles. The $\mathbb{Z}_4$ operators $Z,Z^{\dagger}$ become effective Pauli $\sigma_z$ operators on this subspace and likewise $X^2 \rightarrow \sigma_x$, while $X$ mixes the low-energy subspace with higher-energy ones. Degenerate perturbation theory leads to an effective boundary commuting-projector Hamiltonian whose terms are shown in Fig.~\ref{fig:XCubeZ4WithTC}.

It is straightforward to check that operators acting in the second (next-to-top) layer create the same excitations as they do when the $e^2$ particle in the first (top) layer is not condensed. In the top layer, the 2D excitations form $\mathbb{Z}_2$ toric code topological order, but there are additional 0D or 1D excitations much like the surface of the $\mathbb{Z}_2$ X-cube model.

Condensing $e^2$ on the surface after condensing $e^{1,2,3}_z$ yields a gapped boundary that is physically distinct from the original boundary because it changes the set of 2D particles that are deconfined at the boundary. However, one can check that the set of condensed excitations with BHCs does not change after condensing the $e^2$ excitation on the surface.  Thus, condensing $e^2$ did not change the BLS, although it did lead to a distinct gapped surface theory.

We have shown that condensing 2D particles on the surface yields results expected from standard (2+1)D topological order. Another option is to try to condense 0D or 1D particles that live on the surface. However, in the X-cube model, for every 0D or 1D particle, there exists a bound state of those subdimensional particles that is mobile in the $z$-direction. Therefore, such a condensation procedure always condenses an excitation that \textit{does} have a BHC operator, which (depending on the precise nature of the terms added to the Hamiltonian) either leads to a gapless boundary or drives a transition to a gapped boundary with a different BLS. For example, attempting to condense $e_x$ on the $(mm)$ surface of the $\mathbb{Z}_2$ X-cube model necessarily condenses $q_{xz}$, which is a bound state of two $e_x$ excitations separated in the $y$ direction. Therefore, this process either leads to a gapless boundary or drives the system to $(em)$ boundary conditions.

In the type-I models that we are familiar with, for every boundary that we have checked, every excitation does one of three things: it has a BHC, it has a bound state that has a BHC, or it is a 2D excitation on the boundary. We have accounted for all such cases. It would be interesting to find a type-I model and boundary for which this is not true. (For our present purposes, we count the Sierpinski prism model~\cite{Castelnovo2012,YoshidaFractal} as type-II due to the presence of operators with fractal support that create isolated excitations.)

\section{Discussion}

By studying gapped boundaries of fracton phases, we have greatly clarified the notion of braiding for fractons. In particular, we have shown that the natural definition of bulk braiding is non-reciprocal and clarified the role that geometry plays in fracton braiding. This led to the idea of boundary braiding, which accounts for the boundary geometry. We further proved that boundary braiding has physical significance, in that it restricts the set of excitations that may be simultaneously condensed at gapped boundaries.

In generating our conjecture, several open questions arose, such as: When do BHCs exist? When is bulk braiding non-reciprocal? Is braiding nondegeneracy important in a fracton phase, and if so is there some generalization of modular invariance that justifies it? What (anomalous?) surface topological order is allowed in a fracton system, and how do we describe subdimensional excitations in the surface theory? Such questions could help guide the search for a general theory of fractons.

Another obviously interesting challenge would be to provide further evidence for (or counterexamples to) our conjecture. Analyzing Chamon's model \cite{ChamonGlass} would be a natural step, although the geometry of that model is quite complicated. One could also examine Chern-Simons-like versions of higher-rank gauge theories \cite{SlagleFieldTheory}, which provide a low-energy field-theoretic description  of some fracton models, in the presence of a spatial boundary; perhaps some $K$-matrix-like classification procedure could result from such considerations.

Another interesting direction would be to extend our discussion to type-II models. A priori there is no obvious obstruction, but this does turn out to present a challenge. In the case of Haah's code\cite{HaahsCode}, one can check that there are cage-like operators with well-defined nontrivial (and reciprocal) bulk braiding. However, due to the particle mobility constraints, the natural candidates for BHC operators create a mobile string of excitations instead of isolated point excitations. It is possible that BHC operators as we have defined them actually exist in Haah's code, but we have not successfully found any such operators. Even so, gapped boundaries still exist and have condensed point-like excitations, so there may exist modifications to our arguments such that they apply to Haah's code as well.

However, a type-II model introduced by Halasz and Hsieh~\cite{HsiehFractonsPartons} poses more severe problems. In that case, it is not clear how to define bulk cage operators or BHCs that are associated only to one of the two excitation types in the model, which makes braiding extremely ambiguous. These examples, taken together with the fact that cage operators exist in all type-I models that we are aware of, provide circumstantial evidence that the existence of cage operators and BHC operators as we have defined them is deeply connected to the presence of mobile excitations. We hope that this point is investigated further.

\begin{acknowledgments}
The authors thank Dave Aasen, Maissam Barkeshli, Liang Fu, Abhinav Prem, Kevin Slagle, Nicolas Tarantino, Alex Thomson, and Guanyu Zhu for helpful discussions. This work is supported by the Laboratory for Physical Sciences and Microsoft. T.I.~acknowledges a JQI postdoctoral fellowship.
\end{acknowledgments}

\appendix

\section{Definitions of Bulk Concepts}
\label{app:bulkDefinitions}

\begin{figure}[t!]
\centering
\includegraphics[width=.6\columnwidth]{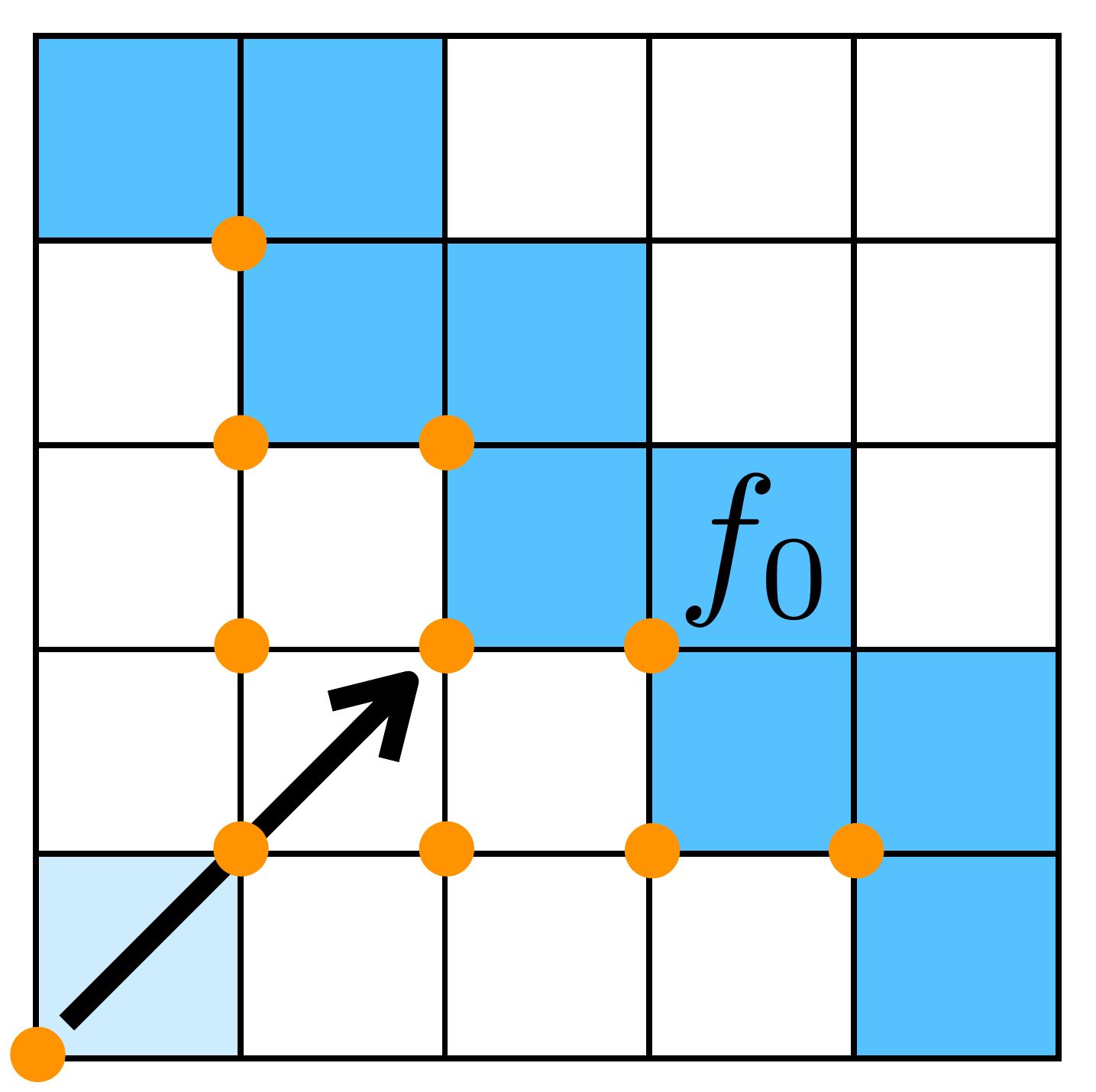}
\caption{Top view of a fracton cascading in the $(110)$ direction (the $z$-direction is perpendicular to the page). Orange circles represent $X$ operators applied to the link sticking out of the page.  If the fracton was initially present in the bottom-left corner (light blue square), then the arrangement of operators shown cascades the fracton in the direction of the arrow.  Note that cascading in a certain direction does not amount to mobility in that direction (although the reverse is true); for example, the depicted cascading process costs finite energy and is thus disallowed at zero temperature.}
\label{fig:f0Cascade}
\end{figure}

In the main text, we alluded to two concepts which are not critical for the development of our BLS criterion but which may be useful for studying the bulk properties of fractons: bulk cage operators and ``cascading." We define these terms presently, assuming for simplicity that the system has no boundary and is in the thermodynamic limit.

\textit{Definition}: A (bulk) \textbf{cage operator} for an excitation $a$ is an operator $\mathcal{C}_a$ such that:
\begin{enumerate}
\item $\mathcal{C}_a$ has finite support
\item $\mathcal{C}_a$ commutes with the Hamiltonian
\item $\mathcal{C}_a = W_a \bar{W}_a$ where $W_a$ and $\bar{W}_a$ obey the following conditions:
\begin{enumerate}
\item Both $W_a$ and $\bar{W}_a$, when acting far away from other excitations, create some (non-empty) set of isolated $a$ and $\bar{a}$ excitations
\item Neither $W_a$ nor $\bar{W}_a$ create any excitation types other than $a$ and $\bar{a}$ when acting far away from other excitations
\item The supports of $W_a$ and $\bar{W}_a$ overlap only near the excitations they create.
\end{enumerate}
\end{enumerate}

It is straightforward to see that this generalizes the notion of Wilson loops, where in particular $W_a$ and $\bar{W}_a$ are (open) Wilson lines. Since $\mathcal{C}_a$ does not create any excitations, the existence of $W_a$ and $\bar{W}_a$, which we call ``bulk half-cages" or just ``half-cages," are required to associate $\mathcal{C}_a$ with a specific excitation type $a$.

To define ``cascading" suppose that we have a state $\ket{\psi}$ that has an excitation of type $a$ at the origin, and further suppose that this excitation is isolated, in that all other excitations are many correlation lengths away. For simplicity of language we will imagine that $a$ is the only excitation in the system and that the correlation length is zero, but these assumptions are not necessary. Let $\hat{\bv{n}}$ be an arbitrary unit vector, and choose coordinates such that $\hat{\bv{n}}$ is a coordinate axis.

\textit{Definition}: The excitation $a$ ``\textbf{cascades}" in the $+\hat{\bv{n}}$ direction if there exists a local operator $\mathcal{O}$ such that
\begin{itemize}
\item $\mathcal{O}\ket{\psi}$ has no excitations at $n \leq 0$.
\item The only excitations present in $\mathcal{O}\ket{\psi}$ are of type $a$ or $\bar{a}$.
\end{itemize}
Obviously a particle that is mobile in the $\hat{\bv{n}}$ direction cascades in both the $\pm \hat{\bv{n}}$ directions.

The ability of an excitation to cascade in the $\hat{\bv{n}}$-direction is related to the existence of a BHC for that excitation on the boundary normal to $\hat{\bv{n}}$. For example, in the X-cube model, $f_0$ does not cascade in the $z$ (resp.~$x$, $y$) direction, and there is no BHC for $f_0$ on the $(001)$ [resp.~$(100)$, $(010)$] boundary. However, it does cascade in the $(110)$ direction, as can be seen in Fig.~\ref{fig:f0Cascade}, and there is a well-defined BHC for $f_0$ on the $(110)$ boundary as shown in Fig.~\ref{fig:f0HalfCage}.

\bibstyle{apsrev4-1} \bibliography{references}

\end{document}